\documentclass[11pt, a4paper]{article}

\usepackage{graphicx}
\usepackage{bm}
\usepackage{amsmath, amssymb,theorem,verbatim}
\usepackage[round]{natbib} 
\usepackage{setspace}
\usepackage{multirow} 
\usepackage{hyperref}
\usepackage{bm}
\usepackage{algpseudocode}
\usepackage{textcomp}

\long\def\comment#1{}

\newtheorem{theorem}{Theorem}[section]

\newtheorem{prop}{Proposition}[section]

\newtheorem{corollary}{Corollary}[section]

\newtheorem{assumption}{Assumption}[section]

\begin{document}

\doublespacing

\author{Piotr Fryzlewicz\thanks{Department of Statistics, London School of Economics, Houghton Street, London WC2A 2AE, UK. Email: \href{mailto:p.fryzlewicz@lse.ac.uk}{p.fryzlewicz@lse.ac.uk}.}}

\title{Narrowest Significance Pursuit: inference for multiple change-points in linear models}





\oddsidemargin=0.25in
\evensidemargin=0in
\textwidth=6in
\headheight=0pt
\headsep=0pt
\topmargin=0in
\textheight=9in

\maketitle

\begin{abstract}
We propose Narrowest Significance Pursuit (NSP), a general and flexible methodology for automatically detecting localised regions in data sequences which each must contain a change-point (understood as an abrupt change in the parameters of an underlying linear model), at a prescribed global significance level. NSP works with a wide range of distributional assumptions on the errors, and guarantees important stochastic bounds which directly yield exact desired coverage probabilities, regardless of the form or number of the regressors.
In contrast to the widely studied ``post-selection inference'' approach, NSP paves the way for the
concept of ``post-inference selection''. An implementation is available in the R package \href{https://CRAN.R-project.org/package=nsp}{{\tt nsp}}.

\vspace{5pt}

\end{abstract}

\section{Introduction}

We propose a new generic methodology for determining, for a given data sequence and at a given global significance
level, localised regions of the data that each must contain a change-point. We define a change-point in $Y_t$ on an interval $[s,e]$ as an abrupt departure, 
on that interval, from a linear model for $Y_t$ with respect to pre-specified regressors. We now give examples of scenarios covered by
the proposed methodology.

\begin{description}
\item[Scenario 1.] {\em Piecewise-constant signal plus noise model.}
\begin{equation}
\label{eq:signoise}
Y_t = f_t + Z_t, \quad t = 1, \ldots, T,
\end{equation}
where $f_t$ is a piecewise-constant vector with an unknown number $N$ and locations $0 = \eta_0 < \eta_1 <  \ldots < \eta_N < \eta_{N+1} = T$ of change-points, and $Z_t$ is zero-centred noise.
The location $\eta_j$ is a change-point if $f_{\eta_j-1} = f_{\eta_j}$ but $f_{\eta_j} \neq f_{\eta_j+1}$.
\item[Scenario 2.] {\em Piecewise-polynomial (e.g. piecewise-constant or piecewise-linear) signal plus noise model.}
In (\ref{eq:signoise}), $f_t$ is a piecewise-polynomial
vector, in which the polynomial pieces have a fixed degree $q \ge 0$, assumed known to the analyst.
The location $\eta_j$ is a change-point if $f_t$ can be described as a polynomial vector of degree $q$ on $[\eta_j-q-1, \eta_j]$, but not on $[\eta_j-q, \eta_j+1]$.
\item[Scenario 3.] {\em Linear regression with piecewise-constant parameters.}
For a given design matrix $X = (X_{t,i})$, $t = 1, \ldots, T$, $i=1,\ldots, p$, the response $Y_t$ follows the model
\begin{equation}
\label{eq:reg}
Y_t = X_{t,\cdot}\beta^{(j)} + Z_t\quad\text{for}\quad t = \eta_j+1, \ldots ,\eta_{j+1},\quad j = 0, \ldots, N,
\end{equation}
where the parameter vectors $\beta^{(j)} = (\beta^{(j)}_1, \ldots, \beta^{(j)}_p)'$ are such that $\beta^{(j)} \neq \beta^{(j+1)}$.

\end{description}
Each of these scenarios is a generalisation of the preceding one. We permit a broad range of distributional assumptions for $Z_t$, from i.i.d. Gaussianity to autocorrelation, heavy tails and heterogeneity.
We now review the existing literature on uncertainty in multiple change-point problems which seeks to make 
confidence statements about the existence or locations of change-points in particular regions of the data,
or significance statements about their importance.

In the i.i.d. Gaussian piecewise-constant model, SMUCE \citep{fms14} estimates the number $N$ of change-points as the minimum among those candidate fits $\hat{f}_t$
for which the empirical residuals pass a certain test at level $\alpha$. An issue for SMUCE, discussed e.g. in \cite{css14}, is that 
the smaller the significance level $\alpha$, the more lenient
the test on the empirical residuals, and therefore the higher the risk of underestimating $N$. This
leads to the counter-intuitive behaviour of the coverage properties of SMUCE illustrated in \cite{css14}.
SMUCE$_2$ \citep{css14} remedies this issue, but still requires that the number of estimated change-points agrees with the truth
for successful coverage, which puts it at risk of being unable to cover the truth with a high nominal probability requested by the user. In the approach taken in this paper, this issue
does not arise as we shift the inferential focus away from $N$. SMUCE is extended to heterogeneous Gaussian noise
in \cite{psm17} and to dependent data in \cite{dsv18}.

Some authors approach uncertainty quantification for multiple change-point problems from the point of view of post-selection inference
(PSI, a.k.a. selective inference); these include \cite{hgst18}, \cite{hlgst18}, \cite{jfw20} and \cite{dtst20}. To ensure valid inference, PSI conditions on
many aspects of the estimation process, which tends to produce $p$-values with somewhat complex definitions. PSI also does not permit
the selection of the tuning parameters of the inference procedure from the same data. Useful as they are in assessing the significance of previously estimated change-points, these PSI approaches share the following 
features: (a) they do not consider uncertainties in estimating change-point locations, (b) they do not provide regions of globally significant change in the data, (c)
they define significance for each change-point separately, as opposed to globally, (d) they rely on a particular base change-point detection method
with its potential strengths or weaknesses.
Our approach contrasts with these features; in particular, in contrast to PSI,
it can be described as enabling ``post-inference selection'', as we argue later on.



Some authors provide simultaneous asymptotic distributional results for the distance between the estimated change-point locations and the truth. In the linear regression context, this is done in \cite{bp98, bp03}, and in the piecewise-constant signal plus noise model -- in \cite{km11}.
These approaches are asymptotic, conditional on the estimated change-point locations, and involve unknown quantities. In contrast,
our methodology has a finite-sample nature, makes no assumptions on the signal, is unconditional and automatic.
A further discussion of the differences between our approach and that of \cite{bp98, bp03} can be found in Section \ref{sec:suppl:lit} of the appendix.

Inference for multiple change-points is also sometimes posed as control of the False Discovery Rate (FDR), see e.g. 
\cite{lm16}, \cite{hnz13} and \cite{chs19}, but this approach is
focused on the number of change-points rather than on their locations.


The objective of our methodology, called ``Narrowest Significance Pursuit" (NSP), is to automatically detect localised regions of the data $Y_t$, each of which
must contain at least one change-point (in a suitable sense determined by the given scenario), at a prescribed global significance level. NSP performs unconditional 
inference without
change-point location estimation, and proceeds as follows. A number $M$ of intervals are drawn from the index domain $[1, \ldots, T]$,
with start- and end-points chosen over an equispaced deterministic grid. On each interval drawn, $Y_t$ is then checked to see whether or not it locally conforms to
the prescribed linear model, with any set of parameters. This check is performed through estimating the parameters of the given linear model locally by minimising a particular multiresolution sup-norm loss, and testing
the residuals from this fit via the same norm; self-normalisation is involved if necessary. In the first greedy stage, the shortest interval (if one exists) is chosen on which the test is violated at a certain global significance level $\alpha$.
In the second greedy stage, the selected interval is searched for its shortest sub-interval on which a similar test is violated. This sub-interval is then chosen
as the first region of global significance, in the sense that it must (at a global level $\alpha$) contain a change-point, or otherwise the local test would not have rejected the linear model.
The procedure then recursively draws $M$ intervals to the left and to the right of the chosen region (with or without overlap), and stops when there are no further local regions of global
significance.

\cite{fls20}, in the piecewise-constant signal plus i.i.d. Gaussian noise model, approximate the tail probability of the maximum CUSUM statistic over all sub-intervals of the data. They then
propose an algorithm, in a few variants, for identifying short, non-overlapping segments of the data on which the local CUSUM exceeds the derived tail bound, and hence the segments identified
must contain at least a change-point each, at a given significance level. \cite{fs20} present results of similar nature for a Gaussian model with lag-one autocorrelation, linear trend, and features
that are linear combinations of continuous, piecewise differentiable shapes.
The most important high-level differences between NSP and these two approaches are that (a) NSP is ready for use with any user-provided design matrix $X$,
and this requires no new calculations or coding, and yields correct coverage probabilities
in finite samples of any length; 
 (b) NSP searches for any deviations from local model linearity with respect to
the regressors provided; (c) 
NSP is able to handle regression with autoregression practically in the same way as without, in a stable manner and on arbitrarily short intervals, and does not need
accurate estimation of the unknown (nuisance) AR coefficients. We expand on these points in Section \ref{sec:suppl:lit} of the appendix.

NSP has other distinctive features in comparison with the existing literature.
It is specifically constructed to target the shortest possible significant intervals at every stage of the procedure, and to explore as many intervals as possible while remaining computationally
efficient.
NSP furnishes exact coverage statements, at a prescribed global significance level, for any finite sample sizes, and works in the same way regardless of the scenario and for any given regressors $X$.
Also, thanks to the fact that the multiresolution sup-norm used in NSP can be interpreted as H{\"o}lder-like norms on certain function spaces, NSP naturally extends to the cases of unknown or heterogeneous distributions of $Z_t$ via self-normalisation.
Finally, 
if simulation needs to be used to determine critical values for NSP, then this can be done in a computationally efficient manner.

Section \ref{sec:nspbasic} introduces the NSP methodology and provides the relevant finite-sample coverage theory. Section \ref{sec:arsn} extends this to NSP
under self-normalisation and in the additional presence of autoregression. 
Section \ref{sec:consist} provides finite-sample and traditional large-sample detection consistency and rate optimality results for NSP in Scenarios 1 and 2.
Section \ref{sec:num} provides comparative simulations and extensive numerical examples under a variety of settings. Section \ref{sec:data}
describes two real-data case studies. Complete R code implementing NSP is available in the R package \href{https://CRAN.R-project.org/package=nsp}{{\tt nsp}}. There is an appendix, whose contents are mentioned at appropriate places in the paper. Proofs of our theoretical results are in the appendix.

\section{The NSP inference framework}
\label{sec:nspbasic}


Throughout the section, we use the language of Scenario 3, which includes Scenarios 1 and 2 as special
cases. In Scenario 1, the matrix $X$ in (\ref{eq:reg}) is of dimensions $T \times 1$ and has all entries equal to 1.
In Scenario 2, the matrix $X$ is of dimensions $T \times (q+1)$ and its $i$th column is given by $(t/T)^{i-1}$, $t = 1, \ldots, T$.
Scenario 4 (for NSP in the additional presence of autoregression), which generalises Scenario 3, is dealt with in Section \ref{sec:ar}.

\subsection{Generic NSP algorithm}
\label{sec:genalgo}

We start with a pseudocode definition of the NSP algorithm, in the form of a recursively defined function NSP.
In its arguments, $[s,e]$ is the current interval under consideration and at the start of the procedure, we have $[s,e] = [1,T]$;
$Y$ (of length $T$) and $X$ (of dimensions $T \times p$) are as in the model formula (\ref{eq:reg}); $M$ is the number of sub-intervals of $[s,e]$ drawn; $\lambda_\alpha$
is the threshold corresponding to the global significance level $\alpha$ (typical values for $\alpha$ would be 0.05 or 0.1) and $\tau_L$ (respectively $\tau_R$) is a functional parameter
used to specify
the degree of overlap of the left (respectively right) child interval of $[s,e]$ with respect to the region of significance identified within $[s,e]$, if any.
The no-overlap case would correspond to $\tau_L = \tau_R \equiv 0$. In each recursive call on a generic interval $[s,e]$, NSP
adds to the set $\mathcal{S}$ any globally significant local regions (intervals) of the data identified within $[s,e]$ on which $Y$ is deemed 
to depart significantly (at global level $\alpha$) from linearity with respect to $X$.
We provide more details underneath the pseudocode below.

\vspace{10pt}

\begin{algorithmic}[1]
\Function{NSP}{$s$, $e$, $Y$, $X$, $M$, $\lambda_\alpha$, $\tau_L$, $\tau_R$}
\If {$e-s < 1$}
\State RETURN
\EndIf
\If {$M \ge \frac{1}{2}(e-s+1)(e-s)$}
\State ${M} := \frac{1}{2}(e-s+1)(e-s)$ 
\State draw all intervals $[s_m, e_m] \subseteq [s, s+1, \ldots, e]$, $m=1, \ldots, {M}$, s.t. $e_m-s_m \ge 1$
\Else
\State draw a representative (see description below) sample of intervals $[s_m, e_m] \subseteq [s, s+1, \ldots, e]$, $m=1, \ldots, {M}$, s.t. $e_m-s_m \ge 1$
\EndIf
\For{$m \gets 1,\ldots, M$}
\State $D_{[s_m, e_m]} :=$ \textsc{DeviationFromLinearity}$(s_m, e_m, Y, X)$
\EndFor
\State $\mathcal{M}_0 := \arg\min_m\{ e_m - s_m\,\,:\,\, m = 1, \ldots, M;\,\,D_{[s_m, e_m]} > \lambda_\alpha    \}$
\If {$|\mathcal{M}_0| = 0$}
\State RETURN
\EndIf
\State $m_0 := $\textsc{AnyOf}$(\arg\max_m\{ D_{[s_m, e_m]}\,\,:\,\, m\in \mathcal{M}_0\})$
\State $[\tilde{s}, \tilde{e}] :=$\textsc{ShortestSignificantSubinterval}$(s_{m_0}, e_{m_0}, Y, X, M, \lambda_\alpha)$
\State add $[\tilde{s}, \tilde{e}]$ to the set $\mathcal{S}$ of significant intervals
\State NSP$(s, \tilde{s}+\tau_L(\tilde{s}, \tilde{e}, Y, X), Y, X, M, \lambda_\alpha, \tau_L, \tau_R)$
\State NSP$(\tilde{e}-\tau_R(\tilde{s}, \tilde{e}, Y, X), e, Y, X, M, \lambda_\alpha, \tau_L, \tau_R)$
\EndFunction
\end{algorithmic}

The NSP algorithm is launched by the pair of calls: $\mathcal{S} := \emptyset$;\,\, NSP$(1, T, Y, X, M, \lambda_\alpha, \tau_L, \tau_R)$.
On completion, the output of NSP is in the variable $\mathcal{S}$. We now comment on the NSP function line by line.
In lines 2--4, execution is terminated for intervals that are too short.
In lines 5--10, a check is performed to see
if $M$ is at least as large as the number of all sub-intervals of $[s,e]$. If so, then $M$ is adjusted accordingly, and all sub-intervals are stored in $\{[s_m, e_m]\}_{m=1}^M$.
Otherwise, a sample of $M$ sub-intervals $[s_m, e_m] \subseteq [s,e]$ is drawn in which $s_m$ and $e_m$ are all possible pairs from an (approximately) equispaced grid on $[s,e]$ which permits at least $M$ such sub-intervals (a random alternative, in which $s_m$ and $e_m$ are obtained uniformly with replacement from $[s, e]$, is possible).

In lines 11--13,
each sub-interval $[s_m, e_m]$ is checked to see to what extent the response on this sub-interval (denoted by $Y_{s_m:e_m}$) conforms to the linear model (\ref{eq:reg}) with
respect to the set of covariates on the same sub-interval (denoted by $X_{s_m:e_m,\cdot}$). 
This core step of the NSP algorithm
is described in more detail in Section \ref{sec:iidg}.

In line 14, the measures of deviation obtained in line 12 are tested against threshold $\lambda_\alpha$, chosen to guarantee global significance level $\alpha$. How to choose
$\lambda_\alpha$ depends (only) on the distribution of $Z_t$; this question is addressed in Section \ref{sec:gauss} below and in Sections \ref{sec:lt} and \ref{sec:suppl:sim} of the appendix. The shortest sub-interval(s) $[s_m, e_m]$ for which the
test rejects the local hypothesis of linearity of $Y$ versus $X$ at global level $\alpha$ are collected in set $\mathcal{M}_0$. In lines
15--17, if $\mathcal{M}_0$ is empty, then the procedure decides that it has not found regions of significant deviations from linearity on $[s,e]$, and stops on this interval as a consequence.
Otherwise, in line
18, the procedure continues by choosing the sub-interval, from among the shortest significant ones, on which the deviation from linearity has been the largest.
The chosen interval is denoted by $[s_{m_0}, e_{m_0}]$.

In line 19, $[s_{m_0}, e_{m_0}]$ is searched for its shortest significant sub-interval, i.e. the shortest sub-interval on which the hypothesis of linearity is rejected locally at a global level $\alpha$.
Such a sub-interval certainly exists, as $[s_{m_0}, e_{m_0}]$ itself has this property. The structure of this search again follows the workflow
of the NSP procedure; more specifically, it proceeds by executing lines 2--18 of NSP, but with $s_{m_0}, e_{m_0}$ in place of $s, e$. The chosen interval is denoted by $[\tilde{s}, \tilde{e}]$.
This two-stage search (identification of $[s_{m_0}, e_{m_0}]$ in the first stage and of $[\tilde{s}, \tilde{e}] \subseteq [s_{m_0}, e_{m_0}]$ in the second stage) is crucial in NSP's
pursuit to force the identified intervals of significance to be as short as possible, without unacceptably increasing the computational cost. The importance of this two-stage solution
is illustrated in Section \ref{sec:imp2stage} of the appendix. In line 20, the selected interval $[\tilde{s}, \tilde{e}]$ is added to the output set $\mathcal{S}$.

In lines 21--22, NSP is executed recursively to the left and to the right of the detected interval $[\tilde{s}, \tilde{e}]$. However, we optionally allow for some overlap with $[\tilde{s}, \tilde{e}]$.
The overlap, if present, is a function of $[\tilde{s}, \tilde{e}]$ and, if it involves detection of the location of a change-point within $[\tilde{s}, \tilde{e}]$, then it is also a function of $Y, X$.
Executing NSP without an overlap, i.e. with $\tau_L = \tau_R = 0$, means that the procedure runs, in each recursive step, wholly on data sections between (and only including the end-points of) the previously detected intervals of significance. This ensures that the intervals of significance returned by NSP are non-overlapping; however, this also reduces the amount of data that the procedure is able to use at each recursive stage, which shows the importance of optionally allowing non-zero overlaps $\tau_L$ and $\tau_R$ in NSP. One possibility is e.g. the following.
\begin{equation}
\label{eq:overlap}
\tau_L(\tilde{s}, \tilde{e}) = \lfloor (\tilde{s} + \tilde{e})/2 \rfloor - \tilde{s};\quad 
\tau_R(\tilde{s}, \tilde{e}) = \lfloor (\tilde{s} + \tilde{e})/2 \rfloor +1 - \tilde{e}.
\end{equation}
This setting means that upon detecting a generic interval of significance $[\tilde{s},\tilde{e}]$ within $[s,e]$, the NSP algorithm continues on the left interval
$[s, \lfloor (\tilde{s} + \tilde{e})/2 \rfloor ]$ and the right interval $[\lfloor (\tilde{s} + \tilde{e})/2 \rfloor +1, e]$ (recall that the no-overlap case results uses the left
interval $[s, \tilde{s}]$ and the right interval $[\tilde{e}, e]$). See Section \ref{sec:csim} for more on the overlap parameters.

In NSP, having $p = p(T)$ growing with $T$ is possible, but we must have $p(T)+1 \le T$ or otherwise no regions of significance will be found.
Section \ref{sec:suppl:dis} of the appendix comments on a few other generic aspects of the NSP algorithm.


\subsection{Measuring deviation from linearity in NSP}
\label{sec:iidg}

This section completes the definition of NSP (in the version without self-normalisation) by describing the \textsc{DeviationFromLinearity} function (NSP algorithm, line 12).
Its basic building block is a scaled partial sum statistic, defined for
an arbitrary input sequence $\{y_t\}_{t=1}^T$ by $U_{s,e}(y) = (e-s+1)^{-1/2} \sum_{t=s}^e y_t$.
We define the scan statistic of an input vector $y$ (of length $T$) with respect to the interval set $\mathcal{I}$ as 
\begin{equation}
\label{eq:multiscan}
\|y\|_{\mathcal{I}} = \max_{[s,e]\in\mathcal{I}} |U_{s,e}(y)|.
\end{equation}
The set $\mathcal{I}$ used in NSP contains
intervals at a range of scales and locations.
For computational efficacy, instead of the set $\mathcal{I}^a$ of all subintervals of $[1,T]$, we use the set $\mathcal{I}^d$ of all intervals of dyadic lengths and arbitrary locations, that is $\mathcal{I}^d = \{ [s,e] \subseteq [1,T]\,\,:\,\, e-s = 2^j-1,\quad j=0, \ldots, \lfloor \log_2 T \rfloor   \}$. A simple pyramid algorithm of complexity $O(T \log\,T)$ is available for the computation of all $U_{s,e}(y)$ for $[s,e] \in \mathcal{I}^d$. We also define restrictions of $\mathcal{I}^a$ and $\mathcal{I}^d$ to arbitrary intervals $[s,e]$ as $\mathcal{I}^d_{[s,e]} = \{  [u,v] \subseteq [s,e]\,\,:\,\, [u,v] \in \mathcal{I}^d  \}$,
and analogously for $\mathcal{I}^a_{[s,e]}$. We refer
to $\|\cdot\|_{\mathcal{I}^d}$, $\|\cdot\|_{\mathcal{I}^a}$ and their restrictions as multiresolution sup-norms (see \cite{n85} and \cite{l16}) or, alternatively, multiscale scan statistics
if they are used as operations on data. If the context requires this, the qualifier ``dyadic'' will be added to these terms when referring to the $\mathcal{I}^d$ versions. The facts that, for any interval $[s,e]$ and any input vector $y$ (of length $T$), we have
\begin{equation}
\label{eq:simpineq}
\|  y_{s:e}   \|_{\mathcal{I}^d_{[s, e]}} \le \|  y_{s:e}   \|_{\mathcal{I}^a_{[s, e]}} \le \| y   \|_{\mathcal{I}^a}\quad\text{and}\quad \|  y_{s:e}   \|_{\mathcal{I}^d_{[s, e]}} \le \|  y   \|_{\mathcal{I}^d} \le \| y   \|_{\mathcal{I}^a}
\end{equation}
are trivial consequences of the facts that $\mathcal{I}^d_{[s, e]} \subseteq \mathcal{I}^a_{[s, e]} \subseteq \mathcal{I}^a$
and $\mathcal{I}^d_{[s, e]} \subseteq \mathcal{I}^d \subseteq \mathcal{I}^a$.
With this notation in place, \textsc{DeviationFromLinearity}$(s_m, e_m, Y, X)$ is defined as follows.
\begin{description}
\item[Step 1.] Find $\beta_0 = \arg\min_{\beta}     \|  Y_{s_m:e_m} - X_{s_m:e_m, \cdot} \beta  \|_{\mathcal{I}^d_{[s_m, e_m]}}$.
This fits the postulated linear model between $X$ and $Y$ restricted to the interval $[s_m,e_m]$. However, we use the multiresolution sup-norm $\|  \cdot  \|_{\mathcal{I}^d_{[s_m, e_m]}}$ as the loss function, rather than the more
usual $L_2$ loss. This has important consequences for the exactness of our significance statements, which we explain later below.
\item[Step 2.]
Compute the same multiresolution sup-norm of the empirical residuals from the above fit, $D_{[s_m, e_m]} := \|    Y_{s_m:e_m} - X_{s_m:e_m, \cdot} \beta_0  \|_{\mathcal{I}^d_{[s_m, e_m]}}$.
\item[Step 3.]
Return $D_{[s_m, e_m]}$.
\end{description}

Steps 1. and 2. above can be carried out in a single step as $D_{[s_m, e_m]} = \min_{\beta}     \|  Y_{s_m:e_m} - X_{s_m:e_m, \cdot} \beta  \|_{\mathcal{I}^d_{[s_m, e_m]}}$,
however, for comparison with other approaches, it will be convenient for us to use the two-stage process in steps 1. and 2.
for the computation of $D_{[s_m, e_m]}$. Computationally, the linear model fit in step 1. can be carried out via simple linear programming; we use 
the R package \verb+lpSolve+. The following important property lies at the heart of NSP.

\begin{prop}
\label{prop:supnorm}
Let the interval $[s, e]$ be such that $\forall\,\, j=1, \ldots, N\,\,\,[\eta_j, \eta_{j}+1] \not\subseteq [s, e]$. We have $D_{[s, e]} \le \|  Z_{s:e}   \|_{\mathcal{I}^d_{[s, e]}}$.
\end{prop}



This is a simple but valuable result, which can be read as follows: ``under the local null hypothesis of no signal on $[s,e]$, the test statistic $D_{[s,e]}$, defined as the multiresolution
sup-norm of the empirical residuals from the same multiresolution sup-norm fit of the postulated linear model on $[s,e]$, is bounded by the 
multiresolution
sup-norm
of the 
true residual process $Z_t$''. This bound is achieved because the same norm is used in the linear model fit and in the residual check, and it is important to note that the corresponding bound 
would not be available if the postulated linear model were fitted with a different loss function, e.g. via OLS. Having such a bound allows us to transfer our statistical significance calculations 
to the domain of the unobserved true residuals $Z_t$, which is much easier than working with the corresponding empirical residuals.
It is also critical to obtaining global coverage guarantees for NSP, as we now show.

\begin{theorem}
\label{th:main}
Let $\mathcal{S} = \{ S_1, \ldots, S_R    \}$ be a set of intervals returned by the NSP algorithm. We have 
$P\left( \exists\,\,{i=1,\ldots, R}\,\,\,\forall\,\,{j=1,\ldots, N}\,\,\,   [\eta_j,\eta_j+1] \not\subseteq S_i    \right)     \le P(\| Z \|_{\mathcal{I}^d} > \lambda_\alpha) \le P(\| Z \|_{\mathcal{I}^a} > \lambda_\alpha)$.
\end{theorem}



Theorem \ref{th:main} should be read as follows. Let $\alpha = P(\| Z \|_{\mathcal{I}^a} > \lambda_\alpha)$. For a set of intervals returned by NSP,
we are guaranteed, with probability of at least $1-\alpha$, that there is at least one change-point in each of these intervals. Therefore, $\mathcal{S} = \{ S_1, \ldots, S_R  \}$
can be interpreted as an automatically chosen set of regions (intervals) of significance in the data. In the no-change-point case ($N=0$), the correct reading
of Theorem \ref{th:main} is that the probability of obtaining one of more intervals of significance ($R\ge 1$) is bounded from above by $P(\| Z \|_{\mathcal{I}^a} > \lambda_\alpha)$.

NSP
uses a multiresolution sup-norm fit to be checked via the same multiresolution sup-norm. This leads to 
exact coverage guarantees for NSP with very simple mathematics. In contrast to the confidence intervals 
in e.g. \cite{bp98}, the NSP regions of significance are not conditional on any particular estimator of $N$ or of the change-point locations, and are in addition of a finite-sample nature.
Still, they have a ``confidence interval'' interpretation in the sense that each must contain at least one change, with a certain prescribed global 
probability.

For $S_i = [s,e]$, we define $S_i^- = [s,e-1]$.
A simple corollary of Theorem \ref{th:main} is that for $\mathcal{S} = \{ S_1, \ldots, S_R  \}$, if the corresponding sets $S_i^-$ are mutually disjoint (as is the case e.g. if $\tau_L = \tau_R \equiv 0$), then
we must have 
$N \ge R$ with probability at least $1-\alpha$. It would be impossible to obtain a similar upper bound on $N$ without order-of-magnitude assumptions on spacings between change-points
and magnitudes of parameter changes; we defer this to Section \ref{sec:consist}.
The result in Theorem \ref{th:main} does not rely on asymptotics
and has a finite-sample character.
$\beta_0$ in Step 1 above does not have to be an accurate estimator of the true local $\beta$ for the bound in Proposition
\ref{prop:supnorm} to hold; it holds unconditionally and for arbitrary short intervals $[s,e]$.

NSP is not automatically equipped with pointwise estimators of change-point locations. This is an important feature, because thanks to this, it can be so general and work in the same way for any $X$. If it were to come with meaningful pointwise change-point location estimators, they would have to be designed for each $X$ separately, e.g. using the maximum likelihood
principle. (However, NSP can be paired up with such pointwise estimators; see immediately below for details.) We now introduce a few new concepts, 
to contrast this feature of NSP with the existing concept of post-selection inference.

\vspace{5pt}

\noindent {\em ``Post-inference selection" and ``inference without selection".} If it can be assumed that an interval $S_i = [s_i, e_i] \in \mathcal{S}$ only contains a single change-point, its location can be estimated e.g. via MLE performed locally on the data subsample living on $[s_i, e_i]$. Naturally, the MLE should be constructed with the specific design matrix $X$ in mind, see \cite{bcf16} for examples in Scenarios 1 and 2. In this construction, ``inference'', i.e. the execution of NSP, occurs before ``selection'', i.e. the estimation of the change-point locations, hence the label of ``post-inference selection''. This avoids the complicated machinery of post-selection inference, as we automatically know that the $p$-value associated with the estimated change-point must be less than $\alpha$. Similarly, ``inference without selection" refers to the use of NSP unaccompanied by a change-point location estimator.

\vspace{5pt}

\noindent {\em ``Simultaneous inference and selection" or ``in-inference selection".} In this construction, change-point location estimation on an interval $[\tilde{s}, \tilde{e}]$ occurs directly after adding it to $\mathcal{S}$. The difference with
``post-inference selection'' is that this then naturally enables appropriate non-zero overlaps $\tau_L$ and $\tau_R$ in the execution of NSP. More specifically, denoting the estimated location within $[\tilde{s}, \tilde{e}]$ by $\tilde{\eta}$, we can set,
for example, $\tau_L(\tilde{s}, \tilde{e}, Y, X)  = \tilde{\eta} - \tilde{s}$ and $\tau_R(\tilde{s}, \tilde{e}, Y, X) = \tilde{e} - \tilde{\eta} - 1$,
so that lines 21--22 of the NSP algorithm become, respectively, NSP$(s, \tilde{\eta}, Y, X, M, \lambda_\alpha, \tau_L, \tau_R)$ and 
NSP$(\tilde{\eta}+1, e, Y, X, M, \lambda_\alpha, \tau_L, \tau_R)$.

\vspace{5pt}

By Theorem \ref{th:main}, the only piece of knowledge required to obtain
coverage guarantees in NSP is the distribution of $\| Z \|_{\mathcal{I}^a}$ (or $\| Z \|_{\mathcal{I}^d}$), regardless of the
form of $X$. Much is known about this distribution 
for various underlying distributions of $Z$; see Section \ref{sec:gauss} below and Section \ref{sec:lt} of the appendix 
for $Z$ Gaussian and following other light-tailed distributions, respectively.
Any future further distributional results of this type would only further enhance the applicability of NSP. However, if the distribution of 
$\| Z \|_{\mathcal{I}^a}$ ($\| Z \|_{\mathcal{I}^d}$) is unknown, then an approximation can also be obtained by simulation, which is particularly
computationally efficient for $\| Z \|_{\mathcal{I}^d}$. See Section \ref{sec:suppl:sim} of the appendix for more details on simulation-based threshold selection.

\subsection{Gaussian $Z_t$}
\label{sec:gauss}

We now recall distributional results for $\| Z \|_{\mathcal{I}^a}$, in the case $Z_t \sim \text{i.i.d.}\,\, N(0, \sigma^2)$ with $\sigma^2$ assumed known, which will permit us to choose $\lambda_\alpha = \lambda_\alpha(T)$
so that $P\{\| Z \|_{\mathcal{I}^a} > \lambda_\alpha(T)\} \to \alpha$ as $T \to \infty$. The resulting $\lambda_\alpha(T)$ can then be used in Theorem \ref{th:main}.
As the result of Theorem \ref{th:main} is otherwise of a finite-sample nature, some users may be uncomfortable resorting to large-sample asymptotics to approximate the
distribution of $\| Z \|_{\mathcal{I}^a}$. However, (a) the asymptotic results outlined below approximate the behaviour of $\| Z \|_{\mathcal{I}^a}$ well even for small samples,
and (b) users not wishing to resort to asymptotics have the option of approximating the distribution of $\| Z \|_{\mathcal{I}^a}$ by simulation (see Section \ref{sec:suppl:sim}
of the appendix), which is computationally fast.
The assumption of
a known $\sigma^2$ is common in the change-point inference literature, see e.g. \cite{hgst18}, \cite{fs20} and \cite{jfw20}. 
Section \ref{sec:lt} of the appendix covers the unknown $\sigma^2$ case.
Results on the distribution of $\| Z \|_{\mathcal{I}^a}$ are given in \cite{sv95} and \cite{k07}. We recall the formulation from \cite{k07} as it is slightly more explicit.

\begin{theorem}[Theorem 1.3 in \cite{k07}]
\label{lem:k07}
Let $\{Z_t\}_{t=1}^T$ be i.i.d. $N(0, 1)$. For every $\gamma \in \mathbb{R}$, we have
$\lim_{T\to\infty} P\left( \max_{1\le s\le e\le T}\,\, U_{s,e}(Z) \le a_T + b_T\, \gamma     \right) = \exp(-e^{-\gamma})$,
where
\[
a_T = \sqrt{2\log\,T} + \frac{\frac{1}{2}\log\log\,T + \log\frac{H}{2\sqrt{\pi}}}{\sqrt{2\log\,T}};\quad
b_T = \frac{1}{\sqrt{2\log\,T}};\quad
H = \int_0^\infty \exp\left( -4 \sum_{k=1}^\infty \frac{1}{k} \Phi\left(  -\sqrt{\frac{k}{2y}}   \right)  \right) dy,
\]
and $\Phi()$ is the standard normal cdf.
\end{theorem}
We use the approximate value $H = 0.82$ in our numerical work. Using the asymptotic independence of the maximum and
the minimum \citep{kw14}, and the symmetry of $Z$, we get the following simple corollary.
\begin{eqnarray}
\lefteqn{P\left(\max_{1\le s\le e\le T}|U_{s,e}(Z)| > a_T + b_T\, \gamma\right) = 1 - P\left(\max_{1\le s\le e\le T}|U_{s,e}(Z)| \le a_T + b_T\, \gamma\right) =}\nonumber\\
&& 1 - P\left(\max_{1\le s\le e\le T} U_{s,e}(Z) \le a_T + b_T\, \gamma\quad\land\quad \min_{1\le s\le e\le T} U_{s,e}(Z) \ge -(a_T + b_T\, \gamma)\right) \to \nonumber\\
\label{eq:expbound}
&& 1 - \exp(-2e^{-\gamma})
\end{eqnarray}
as $T\to\infty$.
In light of (\ref{eq:expbound}), we obtain $\lambda_\alpha$ for use in Theorem \ref{th:main} as follows: (a) equate $\alpha = 1 - \exp(-2e^{-\gamma})$ and obtain $\gamma$, (b) form
$\lambda_\alpha = \sigma(a_T + b_T\, \gamma)$.

We now extend NSP to positively-dependent Gaussian innovations.
Let
$\{\tilde{Z}_t\}_{t=1}^T$ be a stationary, zero-mean, non-negatively autocorrelated process with long-run standard deviation $\sigma_{LR}$. Let 
$\sigma_{s,e} = \mbox{Var}^{1/2}\{ U_{s,e}(\tilde{Z})  \}$, and note $\sigma_{s,e} \le \sigma_{LR}$. In the notation of Theorem \ref{lem:k07},
\begin{eqnarray*}
P\left\{\max_{1\le s\le e\le T} U_{s,e}(\tilde{Z}) \ge \sigma_{LR} (a_T + b_T \gamma)\right\} & \le & P\left\{\max_{1\le s\le e\le T} \frac{U_{s,e}(\tilde{Z})}{\sigma_{s,e}} \ge a_T + b_T \gamma\right\}\\
\text{[Slepian's lemma]} & \le & P\left\{\max_{1\le s\le e\le T} U_{s,e}(Z) \ge a_T + b_T \gamma \right\}.
\end{eqnarray*}
This demonstrates that valid coverage guarantees are obtained for a system with innovations $\tilde{Z}$ by applying the NSP threshold equal to the threshold suitable for i.i.d. $N(0,1)$
innovations times the long-run standard deviation of $\tilde{Z}$. Long-run standard deviation estimation, especially in the presence of change-points, is a difficult problem, but several solutions have been proposed, including one in \cite{dsv18} (in our Scenario 1). See also Section \ref{sec:ai} of the appendix for a related discussion of NSP with
autocorrelated innovations.

\subsection{Tightening the bounds: $X$-dependent thresholds}
\label{sec:simthresh}

We now show how to obtain thresholds lower than those in Theorem \ref{th:main} if the analyst is willing to allow their dependence on the design matrix $X$.
This calls for the re-examination of Proposition \ref{prop:supnorm}. Consider the following alternative version.

\begin{prop}
\label{prop:ub}
Let the interval $[s,e]$ be such that $\forall\,\,j=1,\ldots, N\quad [\eta_j, \eta_j+1] \not\subseteq [s,e]$. We have
$D_{[s,e]} = \min_\beta \|  Z_{s:e} - X_{s:e,\cdot} \beta  \|_{\mathcal{I}^d_{[s,e]}} \le \min_\beta \|  Z - X \beta  \|_{\mathcal{I}^d}$.
\end{prop}

\noindent This leads to a tighter version of Theorem \ref{th:main}.

\begin{theorem}
\label{th:main2}
Let $\mathcal{S} = \{ S_1, \ldots, S_R    \}$ be a set of intervals returned by the NSP algorithm. We have
$P\left( \exists\,\,{i=1,\ldots, R}\,\,\,\forall\,\,{j=1,\ldots, N}\,\,\,   [\eta_j,\eta_j+1] \not\subseteq S_i    \right)     \le P(\min_\beta \|  Z - X \beta  \|_{\mathcal{I}^d} > \lambda_\alpha)$.
\end{theorem}
\comment{
The differences with Theorem \ref{th:main} are as follows.
In Theorem \ref{th:main}, the probability \\
$P\left( \exists\,\,{i=1,\ldots, R}\,\,\,\forall\,\,{j=1,\ldots, N}\,\,\,   [\eta_j,\eta_j+1] \not\subseteq S_i    \right)$ is bounded from above by $P( \|  Z  \|_{\mathcal{I}^d} > \lambda_\alpha)$, which is in turn bounded from above by $P( \|  Z  \|_{\mathcal{I}^a} > \lambda_\alpha)$. These bounds are independent of the covariates $X$ (i.e. independent of the scenario). In the Gaussian case, to choose $\lambda_\alpha$, we approximated the probability $P( \|  Z  \|_{\mathcal{I}^a} > \lambda_\alpha)$ using Theorem 1.3 in \cite{k07} (Theorem \ref{lem:k07} in the main paper).
}
In Theorem \ref{th:main2}, the probability $P\left( \exists\,\,{i=1,\ldots, R}\,\,\,\forall\,\,{j=1,\ldots, N}\,\,\,   [\eta_j,\eta_j+1] \not\subseteq S_i    \right)$ is bounded from above by $P(\min_\beta \|  Z - X \beta  \|_{\mathcal{I}^d} > \lambda_\alpha)$. As $\min_\beta \|  Z - X \beta  \|_{\mathcal{I}^d} \le \|  Z - X 0  \|_{\mathcal{I}^d} = \|  Z \|_{\mathcal{I}^d} \le \|  Z \|_{\mathcal{I}^d}$, the threshold $\lambda_\alpha$ obtained by solving
\begin{equation}
\label{eq:alphadep}
P(\min_\beta \|  Z - X \beta  \|_{\mathcal{I}^d} > \lambda_\alpha) = \alpha
\end{equation}
will be lower than that obtained by solving
$P(\|  Z \|_{\mathcal{I}^d} > \lambda_\alpha) = \alpha$
(which was done in Theorem \ref{th:main}). In addition, unlike the solution to $P(\|  Z \|_{\mathcal{I}^d} > \lambda_\alpha) = \alpha$, the solution to (\ref{eq:alphadep}) accounts for the number and form of the covariates $X$.
To solve (\ref{eq:alphadep}), the distribution of $\min_\beta \|  Z - X \beta  \|_{\mathcal{I}^d}$ can be obtained by simulation, separately for each set of covariates $X$ and sample size $T$;
see Section \ref{sec:suppl:sim} of the appendix for details.
The better localisation properties of the thus-obtained tighter bounds are illustrated, for Scenario 1, in Section \ref{sec:csim}.

\section{NSP with self-normalisation and with autoregression}
\label{sec:arsn}

\subsection{Self-normalised NSP for possibly heavy-tailed, heteroscedastic $Z_t$}
\label{sec:sn}

\cite{kw14} point out that the square-root normalisation used in $U_{s,e}(y)$ is not natural for distributions with tails heavier than Gaussian.
We are interested
in obtaining a universal normalisation in $U_{s,e}(y)$ which would work across a wide range of possibly heavy-tailed distributions without requiring their
explicit knowledge, including under heterogeneity.
One such solution is offered by the self-normalisation framework developed in 
\cite{rs03}
and related papers. We now recall the basics and discuss the necessary adaptations to our context; the less mathematically-inclined reader is 
welcome to skip this description and proceed directly to formula (\ref{eq:thsn}), which gives the oracle self-normalised statistic computed on the true
residuals $Z_t$.

We first discuss the relevant distributional results for the true residuals $Z_t$. We only cover the case of symmetric distributions
of $Z_t$. For the non-symmetric case, which requires a slightly different normalisation, see \cite{rs03}.
In the latter work, the following result is proved. Let
${\rho_{\theta, \nu, c}}(\delta) = \delta^\theta \log^\nu(c/\delta)$, $0 < \theta < 1$, $\nu \in \mathbb{R}$,
where $c \ge \exp(\nu/\theta)$ if $\nu > 0$ and $c > \exp(-\nu/(1-\theta))$ if $\nu < 0$. Further, suppose
$\lim_{j\to\infty} 2^j {\rho_{\theta, \nu, c}}^2(2^{-j})/j = \infty$.
This last condition, in particular, is satisfied if $\theta = 1/2$ and $\nu > 1/2$. The function ${\rho_{\theta, \nu, c}}$ will play the role of a modulus of continuity. Let $Z_1, Z_2, \ldots$ be 
independent and symmetrically distributed with $\mathbb{E}(Z_t) = 0$; note they do not need to be identically distributed. Define
$S_t = Z_1 + \ldots + Z_t$ and $V_t^2 = Z_1^2 + \ldots + Z_t^2$. Assume further
$V_T^{-2} \max_{1 \le t \le T} Z_t^2 \to 0$
in probability as $T \to \infty$. \cite{e97} shows that this last condition is equivalent to $Z_t$ being within the domain of attraction of the normal law.
Therefore, the material of this section applies to a much wider class of distributions
than the heterogeneous extension of SMUCE in \cite{psm17}, which only applies to normally distributed $Z_t$.

Let the random polygonal partial sums process $\zeta_T$ be defined on $[0,1]$ as linear interpolation between the knots $(V_t^2/V_T^2, S_t)$, $t = 0, \ldots, T$,
where $S_0 = V_0 = 0$, and let $\zeta_T^{\text{se}} = \zeta_T/V_T$.
Denote by $H_{\rho_{\theta, \nu, c}}[0,1]$ the set of continuous functions $x\,\,:\,\,[0,1] \to \mathbb{R}$ such that $\omega_{\rho_{\theta, \nu, c}}(x, 1) < \infty$, where
$\omega_{\rho_{\theta, \nu, c}}(x,\delta) = \sup_{u,v\in[0,1],\,\, 0 < |v-u| < \delta}  |x(v) - x(u)|/\rho_{\theta, \nu, c}(|v-u|)$.
$H_{\rho_{\theta, \nu, c}}[0,1]$ is a Banach space in its natural norm
$\| x\|_{\rho_{\theta, \nu, c}} = |x(0)| + \omega_{\rho_{\theta, \nu, c}}(x,1)$.
Define $H_{\rho_{\theta, \nu, c}}^0[0,1]$, a closed subspace of $H_{\rho_{\theta, \nu, c}}[0,1]$, by
$H_{\rho_{\theta, \nu, c}}^0[0,1] = \{ x\in H_{\rho_{\theta, \nu, c}}[0,1]\,\, : \,\, \lim_{\delta\to 0} \omega_{\rho_{\theta, \nu, c}}(x,\delta) = 0       \}$.
$H_{\rho_{\theta, \nu, c}}^0[0,1]$ is a separable Banach space. Under the assumptions above, we have the following convergence
in distribution as $T \to \infty$:
\begin{equation}
\label{eq:tow}
\zeta_T^{\text{se}} \to W
\end{equation}
in $H_{\rho_{\theta, \nu, c}}^0[0,1]$, where $W(u), u\in[0,1]$ is a standard Wiener process.
Define
$I_{\rho_{\theta, \nu, c}}(x, u, v) = |x(v) - x(u)|/\rho_{\theta, \nu, c}(|v-u|)$
and, with $\epsilon > 0$ and $c = \exp(1 + 2\epsilon)$, consider the statistic
\begin{eqnarray}
\lefteqn{\sup_{0\le i < j \le T} I_{\rho_{1/2, 1/2+\epsilon, c}}(\zeta_T^{\text{se}}, V_i^2/V_T^2, V_j^2/V_T^2) = \sup_{0\le i < j \le T} \frac{|\zeta_T^{\text{se}}(V_j^2 / V_T^2) - \zeta_T^{\text{se}}(V_i^2 / V_T^2)|}{{\rho_{1/2, 1/2+\epsilon, c}}(V_j^2/V_T^2 - V_i^2/V_T^2)} =}\nonumber\\ 
&& \sup_{0\le i < j \le T} \frac{|S_j - S_i|}{\sqrt{V_j^2 - V_i^2} \log^{1/2+\epsilon} \{c/(V_j^2/V_T^2 - V_i^2/V_T^2)\}} = \nonumber\\
\label{eq:thsn}
&& \sup_{0\le i < j \le T} \frac{|Z_{i+1}+\ldots+Z_j|}{\sqrt{Z_{i+1}^2+\ldots+Z_j^2} \log^{1/2+\epsilon} \{cV_T^2/(Z_{i+1}^2+\ldots+Z_j^2)\}}.
\end{eqnarray}
In the notation and under the conditions listed above, it is a direct consequence of 
the distributional convergence (\ref{eq:tow}) in the space $H_{\rho_{\theta, \nu, c}}^0[0,1]$
that for any level $\gamma$, we have
\begin{eqnarray}
\lefteqn{P\left( \sup_{0\le i < j \le T} I_{\rho_{1/2, 1/2+\epsilon, c}}(\zeta_T^{\text{se}}, V_i^2/V_T^2, V_j^2/V_T^2) \ge \gamma  \right) \le}\nonumber\\
\label{eq:anylevel}
&& P\left( \sup_{u,v\in[0,1]} I_{\rho_{1/2, 1/2+\epsilon, c}}(\zeta_T^{\text{se}}, u, v) \ge \gamma  \right) \to P\left( \sup_{u,v\in[0,1]} I_{\rho_{1/2, 1/2+\epsilon, c}}(W, u, v) \ge \gamma  \right)
\end{eqnarray}
as $T \to \infty$, and the quantiles of the distribution of $\sup_{u,v\in[0,1]} I_{\rho_{1/2, 1/2+\epsilon, c}}(W, u, v)$, which does not depend on the sample size $T$, can be computed (once) by simulation.

Following the narrative of Sections \ref{sec:iidg} and \ref{sec:gauss}, to make these results operational in a new function \textsc{DeviationFromLinearity.SN} (where `SN' stands for self-normalisation)
for use in line 12 of the NSP algorithm, we need the following development. Assume initially that the global residual sum of squares $V_T^2$ is known. For a generic interval $[s,e]$ containing no change-points, we need to be able to obtain
empirical residuals $\hat{Z}_{i+1}^{(k)}, \ldots, \hat{Z}_j^{(k)}$ for $k = 1, 2$ and 
$\hat{Z}_{s}^{(k)}, \ldots, \hat{Z}_e^{(k)}$ for $k = 3$
for which we can guarantee that 
{\small
\begin{eqnarray}
\label{eq:sntarget}
\lefteqn{\sup_{s-1 \le i < j \le e} \frac{|\hat{Z}^{(3)}_{i+1}+\ldots+\hat{Z}^{(3)}_j|}{\sqrt{(\hat{Z}^{(2)}_{i+1})^2+\ldots+(\hat{Z}^{(2)}_j)^2} \log^{1/2+\epsilon} \{cV_T^2/((\hat{Z}^{(1)}_{i+1})^2+\ldots+(\hat{Z}^{(1)}_j)^2)\}}
\le}\nonumber\\
&& \sup_{s-1 \le i < j \le e} \frac{|Z_{i+1}+\ldots+Z_j|}{\sqrt{Z_{i+1}^2+\ldots+Z_j^2} \log^{1/2+\epsilon} \{cV_T^2/(Z_{i+1}^2+\ldots+Z_j^2)\}}.
\end{eqnarray}}
This provides a self-normalised equivalent of Proposition \ref{prop:supnorm} and requires that the deviation from linearity computed on an interval containing no change-points (left-hand side of (\ref{eq:sntarget})) does not exceed the analogous oracle quantity computed on the true residuals (right-hand side of \ref{eq:sntarget}).
Section \ref{sec:suppl:sn} of the appendix describes  the construction of $\hat{Z}^{(k)}$ for $k = 1, 2, 3$ so that (\ref{eq:sntarget}) is guaranteed, and
introduces a suitable estimator of $V_T^2$ for use in (\ref{eq:sntarget}).

\subsection{NSP with autoregression}
\label{sec:ar}

To accommodate autoregression while retaining the serial independence of $Z_t$, we introduce the following additional scenario.

\begin{description}
\item[Scenario 4.] {\em Linear regression with autoregression, with piecewise-constant parameters.}

For a given design matrix $X = (X_{t,i})$, $t = 1, \ldots, T$, $i=1,\ldots, p$, the response $Y_t$ follows the model
\begin{equation}
\label{eq:regar}
Y_t = X_{t,\cdot}\beta^{(j)} + \sum_{k=1}^{r} a_k^{(j)} Y_{t-k}  + Z_t\quad\text{for}\quad t = \eta_j+1, \ldots ,\eta_{j+1},
\end{equation}
for $j = 0, \ldots, N$, where the regression parameter vectors $\beta^{(j)} = (\beta^{(j)}_1, \ldots, \beta^{(j)}_p)'$
and the autoregression parameters $a_k^{(j)}$ are such that either 
$\beta^{(j)} \neq \beta^{(j+1)}$ or $a_k^{(j)} \neq a_k^{(j+1)}$ for some $k$ (or both types of changes occur).
\end{description}

In this work, we treat the autoregressive order $r$ as fixed and known to the analyst.
\cite{fs20} consider $r = 1$ and treat the autoregressive parameter as known, but acknowledge that in
practice it is estimated from the data; however, they add that ``[it] would also be possible to estimate [the autoregressive parameter] from the currently
studied subset of the data, but this estimator appears to be unstable''. NSP circumvents this instability issue, as explained below.
NSP for Scenario 4 proceeds as follows.

\begin{enumerate}
\item
Supplement the design matrix $X$ with the lagged versions of the variable $Y$, or in other words substitute
$X := \begin{bmatrix} 
X & Y_{\cdot - 1} & \cdots & Y_{\cdot - r}\\
\end{bmatrix}$,
where $Y_{\cdot - k}$ denotes the respective backshift operation. Omit the first $r$ rows of the thus-modified $X$, and the first $r$ elements of $Y$.
\item
Run the NSP algorithm of Section \ref{sec:genalgo} with the new $X$ and $Y$ (with a suitable modification to line 12 if using the self-normalised version), with the following single difference.
In lines 21 and 22, recursively call the NSP routine on the intervals $[s, \tilde{s}+\tau_L(\tilde{s}, \tilde{e}, Y, X)-r]$ and $[\tilde{e}-\tau_R(\tilde{s}, \tilde{e}, Y, X)+r, e]$, respectively. As each
local regression is now supplemented with autoregression of order $r$, we insert the extra ``buffer'' of size $r$ between the detected interval $[\tilde{s}, \tilde{e}]$ and the next children intervals
to ensure that we do not process information about the same change-point in both the parent call and one of the children calls, which prevents double detection.

\end{enumerate}

 The result
of Theorem \ref{th:main} applies to the output of NSP for Scenario 4 too.
The NSP algorithm offers a new point of view on change-point analysis in the presence of autocorrelation. Unlike \cite{fs20}, who require accurate
estimation of the autoregressive parameters for successful change-point detection, NSP circumvents the issue by using the same multiresolution norm in the local regression
fits on each $[s,e]$, and in the subsequent tests of the local residuals. In this way, the autoregression parameters do not have to be estimated accurately for the relevant stochastic bound in Proposition
\ref{prop:supnorm} to hold; it holds unconditionally and for arbitrary short intervals $[s,e]$. Therefore, NSP is able to deal with autoregression, stably, on arbitrarily short intervals.
We illustrate the performance of this version of NSP in Section \ref{sec:arex} of the appendix.


\section{Detection consistency and lengths of NSP intervals}
\label{sec:consist}

We now study the consistency of NSP in detecting change-points, and the rates at which the lengths of the NSP intervals contract, as the sample size increases. We consider a version of the NSP algorithm that considers all sub-intervals of $[1,T]$, and we provide results in Scenario 1 as well as in Scenario 2 with continuous piecewise-linearity (this parallels the scenarios for which consistency is shown in \cite{bcf16}).

So far in the paper, we avoided introducing any assumptions on the signal: our coverage guarantees in Theorem \ref{th:main} held under no conditions on the number of change-points, their spacing, or the sizes of the breaks. This was unsurprising as they amounted to statistical size control. By contrast, the results of this section relate to detection consistency (and therefore `power' rather than size) and as such, require minimum signal strength assumptions.

\subsection{Scenario 1 -- piecewise constancy}
\label{sec:sc1}

In this section, $f_t$ falls under Scenario 1.
We start with assumptions on the strength of the change-points. For each change-point $\eta_j$, $j=1,\ldots, N$, define
\begin{equation}
\label{eq:dj}
\bar{d}_j = \left\lceil \frac{16 \lambda_\alpha^2}{|f_{\eta_j+1} - f_{\eta_j}|^2} \right\rceil + 1.
\end{equation}
Recalling that $\eta_0 = 0$ and $\eta_{N+1} = T$, we require the following assumption.

\begin{assumption}
\label{ass:dist}
$\eta_{j+1} - \eta_j \ge 2\bar{d}_{j+1} + 2 \bar{d}_j - 2\,\,\, (j = 1,\ldots, N-1);\qquad
\eta_{1} - \eta_0 \ge 2\bar{d}_1 - 1;\qquad
\eta_{N+1} - \eta_N \ge 2\bar{d}_{N} - 1$.
\end{assumption}

\noindent We have the following theorem.

\begin{theorem}
\label{th:mainconsist}
Let Assumption \ref{ass:dist} hold, with $\bar{d}_j$ defined in (\ref{eq:dj}).
On the set $\|  Z   \|_{\mathcal{I}^a} \le \lambda_\alpha$, a version of the NSP algorithm that considers all sub-intervals, executed with no overlaps and with threshold $\lambda_\alpha$, returns exactly $N$ intervals of significance $[s_1, e_1] < \ldots < [s_N, e_N]$ such that
$\eta_j \in [s_j, e_j-1]$ and 
$e_j - s_j + 1 \le 2\bar{d}_j$, for $j = 1, \ldots, N$.
\end{theorem}
Theorem \ref{th:mainconsist} leads to the following corollary.
\begin{corollary}
\label{cor:consist}
Let the assumptions of Theorem \ref{th:mainconsist} hold, and in addition let $Z_t \sim N(0, \sigma^2)$. Let $\lambda_\alpha = \sigma (1 + \Delta) \sqrt{2\log\,T}$ for any $\Delta > 0$. Let $\mathcal{S}$ denote the set of intervals of significance $[s_1, e_1] < [s_2, e_2] < \ldots$ returned by a version of the NSP algorithm that considers all sub-intervals, executed with no overlaps and with threshold $\lambda_\alpha$. Let $\mathcal{A} = \{  |\mathcal{S}| = N\,\,\,  \land\,\,\, \forall j=1, \ldots, N \,\,\, \eta_j \in [s_j, e_j-1] \,\,\, \land \,\,\, e_j - s_j + 1 \le 2\bar{d}_j\}$.
We have $P(\mathcal{A}) \to 1$ as $T \to \infty$.
\end{corollary}

Corollary \ref{cor:consist} is a traditional, large-sample consistency result for NSP. Consider first Assumption \ref{ass:dist}, under which it operates. With $\lambda_\alpha$ as in Corollary \ref{cor:consist}, Assumption \ref{ass:dist} permits $\min_j \{|\eta_{j+1} - \eta_j|^{1/2} \min(  |f_{\eta_j+1} - f_{\eta_j}|, |f_{\eta_{j+1}+1} - f_{\eta_{j+1}}|  )\}$, a quantity that characterises the difficulty of the multiple change-point detection problem, to be of order $O(\log^{1/2}T)$, which is the same as in \cite{bcf16} and minimax-optimal as argued in \cite{cw13}.
Further, the statement of Corollary \ref{cor:consist} implies statistical consistency of NSP in the sense that with probability tending to one with $T$, NSP estimates the correct number of change-points and each NSP interval contains exactly one true change-point. Moreover, the length of the NSP interval around each $\eta_j$ is of order $O(\log\,T / |f_{\eta_j+1} - f_{\eta_j}|^2)$, which is near-optimal and the same as in \cite{bcf16}. Finally, this also implies that this consistency rate is inherited by {\em any} pointwise estimator of $\eta_j$ that takes its value in the $j$th NSP interval of significance; this applies even to naive estimators constructed e.g. as the middle points of their corresponding NSP intervals $[s_j, e_j]$, i.e. $\hat{\eta}_j = \lfloor (s_j + e_j)/2  \rfloor$. More refined estimators, e.g. one based on CUSUM maximisation within each NSP interval, can also be used and will also automatically inherit the consistency and rate.

\subsection{Scenario 2 -- continuous piecewise linearity}

In this section, $f_t$ falls under Scenario 2 and is piecewise linear and continuous.
Naturally, the definition of change-point strength has to be different from that in Section \ref{sec:sc1}. For each change-point $\eta_j$, $j=1,\ldots, N$, let
\begin{equation}
\label{eq:djlin}
\bar{d}_j = \left\lceil C_2 \lambda_\alpha^{2/3} \xi_j^{-2/3}  \right\rceil,
\end{equation}
where $\xi_j = |\xi_{j,1} - \xi_{j,2}|/2$ and $\xi_{j,1}, \xi_{j,2}$ are, respectively, the slopes of $f_t$ immediately to the left and to the right of $\eta_j$, and $C_2$ is a certain universal constant (i.e. valid for all $f_t$), suitably large. The following theorem holds.

\begin{theorem}
\label{th:linconsist}
Let Assumption \ref{ass:dist} hold, with $\bar{d}_j$ defined in (\ref{eq:djlin}).
On the set $\|  Z   \|_{\mathcal{I}^a} \le \lambda_\alpha$, a version of the NSP algorithm that considers all sub-intervals, executed with no overlaps and with threshold $\lambda_\alpha$, returns exactly $N$ intervals of significance $[s_1, e_1] < \ldots < [s_N, e_N]$ such that
$\eta_j \in [s_j, e_j-1]$ and 
$e_j - s_j + 1 \le 2\bar{d}_j$,
for $j = 1, \ldots, N$.
\end{theorem}

We note that Assumption \ref{ass:dist} is model-independent: we require it as much in the piecewise-constant Scenario 1 as in the piecewise-linear Scenario 2 (and in any other scenario), but with $\bar{d}_j$ defined separately for each scenario. Theorem \ref{th:linconsist} leads to the following corollary.

\begin{corollary}
\label{cor:consistlin}
Let the assumptions of Theorem \ref{th:linconsist} hold, and in addition let $Z_t \sim N(0, \sigma^2)$. Let $\lambda_\alpha = \sigma (1 + \Delta) \sqrt{2\log\,T}$ for any $\Delta > 0$. Let $\mathcal{S}$ denote the set of intervals of significance $[s_1, e_1] < [s_2, e_2] < \ldots$ returned by a version of the NSP algorithm that considers all sub-intervals, executed with no overlaps and with threshold $\lambda_\alpha$. Let $\mathcal{A} = \{  |\mathcal{S}| = N\,\,\,  \land\,\,\, \forall j=1, \ldots, N \,\,\, \eta_j \in [s_j, e_j-1] \,\,\, \land \,\,\, e_j - s_j + 1 \le 2\bar{d}_j\}$.
We have $P(\mathcal{A}) \to 1$ as $T \to \infty$.
\end{corollary}

Corollary \ref{cor:consistlin} implies that with $\lambda_\alpha$ as defined therein, and if $\xi_j \sim T^{-1}$ (a case in which $f_t$ is bounded; see \cite{bcf16}), we have that the accuracy of change-point localisation via NSP (measured by $e_j-s_j$) is $O(T^{2/3} \log^{1/3}T)$, the same as in \cite{bcf16} and within a logarithmic factor of \cite{r98}. Our comment (made in Section \ref{sec:sc1}) regarding this rate being inherited by any pointwise estimator of $\eta_j$, as long as it falls within $[s_j, e_j]$, applies equally in this case. 


\section{Numerical illustrations}
\label{sec:num}


\subsection{Scenario 1 -- piecewise constancy}
\label{sec:csim}

\begin{table}
{\small
\centering
\begin{tabular}{ |c|c|l| } 
\hline
model name & no. of cpts & sample path execution in R \\
\hline\hline
Noise 100 & 0 & \verb+rnorm(100)+\\
Noise 300 & 0 & \verb+rnorm(300)+\\
Single 100 & 1 & \verb|c(rep(0, 50), rep(1, 50)) + rnorm(100)|\\
Single 300 & 1 & \verb|c(rep(0, 150), rep(1, 150)) + rnorm(300)|\\
Wave & 3 & \verb|rep(rep(c(0, 100), each = 100), 2) + 100 * rnorm(400)|\\
Wide Teeth & 9 & \verb|rep(rep(c(0, 1), each = 30), 5) + rnorm(300)|\\
Teeth 10 & 13 & \verb|rep(rep(c(0, 1), each = 10), 7) + 0.4 * rnorm(140)|\\
Blocks & 11 & signal defined in \cite{f14a}; noise \verb+10 * rnorm(2048)+\\
 \hline
\end{tabular}
\caption{Models for the comparative simulation study in Section \ref{sec:csim}; ``no. of cpts'' means ``number of change-points''. \label{tab:models}}
}
\end{table}

\begin{table}
{\small
\centering
\begin{tabular}{ |c|c|c|c|c|c|c|c| } 
\hline
model & NSP & NSP-SIM & NSP-O & NSP-SIM-O & BP & BP-LIM & SMUCE\\
\hline\hline
Noise 100 & 96 & 86 & 96 & 86 & 96 & 97 & 97\\
Noise 300 & 99 & 89 & 99 & 89 & 99 & 99 & 98\\
 \hline
\end{tabular}
\caption{Numbers of times, out of 100 simulated sample paths of each null model, that the respective method indicated no intervals of significance. Throughout the paper,
all batches of 100 sample paths are simulated with the random seed initially set to 1.\label{tab:nullres}}
}
\end{table}

\begin{table}
{\scriptsize
\centering
\begin{tabular}{ |c|c|c|c|c|c|c|c|c| } 
\hline
model & attribute & NSP & NSP-SIM & NSP-O & NSP-SIM-O & BP & BP-LIM & SMUCE \\
\hline\hline
                  & coverage & 96 & 90 &  95 & 90 & 78 & 84 & 98\\
Single 100 & prop. gen. int. & 0.95  & 0.91 & 0.94 & 0.92 & 0.8 &  0.84 &  0.98\\
                  & no. gen. int. & 0.48  & 0.74 & 0.48 & 0.77 & 0.82 & 0.83 & 0.8\\
                  & no. all int. & 0.54 & 0.92 & 0.55 & 0.97 & 1.15 & 0.99 & 0.82 \\
                  & av. gen. int. len. & 48.17 &  44.64 & 48.17 & 43.93 & 15.91 &15.66 & 48.71 \\
 \hline\hline
                  & coverage & 99 & 92 & 99 & 92 & 89 & 91 & 100\\
Single 300 & prop. gen. int. & 0.99 & 0.94 & 0.99 & 0.95 & 0.89 & 0.91 & 1 \\
                  & no. gen. int. & 0.99  & 0.97  &  1.02  & 1.16 & 0.9 & 0.91 &  1 \\
                  & no. all int. & 1.01 & 1.13 & 1.05 & 1.34 & 1.02 & 1 & 1 \\
                  & av. gen. int. len. & 118.95 & 81.7 & 119.17 & 82.6 & 15.68 & 15.81 & 55.7\\
 \hline\hline
            & coverage & 100 & 96 & 100 & 96 & 84 & 86 & 75 \\
Wave & prop. gen. int. & 1 & 0.99 & 1 & 0.99 & 0.94 & 0.93 & 0.81 \\
            & no. gen. int. & 1.87  & 2.49  & 2.57 & 3.03 & 2.87 & 1.75 & 2.27\\
            & no. all int. & 1.87 & 2.53 & 2.57 & 3.07 & 3.05 & 1.89 & 2.65 \\
            & av. gen. int. len. & 104.78 & 86.01 & 113.07 & 90.09 & 26.3 & 40.02 & 75.71\\
 \hline\hline
            & coverage & 100 & 100 & 100 & 100 & 77 & 95 & 75 \\
Wide Teeth & prop. gen. int. & 1 & 1 & 1 & 1 & 0.87 & 0.92 & 0.62\\
            & no. gen. int. & 0.77 & 1.78  & 1 & 2.49 & 2.88 & 0.65 & 0.53 \\
            & no. all int. & 0.77 & 1.78 & 1 & 2.49 & 3.23 & 0.7 & 0.79 \\
            & av. gen. int. len. & 84.61 & 59.67 & 93.65 & 65.48 & 24.77 & 29.95 & 82.7\\
 \hline\hline
            & coverage & 100 & 100 & 100 & 100 & 50 & 88 & 24 \\
Teeth 10 & prop. gen. int. & 1 & 1 & 1 & 1 & 0.94 & 0.95 & 0.46\\
            & no. gen. int. & 3.34 & 6.76  & 5.08 & 9.18 & 11.44 & 1.92 & 1.66 \\
            & no. all int. & 3.34 & 6.76 & 5.08 & 9.18 & 12.24 & 2.1 & 3 \\
            & av. gen. int. len. & 20.74 & 12.41 & 23.01 & 13.62 & 6.94 & 8.19 & 21.24\\
 \hline\hline
            & coverage & 100 & 100 & 100 & 100 & -- & -- & 52 \\
Blocks & prop. gen. int. & 1 & 1 & 1 & 1 & -- & -- & 0.89\\
            & no. gen. int. & 7.25 & 8.24  & 9.42 & 10.41 & -- & -- & 7.56 \\
            & no. all int. & 7.25 & 8.24 & 9.42 & 10.41 & -- & -- & 8.42 \\
            & av. gen. int. len. & 79.5 & 69.74 & 92.64 & 80.7 & -- & -- & 76.46\\
 \hline
 
 \end{tabular}
\caption{Results for each model+method combination: ``coverage'' is the number of times, out of 100 simulated sample paths, that the respective model+method combination did not return a spurious interval of significance; ``prop. gen. int." is the average (over 100 simulated sample paths) proportion of genuine intervals out of all intervals returned, if any (if none are returned, the corresponding 0/0 ratio is ignored in the average); ``no. gen. int." is the average (over 100 sample paths) number of genuine intervals returned; ``no. all int.'' is the average (over 100 sample paths) number of all intervals returned; ``av. gen. int. len." is the average (over 100 sample paths) length of a genuine interval returned in the respective model+method combination. Note 1: for the Teeth 10 signal only, the corresponding averages are over 50 simulated sample paths as the BP method crashed for sample path indexed 52. Note 2: the BP methods were too slow to execute for the Blocks model.\label{tab:models2res}}
}
\end{table}

\begin{table}
{\small
\centering
\begin{tabular}{ |c|c|c|c|c| } 
\hline
model & NSP & NSP-SIM & NSP-O & NSP-SIM-O \\
\hline\hline
Noise 300 (0.1) & 100 & 97 & 100 & 97\\
Noise 300 (0.3) & 100 & 99 & 100 & 99\\
Noise 300 (0.5) & 100 & 100 & 100 & 100\\
Noise 300 (0.7) & 100 & 100 & 100 & 100\\
\hline
\end{tabular}
\caption{Numbers of times, out of 100 simulated sample paths of each null model, that the respective method indicated no intervals of significance. Here, the process $Z_t$ is autocorrelated and the $\sigma$ is set to its true long-run standard deviation, rather than being estimated via MAD. ``Noise 300 ($a$)'' means a sample path of length 300 with marginal variance 1 and AR(1) autocorrelation structure with AR coefficient equal to $a$.\label{tab:nullres2}}
}
\end{table}

\begin{table}
{\scriptsize
\centering
\begin{tabular}{ |c|c|c|c|c|c| } 
\hline
model & attribute & NSP & NSP-SIM & NSP-O & NSP-SIM-O \\
\hline\hline
                  & coverage & 100 & 97 & 100 & 97 \\
Single 300 (0.1) & prop. gen. int. & 1 & 0.98 & 1 & 0.98 \\
                  & no. gen. int. & 0.96  & 1  &  0.97  & 1.05  \\
                  & no. all int. & 0.96 & 1.03 & 0.97 & 1.08  \\
                  & av. gen. int. len. & 128.91 & 94.25 & 128.89 & 95.11 \\
 \hline\hline
                  & coverage & 100 & 100 & 100 & 100 \\
Single 300 (0.3) & prop. gen. int. & 1 & 1 & 1 & 1 \\
                  & no. gen. int. & 0.82  & 0.96  &  0.83  & 0.98  \\
                  & no. all int. & 0.82 & 0.96 & 0.83 & 0.98  \\
                  & av. gen. int. len. & 192.72 & 142.61 & 192.76 & 142.86 \\
 \hline\hline
                  & coverage & 100 & 100 & 100 & 100 \\
Single 300 (0.5) & prop. gen. int. & 1 & 1 & 1 & 1 \\
                  & no. gen. int. & 0.42  & 0.74  &  0.42  & 0.74  \\
                  & no. all int. & 0.42 & 0.74 & 0.42 & 0.74  \\
                  & av. gen. int. len. & 228.43 & 194.41 & 228.43 & 194.41 \\
 \hline\hline
                  & coverage & 100 & 100 & 100 & 100 \\
Single 300 (0.7) & prop. gen. int. & 1 & 1 & 1 & 1 \\
                  & no. gen. int. & 0.04  & 0.12  &  0.04  & 0.12  \\
                  & no. all int. & 0.04 & 0.12 & 0.04 & 0.12  \\
                  & av. gen. int. len. & 263.25 & 227.25 & 263.25 & 227.25 \\
 \hline 
 \end{tabular}
\caption{Results for each model+method combination under auto-correlation: 
the process $Z_t$ is autocorrelated and the $\sigma$ is set to its true long-run standard deviation, rather than being estimated via MAD. ``Single 300 ($a$)'' means the Single 300 signal plus a sample path of length 300 with marginal variance 1 and AR(1) autocorrelation structure with AR coefficient equal to $a$.
\label{tab:models2res2}}
}
\end{table}

In this section, we demonstrate numerically that the guarantee offered by Theorem \ref{th:main} holds for NSP in practice over a variety of Gaussian models with and without change-points in Scenario 1.
We start by describing the competing methods. ``NSP'' is the NSP method executed with a deterministic grid using $M = 1000$ intervals, with the threshold chosen as in Section \ref{sec:gauss} and no interval overlaps, i.e. $\tau_L = \tau_R = 0$; $\sigma$ is estimated via MAD. ``NSP-SIM'' is like ``NSP'' but uses the simulation-based thresholds of Section \ref{sec:simthresh}. ``NSP-O'' is like ``NSP'' but uses the overlap functions defined in (\ref{eq:overlap}). ``NSP-SIM-O'' is like ``NSP-SIM'' but uses the overlap functions as in ``NSP-O''. ``BP'' is the method of \cite{bp03} as implemented in the routine \verb+breakpoints+ of R package \verb+strucchange+ (version 1.5-3) with the minimum segment size set to 2; the number of change-points is chosen by BIC, and confidence intervals are then formed conditionally on the estimated model by using the \verb+confint.breakpointsfull+ routine, with the significance level Bonferroni-corrected for the estimated number of change-points. ``BP-LIM'' is like ``BP'' but with the number of change-points limited from above by the number of intervals returned by NSP (or one if NSP returns no intervals). ``SMUCE'' is the method of \cite{fms14}, for which the execution is \verb+stepR::stepFit(data, alpha, confband=TRUE)+; we use version 2.1-3 of \verb+stepR+.

We begin with null models, by which we mean models (\ref{eq:signoise}) for which $f_t$ is constant throughout, i.e. $N = 0$. For null models, Theorem \ref{th:main} promises that NSP at level $\alpha$ returns no intervals of significance with probability at least $1-\alpha$. In this section, we use $\alpha = 0.1$. There are similar parameters in BP, BP-LIM and SMUCE, and they are also set to 0.1. All models used are listed in Table \ref{tab:models}.

Table \ref{tab:nullres} shows the null model results. All methods tested keep the nominal size well for both null signals; note that the empirical binomial proportion of $0.86$, observed in NSP-SIM and NSP-SIM-O, is only insignifcantly (in the sense of the binomial $Z$-test) different from the nominal value of 0.9, with the sample size used (100 simulated sample paths).

We now discuss performance for signals with change-points ($N > 0$). 
For each model and method tested, we evaluate the following aspects: the empirical coverage (i.e. whether at least $(1-\alpha)100\%$ of the simulated sample paths are such that any intervals of significance returned contain at least one true change-point each); if any intervals are returned, the proportion of those that are genuine (i.e. the proportion of those intervals returned that contain at least one true change-point); the number of genuine intervals; the number of all intervals; and the average length of genuine intervals.
Table \ref{tab:models2res} shows the results; note that the Wide Teeth model is challenging from the point of view of detection for all methods tested, but this should not surprise on visual inspection of its sample paths.

The BP method suffers from under-coverage in all models tested with the exception of Single 300; this is the most pronounced for Teeth 10, for which the empirical coverage is only 50 (to the nominal 90). BP-LIM (a method designed not to over-detect the true number of change-points) does not suffer from the same problem (with the exception of Single 100, for which it under-covers slightly); however, the price to pay for the mostly satisfactory coverage performance of BP-LIM is the fact that it only detects a small proportion of the true change-points: for example, on average 1.75 out of 3 for Wave, and 1.92 out of 13 for Teeth 10. The message is that in the presence of under-detection (as in BP-LIM), conditional confidence intervals can be capable of offering correct unconditional coverage; but this advantage disappears if more realistic change-point models are chosen and post-equipped with conditional confidence intervals (as in BP).
SMUCE suffers from under-coverage in most of the models tested, most notably in Teeth 10 (coverage 24) and Blocks (52).

All of the NSP-* methods offer correct coverage for all the signals tested (empirical coverage of $\ge 90$ to the nominal 90). As expected, the coverage of the -SIM versions does not exceed that of their theoretical threshold counterparts. Being based on lower thresholds, the -SIM versions also return more genuine intervals on average, which are in addition on average shorter. Also as expected, the -O versions return more intervals on average than the corresponding non-O versions.

We further test the NSP-* in the presence of noise autocorrelation as follows. We modify the Noise 300 and Single 300 signals of Table \ref{tab:models} so that the innovations used are simulated from an AR(1) process with the marginal variance set to 1 and the autocorrelation coefficient spanning the set $0.1, 0.3, 0.5$ and $0.7$. Instead of estimating $\sigma$ via MAD (which would lead to incorrect behaviour for autocorrelated noise), we set it to the true long-run standard deviation of the relevant noise process, as per the discussion of Section \ref{sec:gauss},  . Tables \ref{tab:nullres2} and \ref{tab:models2res2} confirm the correct coverage behaviour of all NSP-* methods in these settings. Note, in Table \ref{tab:models2res2}, the increasing detection challenge in the Single 300 ($a$) model as $a$ increases to 0.7. Satisfactory estimation of the long-run standard deviation, especially in the presence of change-points, is a difficult problem but several solutions exist; we refer the reader in particular to \cite{dsv18}.

We now illustrate NSP and NSP-SIM-O on the Blocks model (simulated with random seed set to 1).
This represents a difficult setting for change-point detection, with practically all state of the art multiple change-point detection
 methods failing to estimate all 11 change-points 
with high probability \citep{af18}. A high degree of uncertainty with regards to the existence and locations of change-points can be expected.

\begin{figure}[t]
\centering
\begin{minipage}{.33\textwidth}
  \centering
  \includegraphics[width=\linewidth]{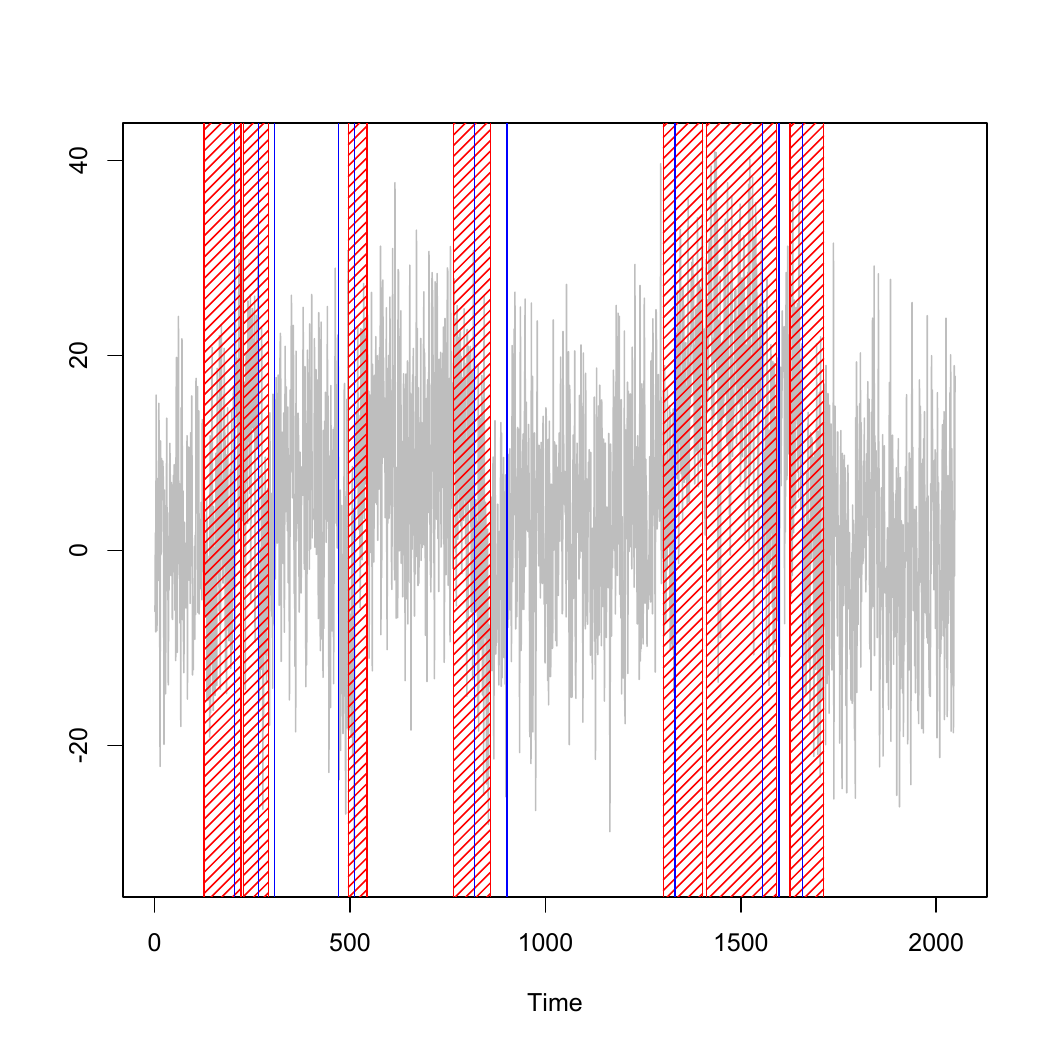}
\end{minipage}%
\begin{minipage}{.33\textwidth}
  \centering
  \includegraphics[width=\linewidth]{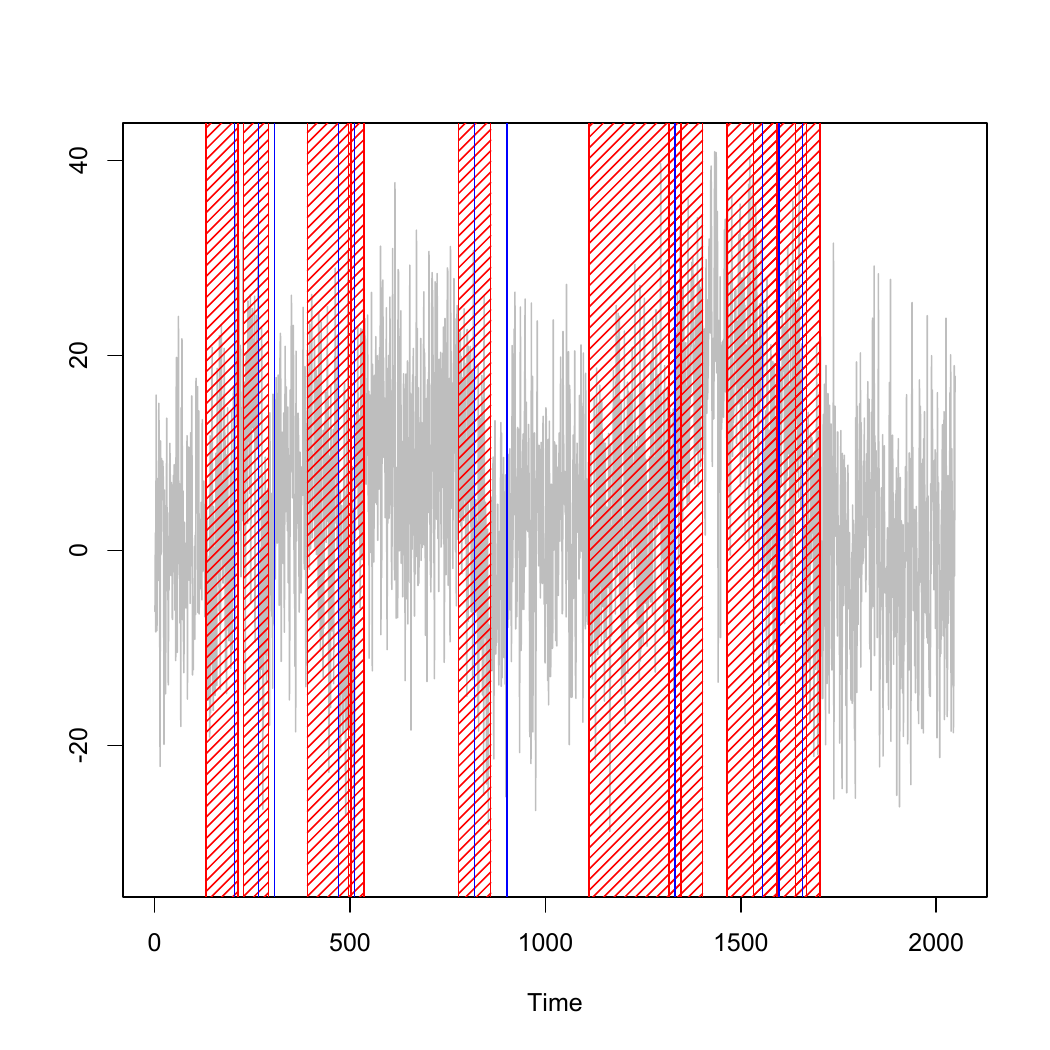}
\end{minipage}
\begin{minipage}{.33\textwidth}
  \centering
  \includegraphics[width=\linewidth]{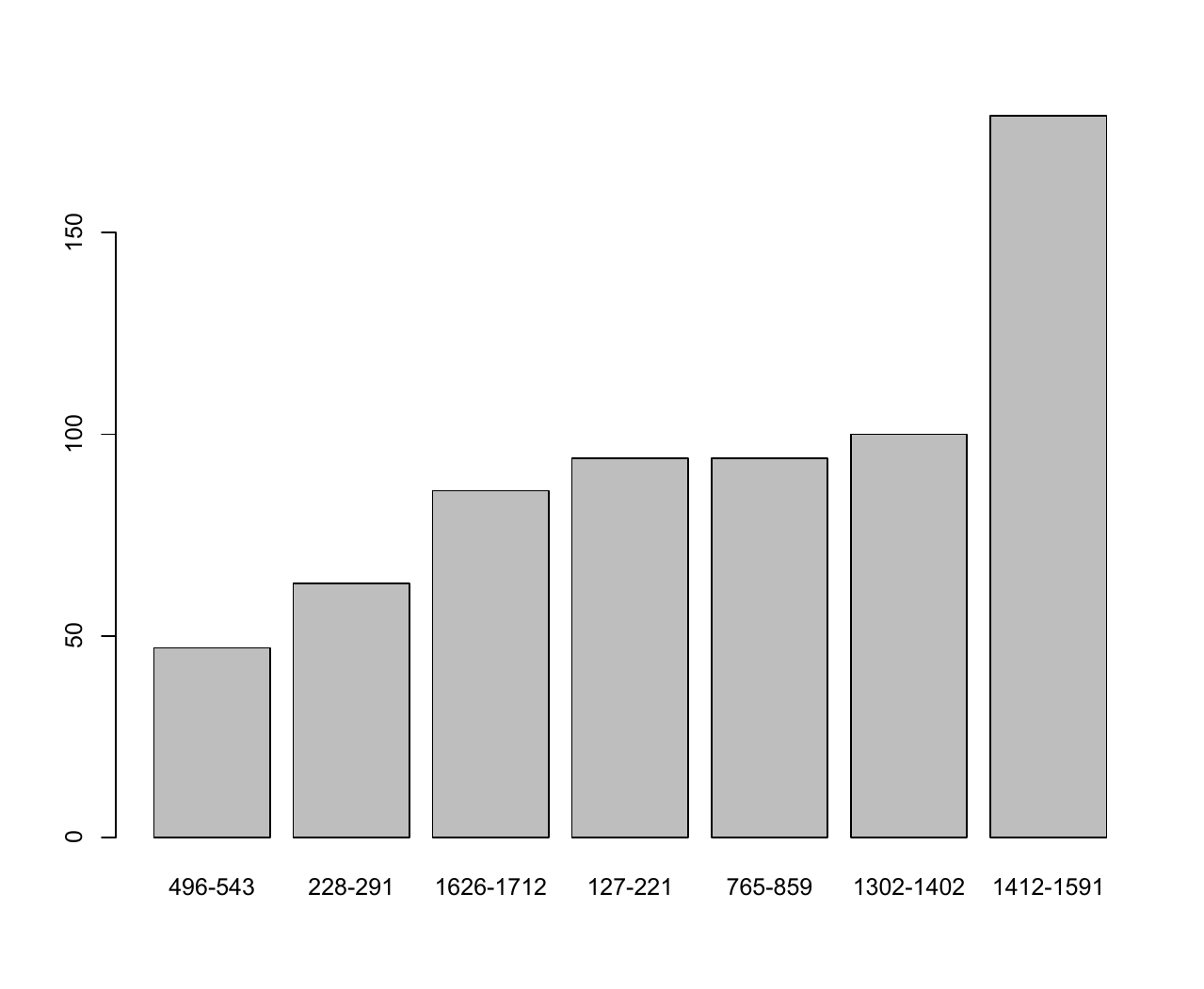}
\end{minipage}%
\caption{Left: realisation $Y_t$ of noisy {\tt blocks} with $\sigma=10$ (light grey), true change-point locations (blue), NSP intervals of significance ($\alpha=0.1$, shaded red). Middle: the same for NSP-SIM-O. Right: ``prominence plot" -- bar plot of $\tilde{e}_i-\tilde{s}_i$, $i=1,\ldots,7$, plotted in increasing order, where $[\tilde{s}_i, \tilde{e}_i]$ are the NSP significance intervals; the labels are ``$\tilde{s}_i$--$\tilde{e}_i$''. See Section \ref{sec:csim} for more details.\label{fig:blocks}}
\end{figure}

NSP returns 7 intervals of significance, shown in the left-hand plot of Figure \ref{fig:blocks}. We recall that at a fixed significance level, it is not the aim of the NSP procedure to detect all change-points. The correct interpretation of the result is that we can be at least $100(1 - \alpha)\% = 90\%$ certain that each of the intervals returned by NSP covers at least one true change-point. This coverage holds for this particular sample path, with exactly one true change-point being located within each interval of significance.

NSP enables the following definition of a change-point hierarchy. A hypothesised change-point contained in the detected interval of significance $[\tilde{s}_1, \tilde{e}_1]$ is considered more prominent
than one contained in $[\tilde{s}_2, \tilde{e}_2]$ if $[\tilde{s}_1, \tilde{e}_1]$ is shorter than $[\tilde{s}_2, \tilde{e}_2]$. The right-hand plot of Figure \ref{fig:blocks} shows a ``prominence plot'' for this output of the NSP procedure.

The output of NSP-SIM-O is in the middle plot of Figure \ref{fig:blocks}. This version of the procedure returns 10 intervals of significance, such that (a) each interval
covers at least one true change-point, and (b) they collectively cover 9 of the signal's $N = 11$ change-points, the only exceptions being $\eta_3 = 307$ and $\eta_7 = 901$.

Finally, we mention computation times for this particular example, on a standard 2015 iMac: 14 seconds (NSP, $M=1000$), 24 seconds (NSP-O, $M=1000$),
1.6 seconds (NSP, $M=100$), and 2.6 seconds (NSP-O, $M=100$).

\subsection{Scenario 2 -- piecewise linearity}
\label{sec:pl}

\begin{figure}
\centering
\begin{minipage}{.33\textwidth}
\centering
\includegraphics[width=\linewidth]{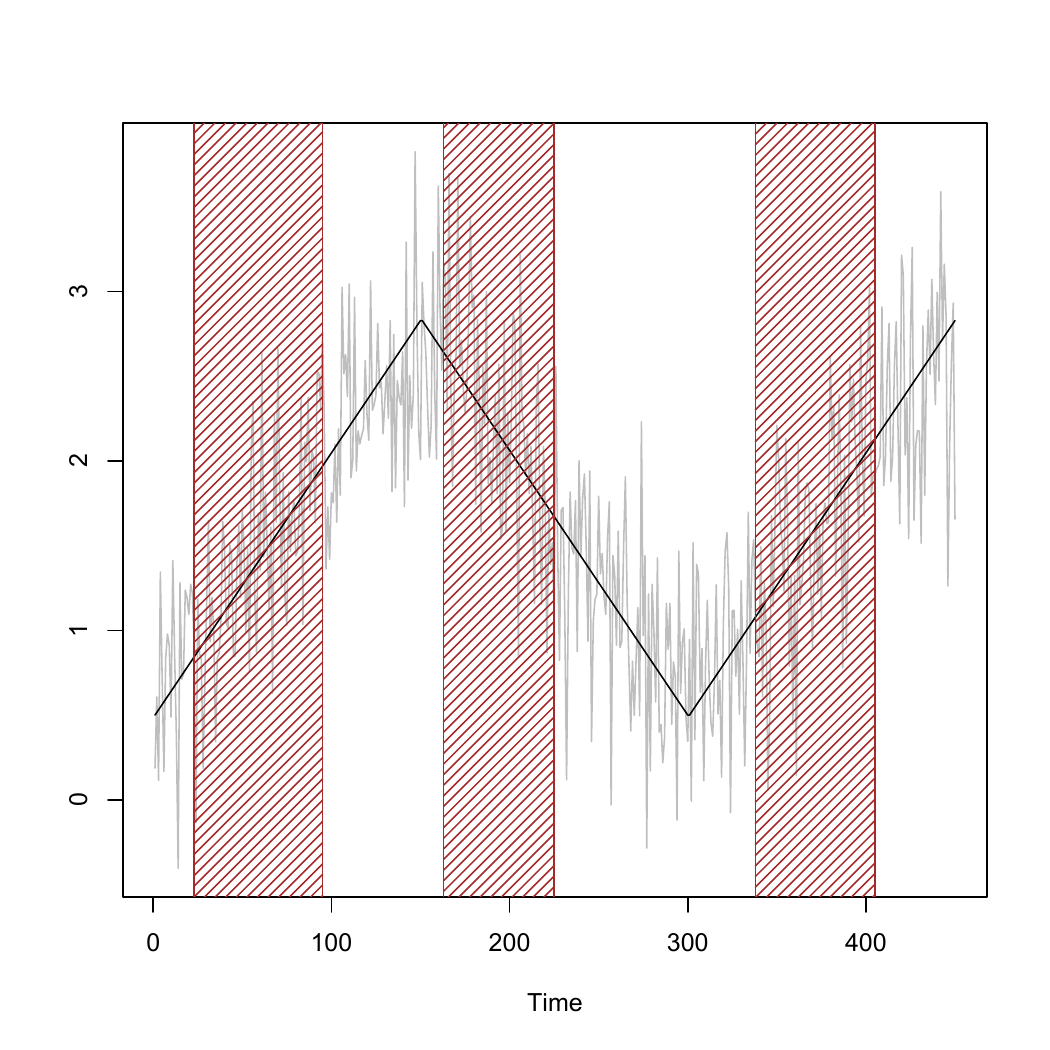}
\end{minipage}%
\begin{minipage}{.33\textwidth}
\centering
\includegraphics[width=\linewidth]{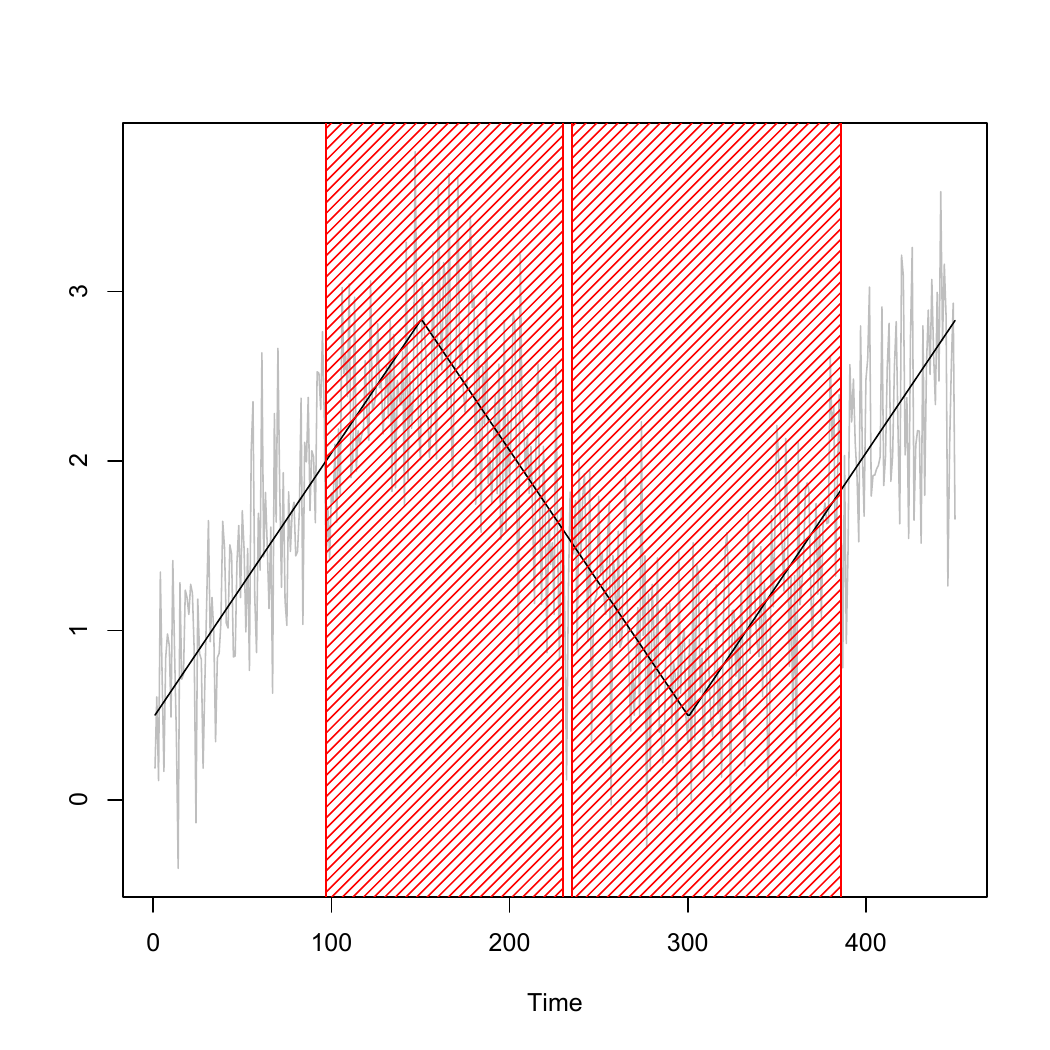}
\end{minipage}%
\begin{minipage}{.33\textwidth}
\centering
\includegraphics[width=\linewidth]{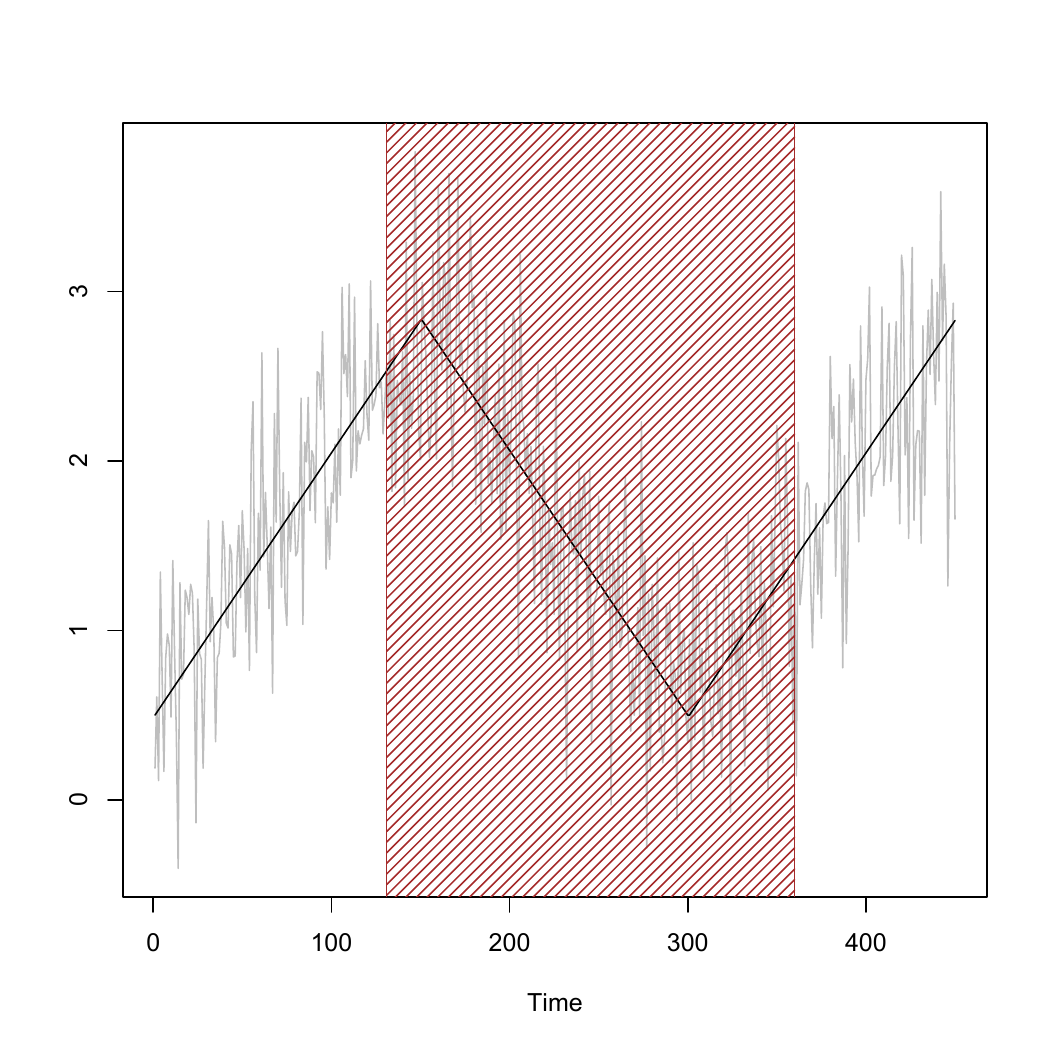}
\end{minipage}%
\caption{Noisy (light grey) and true (black) {\tt wave2sect} signal, with NSP$_q$ significance intervals for $q=0$ (left, misspecified model), $q=1$ (middle, well-specified model), $q=2$ (right, over-specified model). See Section \ref{sec:pl} for more details. \label{fig:wave}}
\end{figure}

We consider the continuous, piecewise-linear \verb+wave2sect+ signal, defined as the first 450 elements of the \verb+wave2+ signal from \cite{bcf16}, contaminated with i.i.d. Gaussian noise with $\sigma = 0.5$. The signal and a sample path are shown in Figure \ref{fig:wave}.
In this model, we run the NSP procedure, with no overlaps and with the other parameters set as in Section \ref{sec:csim}, (wrongly or correctly) assuming the following, where $q$ denotes the postulated degree of the underlying piecewise polynomial: (a) $q = 0$, which wrongly assumes that the true signal is piecewise constant; (b) $q = 1$, which 
assumes the correct degree of the polynomial pieces making up the signal; (c) $q = 2$, which over-specifies the degree.
We denote the resulting versions of the NSP procedure by NSP$_q$ for $q = 0, 1, 2$. The intervals of significance returned by all three NSP$_q$ methods are shown in Figure \ref{fig:wave}. Theorem
\ref{th:main} guarantees that the NSP$_1$ intervals each cover a true change-point with probability of at least $1-\alpha = 0.9$ and this behaviour occurs in this particular realisation. The same guarantee
holds for the over-specified situation in NSP$_2$, but there is no performance guarantee for NSP$_0$.

\comment{
The total length of the intervals of significance returned by NSP$_q$ for a range of $q$ can potentially be used to aid the selection of the `best' $q$. To illustrate this potential use, note that the total length of the NSP$_0$
intervals of significance is much larger than that of NSP$_1$ or NSP$_2$, and therefore the piecewise-constant model would not be preferred here on the grounds that the data deviates from it over a large proportion of its domain.
The total lengths of the intervals of significance for NSP$_1$ and NSP$_2$ are very similar, and hence the piecewise-linear model might (correctly) be preferred here as offering a good description of a similar portion of the data, with fewer parameters than the piecewise-quadratic model.}

\subsection{Self-normalised NSP}
\label{sec:snex}

\begin{figure}
\centering
\begin{minipage}{.5\textwidth}
  \centering
  \includegraphics[width=.7\linewidth]{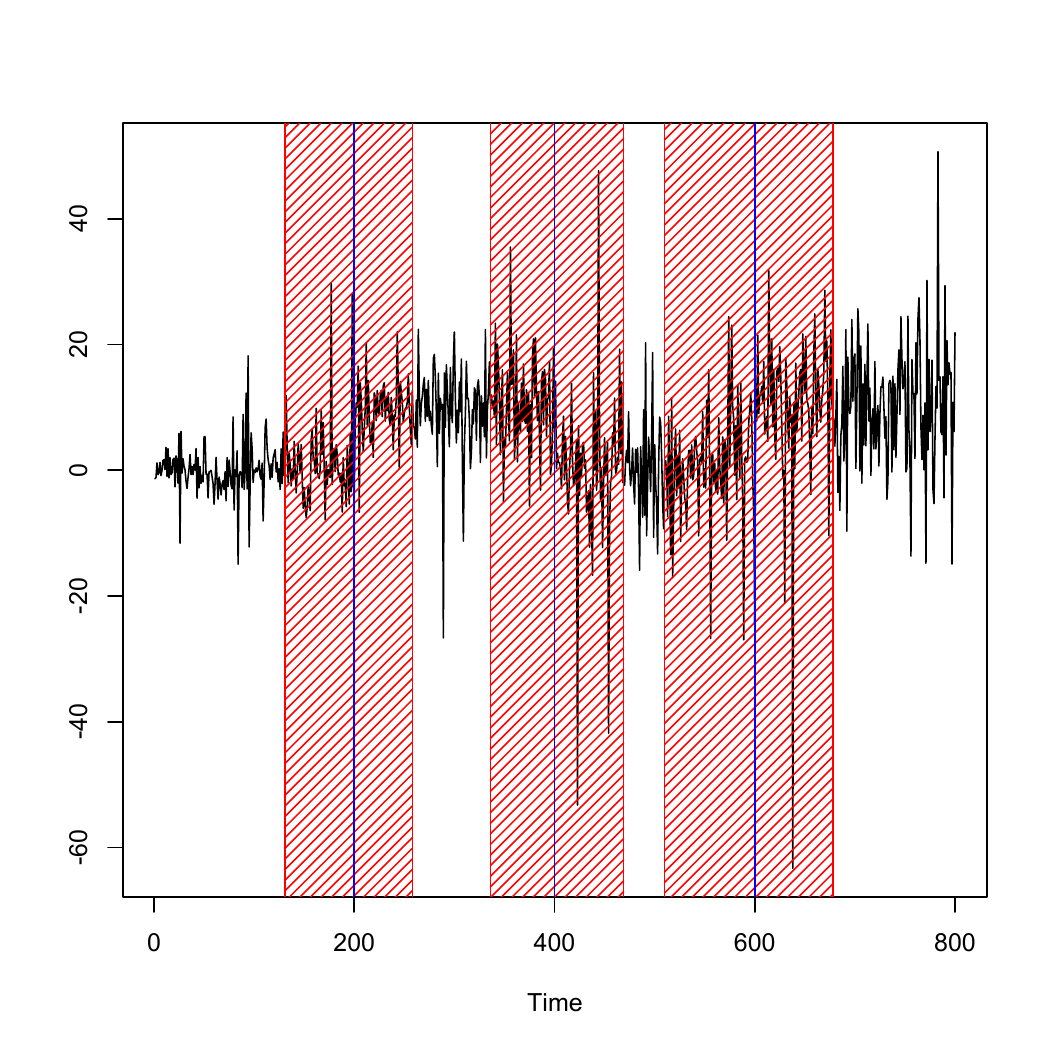}
\end{minipage}%
\begin{minipage}{.5\textwidth}
  \centering
  \includegraphics[width=.7\linewidth]{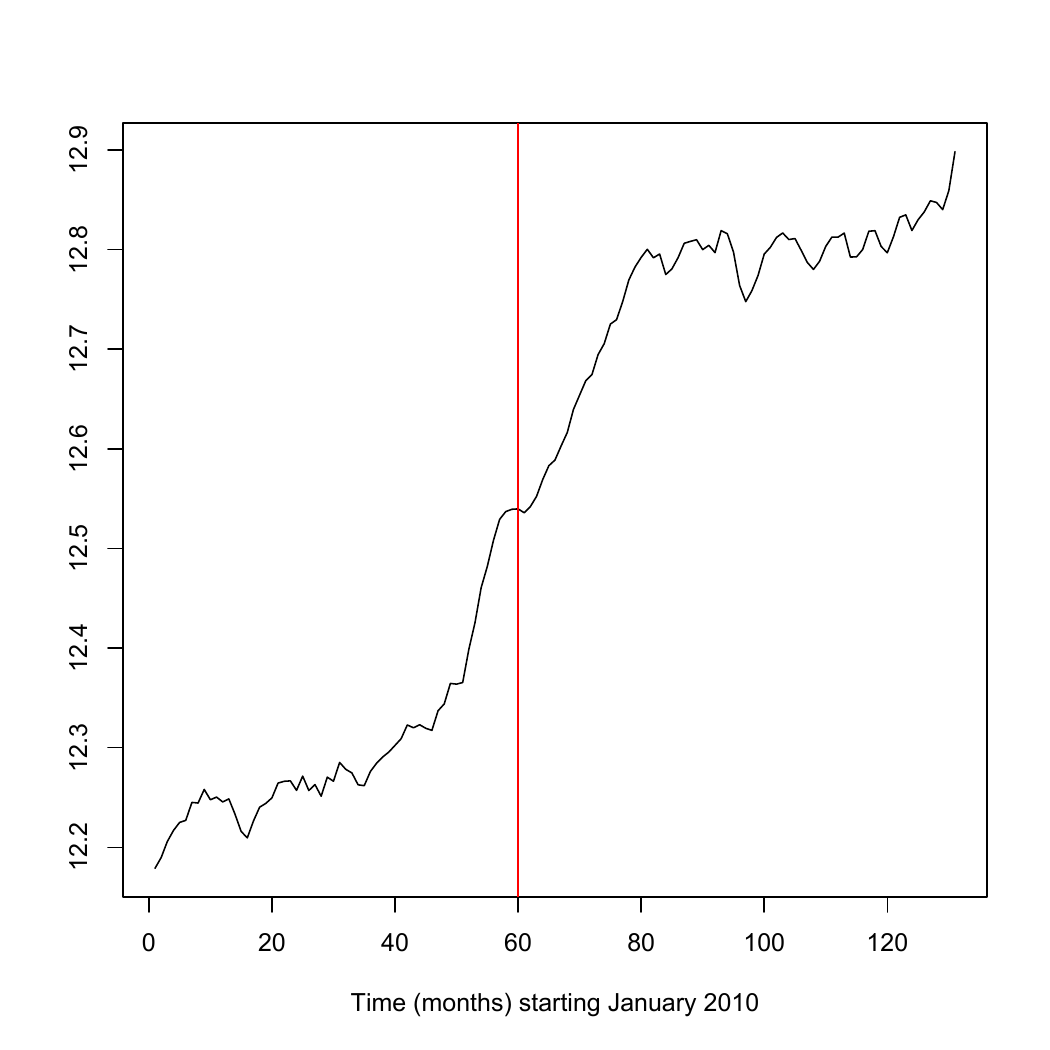}
\end{minipage}
\caption{Left: {\tt squarewave} signal with heterogeneous $t_4$ noise (black), self-normalised NSP intervals of significance (shaded red), true change-points (blue); see Section \ref{sec:snex} for details. Right: time series $Q_t$ for $t = 1, \ldots, 131$. Red: the centre of the (single) NSP interval of significance. See Section \ref{sec:nhp} for details.\label{fig:snex}}
\end{figure}

We briefly illustrate the performance of the self-normalised NSP. We define the piecewise-constant \verb+squarewave+ signal as taking the values of $0, 10, 0, 10$, each over a stretch of 200 time points. With the random seed set to 1, we contaminate it with a sequence of independent $t$-distributed random variables with 4 degrees of freedom, with the standard deviation changing linearly from $\sigma_1 = 2\sqrt{2}$ to $\sigma_{800} = 8\sqrt{2}$. The simulated dataset, showing the ``spiky'' nature of the noise, is in the left plot of Figure \ref{fig:snex}.

We run the self-normalised version of NSP with the following parameters: a deterministic equispaced interval sampling grid, $M = 1000$, $\alpha = 0.1$, $\epsilon = 0.03$, no overlap; the outcome is in the left
plot of Figure \ref{fig:snex}. Each true change-point is correctly contained within a (separate) NSP interval of significance, and we note that no spurious intervals get detected despite the heavy-tailed and heterogeneous
character of the noise.

In addition, we run the self-normalised NSP, with the parameters as above, on heavy-tailed versions of the Noise 300 and Single 300 models from Table \ref{tab:models}, in which the Gaussian innovations have been replaced with $t_3$-distributed innovations scaled to have marginal variance 1. For the thus-modified Noise 300 model, self-normalised NSP correctly identifies no intervals of significance in 100 out of 100 simulated sample paths. For the modified Single 300 model, self-normalised NSP correctly identifies one interval of significance in 100/100 simulated sample paths, with the average interval length of 124.54.


\section{Data examples}
\label{sec:data}

\subsection{The US ex-post real interest rate}
\label{sec:ex1}

\begin{figure}
\centering
\begin{minipage}{.5\textwidth}
  \centering
  \includegraphics[width=.7\linewidth]{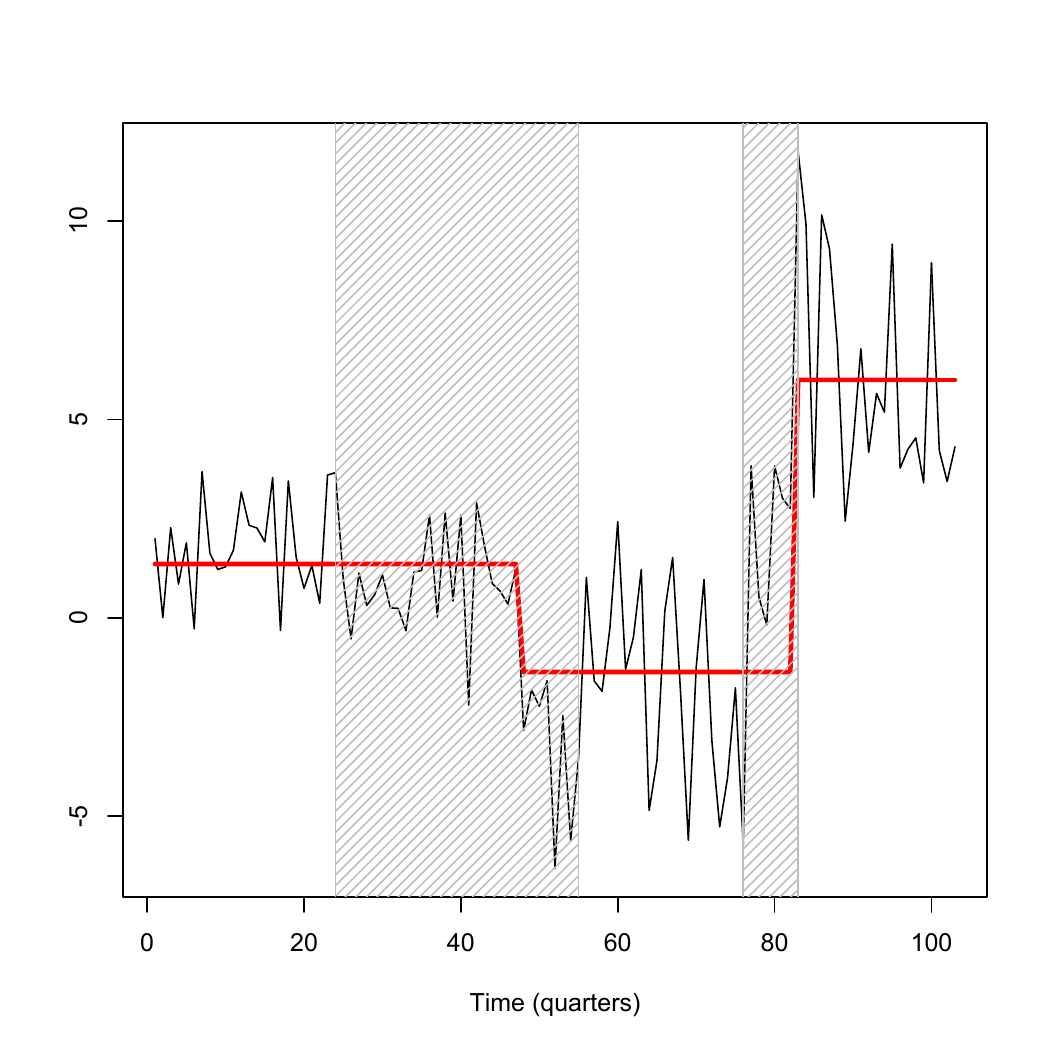}
\end{minipage}%
\begin{minipage}{.5\textwidth}
  \centering
  \includegraphics[width=.7\linewidth]{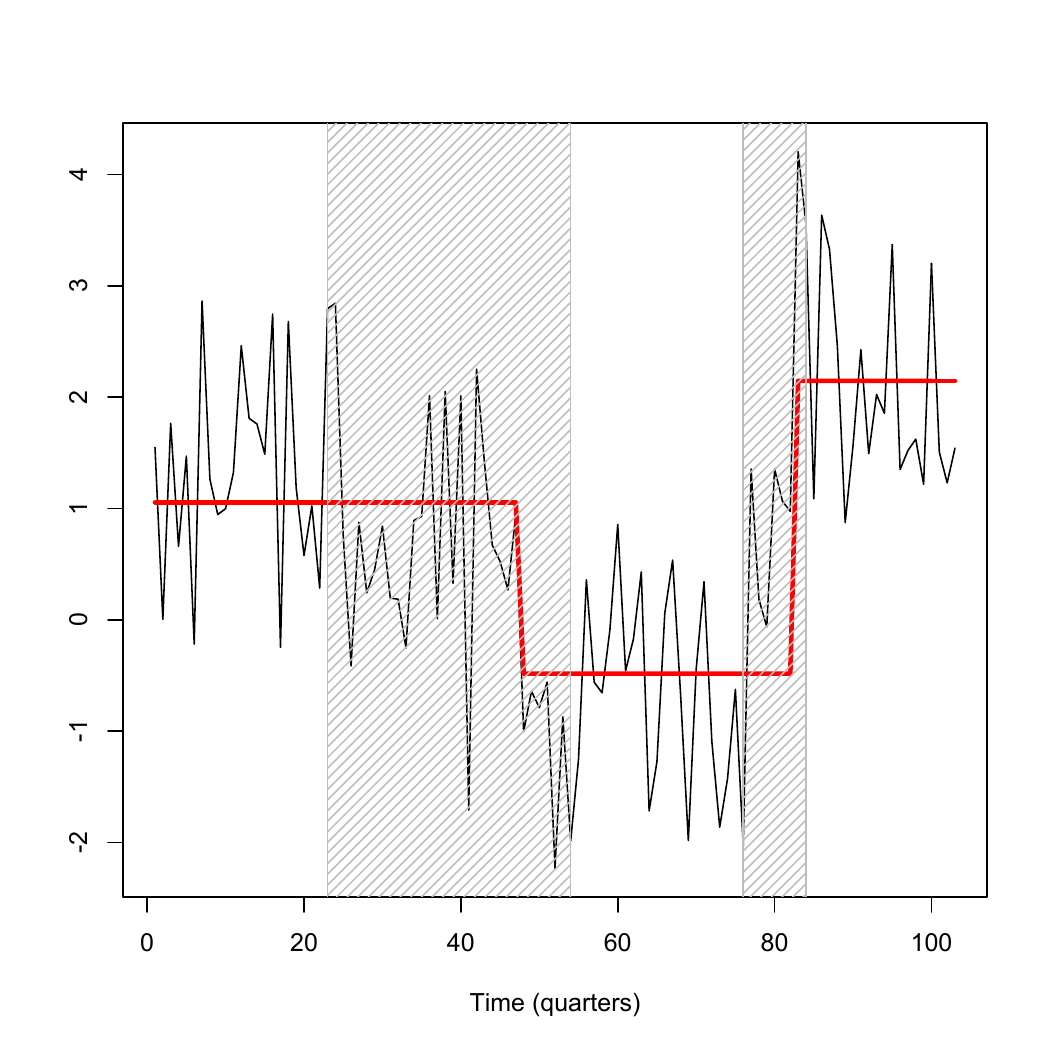}
\end{minipage}
\caption{Left plot: time series $Y_t$; right plot: time series $\tilde{Y}_t$; both with piecewise-constant fits (red) and intervals of significance returned by NSP (shaded grey). See Section \ref{sec:ex1} for a detailed description.\label{fig:real_dat}}
\end{figure}

We re-analyse the time series of US ex-post real interest rate (the three-month treasury bill rate deflated by the CPI inflation rate) considered in \cite{gp96} and
\cite{bp03}. The dataset is available at \url{http://qed.econ.queensu.ca/jae/datasets/bai001/}.
The dataset $Y_t$, shown in the left plot of Figure \ref{fig:real_dat}, is quarterly and the range is 1961:1--1986:3, so $t = 1, \ldots, T=103$. The arguments outlined in Section \ref{sec:suppl:data} of the appendix justify the applicability of NSP in this context.

We first perform a naive analysis in which we assume our Scenario 1 (piecewise-constant mean) plus i.i.d. $N(0, \sigma^2)$ innovations. This is only so we can obtain a rough segmentation
which we can then use to adjust for possible heteroscedasticity of the innovations in the next stage. We estimate $\sigma^2$ via $\hat{\sigma}^2_{MAD}$
and run the NSP algorithm with the following parameters: $M = 1000$, $\alpha = 0.1$, $\tau_L = \tau_R = 0$.
This returns the set $\mathcal{S}_0$ of two significant intervals: $\mathcal{S}_0 = \{ [24, 55], [76, 83]  \}$. We estimate the locations of the change-points within these two intervals via CUSUM fits on $Y_{24:55}$ and $Y_{76:83}$;
this returns $\hat{\eta}_1 = 47$ and $\hat{\eta}_2 = 82$. The corresponding fit is in the left plot of Figure \ref{fig:real_dat}. We then produce an adjusted dataset, in which we divide $Y_{1:47}, Y_{48:82}, Y_{83:103}$ by the respective estimated standard deviations of these sections of the
data. The adjusted dataset $\tilde{Y}_t$ is shown in the right plot of Figure \ref{fig:real_dat} and has a visually homoscedastic appearance. NSP run on the adjusted dataset with the same parameters produces the significant interval set $\tilde{\mathcal{S}}_0 = \{ [23, 54], [76, 84]  \}$. CUSUM fits on the corresponding data sections $\tilde{Y}_{23:54}, \tilde{Y}_{76:84}$ produce
identical estimated change-point locations $\tilde{\eta}_1 = 47$, $\tilde{\eta}_2 = 82$. The fit is in the right plot of Figure \ref{fig:real_dat}.

We could stop here and agree with \cite{gp96}, who also conclude that there are two change-points in this dataset, with locations within our detected intervals of significance. However, we note that the first interval, $[23, 54]$,
is relatively long, so one question is whether it could be covering another change-point to the left of $\tilde{\eta}_1 = 47$. To investigate this, we re-run NSP with the same parameters on $\tilde{Y}_{1:47}$ but find no intervals
of significance (not even with the lower thresholds induced by the shorter sample size $T_1 = 47$ rather than the original $T = 103$). Our lack of evidence for a third change-point contrasts with \cite{bp03}'s preference for a model with three change-points. 

However, the fact that the first interval of significance $[23,54]$ is relatively long could also be pointing to model misspecification. If the change of level over the first portion of the data were gradual rather than abrupt, we could naturally expect longer intervals of significance under the misspecified piecewise-constant model. To investigate this further, we now run NSP on $\tilde{Y}_{t}$ but in Scenario 2, initially in the piecewise-linear model ($q = 1$), which leads to one interval of significance: $\mathcal{S}_1 = \{  [57,   84]   \}$.

\begin{figure}
\centering
\begin{minipage}{.5\textwidth}
  \centering
  \includegraphics[width=.7\linewidth]{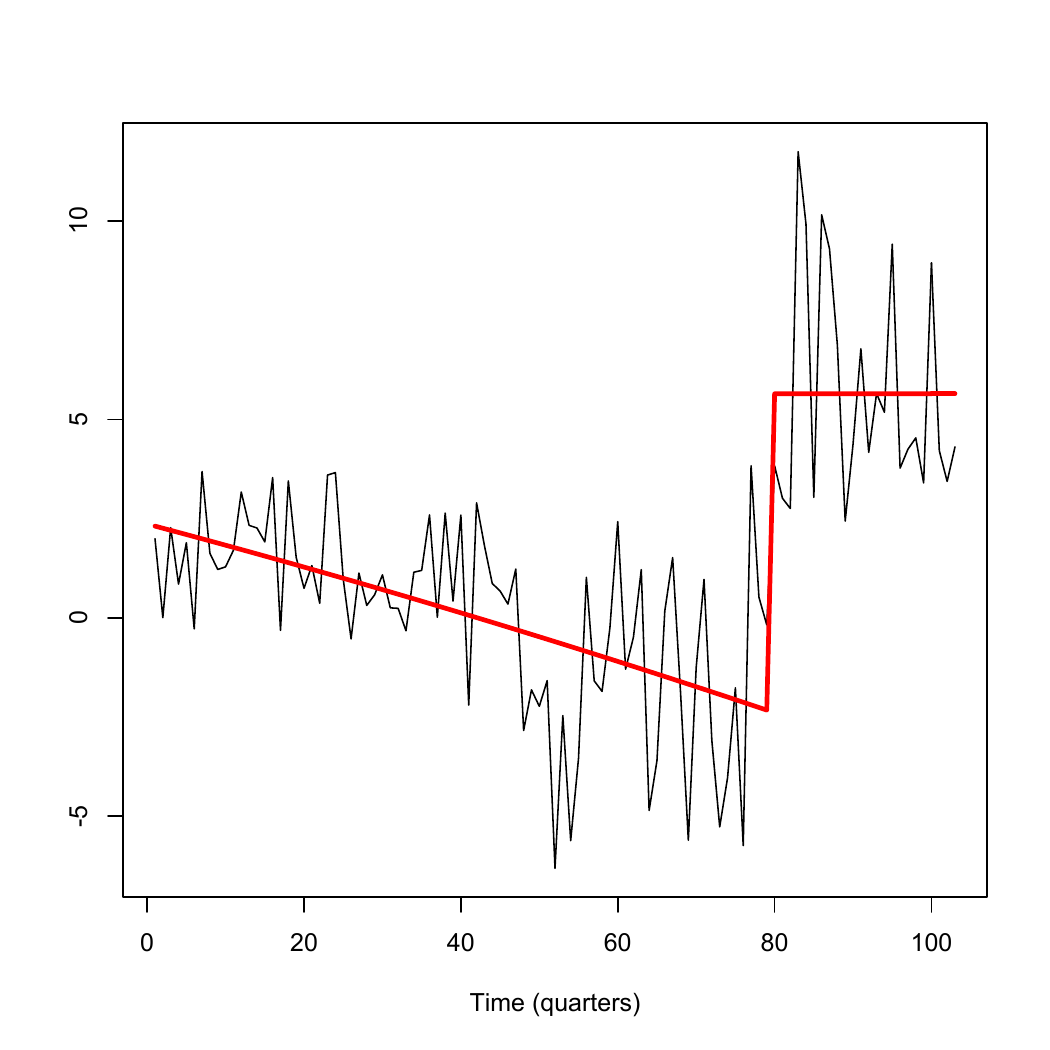}
\end{minipage}%
\begin{minipage}{.5\textwidth}
  \centering
  \includegraphics[width=.7\linewidth]{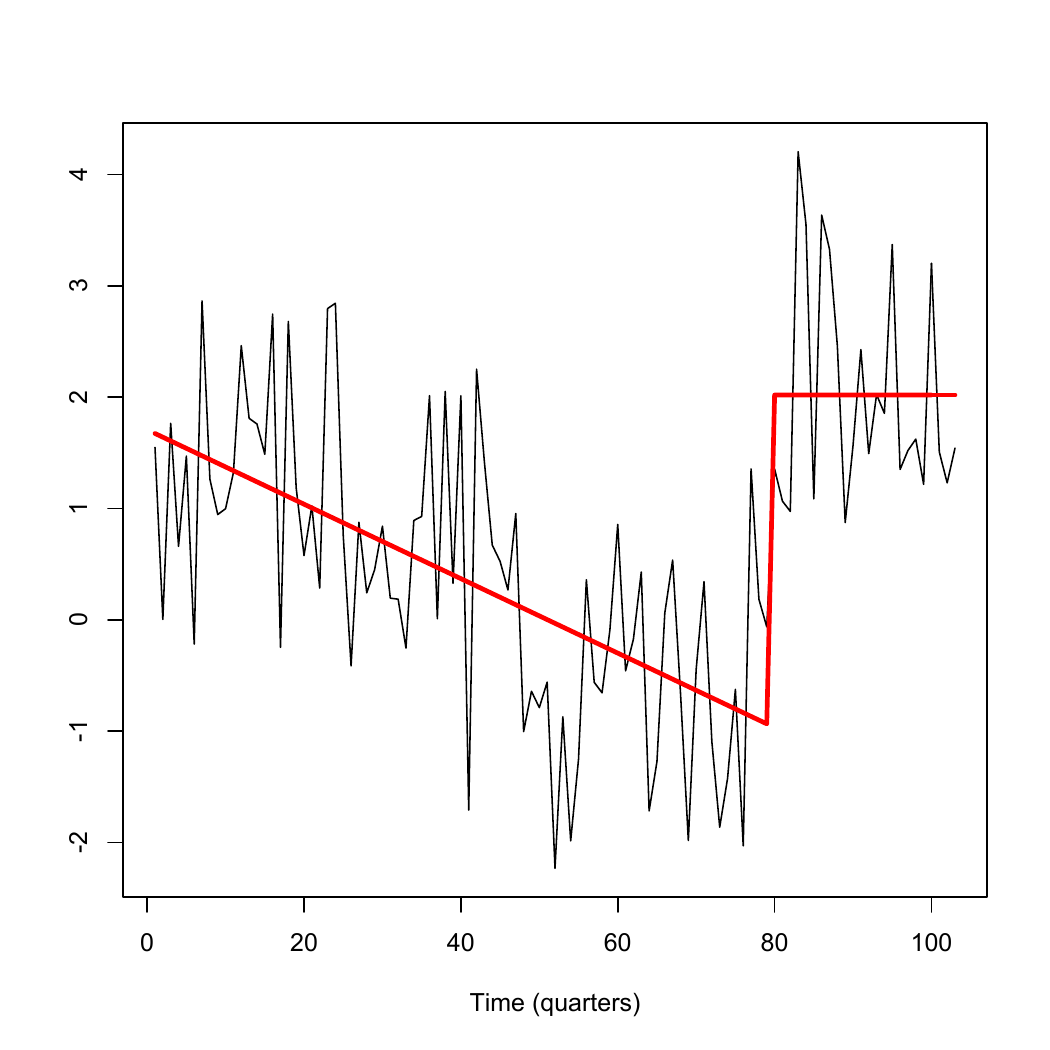}
\end{minipage}
\caption{Left plot: $Y_t$ with the quadratic+constant fit; right plot: $\tilde{Y}_t$ with the linear+constant fit. See Section \ref{sec:ex1} for a detailed description.\label{fig:real_dat_fits}}
\end{figure}

This raises the prospect of modelling the mean of $\tilde{Y}_{1:57}$ as linear. We produce such a fit, in which in addition the mean of $\tilde{Y}_{58:103}$ is modelled as piecewise-constant, with the change-point location $\tilde{\eta}_2 = 79$ found via a CUSUM fit on $\tilde{Y}_{58:103}$. We also produce an alternative fit in which the mean of $\tilde{Y}_{1:79}$ (up to the change-point) is modelled as linear, and the mean of $\tilde{Y}_{80:103}$ (post-change-point) as constant. This is in the right plot of Figure \ref{fig:real_dat_fits} and has a lower BIC value (9.52) than the piecewise-constant fit from the right plot of Figure \ref{fig:real_dat} (10.57). This is because the linear+constant fit uses four parameters, whereas the piecewise-constant fit uses five.

The viability of the linear+constant model for the scaled data $\tilde{Y}_t$ is encouraging because it raises the possibility of a model for the original data $Y_t$ in which the mean of $Y_t$ evolves smoothly in the initial part of the data. We construct a simple example of such a model by fitting the best quadratic on $Y_{1:79}$ (resulting in a strictly decreasing, slightly concave fit), followed by a constant on $Y_{80:103}$. The change-point location, 79, is the same as in the linear+constant fit for $\tilde{Y}_t$. The fit is in the left plot of Figure \ref{fig:real_dat_fits}. It is interesting to see that the quadratic+constant model for $Y_t$ leads to a slightly lower residual variance than the piecewise-constant model (4.9 to 4.94). Both models use five parameters. We conclude that more general piecewise-polynomial modelling of this dataset can be a viable alternative to the piecewise-constant modelling used in \cite{gp96} and \cite{bp03}. This example shows how NSP, beyond its usual role as an automatic detector of regions of significance, can also serve as a useful tool in achieving improved model selection.

\subsection{House prices in London Borough of Newham}
\label{sec:nhp}
We consider the average monthly property price $P_t$ in the London Borough of Newham, for all property types, recorded from January 2010 to November 2020 ($T = 131$) and accessed on 1st February 2021. The data is available on \url{https://landregistry.data.gov.uk/}. We use the logarithmic scale $Q_t = \log\,P_t$ and are interested in the stability of the autoregressive model $Q_t = b + a Q_{t-1} + Z_t$. Again, the arguments of Section \ref{sec:suppl:data} of the appendix justify the applicability of NSP here.

NSP, run on a deterministic equispaced interval sampling grid, with $M = 1000$ and $\alpha = 0.1$, with the $\hat{\sigma}^2_{MOLS}$ estimator of the residual variance (see Section \ref{sec:lt} of the appendix) and both with no overlap and with an overlap as defined in 
formula (\ref{eq:overlap}), returns a single interval of significance $[24, 96]$, which corresponds to a likely change-point location between December 2011 and December 2017. Assuming a possible change-point in the middle of this interval, i.e. in December 2014, we run two autoregressions (up to December 2014 and from January 2015 onwards) and compare the coefficients. Table \ref{tab:hp} shows the estimated regression coefficients (with their standard errors) over the two sections.

\begin{table}
\caption{Parameter estimates (standard error in brackets) in the autoregressive model of Section \ref{sec:nhp}.\label{tab:hp}}
\centering
\begin{tabular}{ |c|c|c| } 
\hline
Parameter & Jan 2010 -- Dec 2014 & Jan 2015 -- Nov 2020 \\
\hline
$b$ & -0.35 (0.2) & 0.66 (0.23) \\
$a$ & 1.03 (0.02) & 0.95 (0.02) \\
 \hline
\end{tabular}
\end{table}

It appears that both the intercept and the autoregressive parameter change significantly at the change-point. In particular, the change in the autoregressive parameter from 1.03 (standard error 0.02) to 0.95 (0.02) suggest a shift from a unit-root process to a stationary one. This agrees with a visual assessment of the character of the process in the right plot of Figure \ref{fig:snex}, where it appears that the process is more `trending' before the change-point than it is
after, where it exhibits a conceivably stationary behaviour, particularly from the middle of 2016 or so. Indeed, the average monthly change in $Q_t$ over the time period Jan 2010 -- Dec 2014 is $0.0061$, larger than the corresponding average change of $0.0052$ over Jan 2015 -- Nov 2020.




\appendix

\section*{Appendix}

\section{Additional literature review}
\label{sec:suppl:lit}

We first comment in more detail on the UD max and WD max tests of \cite{bp98} and \cite{bp03} and their relationship to NSP. \cite{bp03} write:
\begin{quote}
A useful strategy is to first look at the UD max or WD max tests to see if at least one break is present. If these indicate the presence of at least one break, then the number of breaks can be decided based upon a sequential examination of the sup $F(l+1 | l)$ statistics constructed using global minimizers for the break dates (i.e. ignore the test $F(1|0)$ and select $m$ such that the tests sup $F(l+1|l)$ are insignificant for $l \ge m$. This method leads to the best results and is recommended for empirical applications.
\end{quote}

For the purpose of this discussion, we label the process above the `Improved Sequential Procedure' (ISP). \cite{bp03} do not formulate or prove the inferential properties of the $m$ selected by ISP. For a procedure that selects the number of change-points, the control of global significance would have to mean, in particular, a guarantee that the true number of change-points is at least as high as the estimated number, with at least $1-\alpha$ probability. NSP provides such a statement as a simple corollary of Theorem \ref{th:main} in the main paper, but ISP is a complex sequential process put together from separate, non-independent, conditionally applied tests, and the exact guarantees for the resulting output ($m$) have not been shown.

The next difference is that the UD max and WD max tests require the provision of the maximum number of change-points, but NSP does not require this, thereby eliminating the risk of providing too low a maximum by the user.

Furthermore, the ISP test only concerns the number of change-points, but not their locations: inference for locations in \cite{bp98} and \cite{bp03} is carried out later, {\em conditionally} on the number of change-points and on their estimated locations. Not only that, but also the obtained conditional confidence intervals are asymptotic in nature and are only valid for large sample sizes (unknown to the user). By contrast, NSP provides a single, clear, joint, finite-sample guarantee for the number of change-points and for their locations: it flags up disjoint regions in the data, each of which must contain at least one change-point with a global probability specified by the user. The NSP intervals of significance serve as ``unconditional'' confidence intervals (in contrast to the conditional CIs of \cite{bp98} and \cite{bp03}, whose conditionality on the number of estimated change-points and the estimated locations means that the user cannot be sure whether they contain change-points with a certain probability). The NSP guarantees are valid for any, even small, sample sizes.

Next, we discuss in more detail the most important
high-level differences between NSP and the approaches of \cite{fls20} and \cite{fs20}.
\begin{enumerate}
\item[(a)]
While \cite{fls20} and \cite{fs20} perform change-point location estimation as well as inference, NSP works on the principle of ``inference without location estimation''. This is a key property of NSP, which enables it to use an all-purpose multiscale test, whose distribution under the null is stochastically bounded by the scan
statistic of the corresponding true residuals $Z_t$, and is therefore independent of the scenario and of the design matrix $X$ used. This means that NSP is ready for use with any user-provided design matrix $X$,
and this will require no new calculations or coding, and will yield correct coverage probabilities. This is in contrast to the approach taken in \cite{fls20} and \cite{fs20}, in which,
because of their focus on location estimation, each
new scenario not already covered would involve new and fairly complicated approximations of the null distribution.
(We note that outside the change-point context,
the method for constructing 
confidence intervals for groups of variables in sparse high dimensional regression by \cite{m15} shares with NSP the attractive property of providing valid error control without assumptions on the design matrix.)
\item[(b)]
While in \cite{fls20} and \cite{fs20}, the user needs to be able to specify the significant signal shapes to look for, NSP searches for any deviations from local model linearity with respect to
specific regressors.
\item[(c)]
Out of our scenarios, \cite{fls20} and \cite{fs20} provide results under our Scenario 1 and Scenario 2 with linearity and continuity. Their results do not cover our Scenario 3 (linear regression with arbitrary $X$)
or Scenario 2 with linearity but not necessarily continuity, or Scenario 2 with higher-than-linear polynomials.
\item[(d)]
Thanks to its double use of the multiresolution sup-norm (in the local linear fit, and then in the test of this fit), NSP is able to handle regression with autoregression practically in the same way as without, in a stable manner and on arbitrarily short intervals, and does not
suffer from having to estimate the unknown (nuisance) AR coefficients accurately.
This is of importance, as change-point analysis under serial dependence in the data is a problem known to be difficult, and NSP offers a new approach to it, thanks to this feature.
\end{enumerate}

Finally, we provide additional references on the use of scan statistics.
In the literature, scaled partial sum statistics acting directly on the data are often combined into variants of scan statistics \citep{sv95, adh05, jcl10, w10, cw13, sa16, kmw20, mplw20}.
They are also used in estimators represented as the simplest (from the point of view of a certain regularity or smoothness functional)
fit to the data for which the empirical residuals are deemed to behave like the true residuals
\citep{fms14, dk01, dkm09, l16}.

\section{Discussion of the NSP algorithm}
\label{sec:suppl:dis}

We now comment on a few generic aspects of the NSP
algorithm as defined in the main paper.

\paragraph{Length check for $[s,e]$ in line 2}

Consider an interval $[s,e]$ with $e-s<p$. If it is known that the matrix $X_{s:e,\cdot}$ is of rank $e-s+1$ (as is the case,
for example, in Scenario 2, for all such $s, e$) then it is safe to disregard $[s,e]$, as the response $Y_{s:e}$ can then be explained exactly as a linear combination 
of the columns of $X_{s:e,\cdot}$, so it is impossible to assess any deviations from linearity of $Y_{s:e}$ with respect to $X_{s:e,\cdot}$. Therefore, if this rank condition holds,
the check in line 2 of NSP can be replaced with $e - s < p$, which (together with the corresponding modifications in lines 5--10) will reduce the computational effort if $p > 1$.
Having $p = p(T)$ growing with $T$ is possible in NSP, but by the above discussion, we must have $p(T)+1 \le T$ or otherwise no regions of significance will be found.

\paragraph{Sub-interval sampling}

Sub-interval sampling in lines 5--10 of the NSP algorithm is done to reduce the computational effort.
In the change-point detection literature (without inference considerations), Wild Binary Segmentation (WBS, \citeauthor{f14a}, \citeyear{f14a}) uses a random interval sampling
mechanism in which all or almost all intervals are sampled at the start of the procedure, i.e. with all or most intervals not being sampled recursively. The same style of interval sampling
is used in the Narrowest-Over-Threshold change-point detection (note: not change-point inference) algorithm \citep{bcf16} and is mentioned in passing in \cite{fls20}. Instead, NSP uses a different, recursive interval sampling mechanism, introduced in the change-point detection (not inference) context in Wild Binary Segmentation 2 (WBS2, \citeauthor{f20}, \citeyear{f20}).
In NSP (lines 5--10), intervals are sampled separately in each recursive call of the NSP routine. As argued in \cite{f20}, this enables more thorough exploration of the domain
$\{1, \ldots, T\}$ and hence better feature discovery than the non-recursive sampling style. We note that NSP can equally use random or deterministic interval selection mechanisms;
a specific example of a deterministic interval sampling scheme in a change-point detection context can be found in \cite{klbm20}. Our general preference is for NSP to be used with deterministic
sampling as it leads to reproducible results without the user having to fix the random seed.

\paragraph{Relationship to NOT}

The Narrowest-Over-Threshold (NOT) algorithm of \cite{bcf16} is a change-point detection procedure (valid in Scenarios 1 and 2) and comes with no
inference considerations. The common feature shared by NOT and NSP is that in their respective aims (change-point detection for NOT; locating regions of global significance for NSP)
they iteratively focus on the narrowest intervals on which a certain test (a change-point locator for NOT; a multiscale scan statistic on multiresolution sup-norm fit residuals for NSP) exceeds a threshold, but this is where
similarities end: apart from this common feature, the objectives, scopes and modi operandi of both methods are different.

\paragraph{Focus on the smallest significant regions}

Some authors in the inference literature also identify the shortest intervals (or smallest regions) of significance
in data. For example, \cite{dw08} plot minimal intervals on which a density function significantly decreases or increases. \cite{w10} plots minimal significant rectangles
on which the probability of success is higher than a baseline, in a two-dimensional spatial model. \cite{fls20} mention the possibility of using the interval sampling scheme from
\cite{f14a} to focus on the shortest intervals in their CUSUM-based determination of regions of significance in Scenario 1.
In addition to NSP's new definition of significance involving the multiresolution sup-norm fit (whose benefits are explained in Section \ref{sec:iidg} of the main paper), NSP is also different from these approaches
in that its pursuit
of the shortest significant intervals is at its algorithmic core and is its main objective. To achieve it, NSP uses a number of solutions which, to the best of our knowledge, either
are new or have not been considered in this context before. These include
the two-stage search for the shortest significant subinterval (NSP routine, line 19) and 
the recursive sampling (lines 5--10, proposed previously but in a non-inferential context by \cite{f20}).

\paragraph{Lack of penalisation for fine scales.}

Instead of using multiresolution sup-norms (multiscale scan statistics) as defined in the main paper, some authors, including
\cite{w10} and \cite{fms14}, use alternative definitions which penalise fine scales (i.e. short intervals) in order to enhance detection power at coarser scales. We do not pursue
this route, as NSP aims to discover significant intervals that are as short as possible, and hence we are interested in retaining good detection power at fine scales. However, some
natural penalisation of fine scales necessarily occurs in the self-normalised case; see Section \ref{sec:sn} of the main paper.

\paragraph{Upper bounds for $p$-values on non-detection intervals.}

By calculating the quantity $D_{[s,e]}$ on each data section $[s,e]$ delimited by the detected intervals of significance,
an upper bound on the $p$-value for the existence of a change-point in $[s,e]$ can be obtained as 
$P(\|  Z   \|_{\mathcal{I}^a} > D_{[s,e]})$. If the interval $[s,e]$ were considered by NSP before (as would be the case e.g. if $\tau_L = \tau_R = 0$ and the deterministic sampling grid were used),
from the non-detection on $[s,e]$, we would necessarily have $P(\|  Z   \|_{\mathcal{I}^a} > D_{[s,e]}) \ge \alpha$.

\paragraph{Bottom-up implementation of NSP}

Our implementation of NSP is ``bottom-up'', in the sense that at each recursive stage, we consider the intervals $[s_m, e_m]$ in non-decreasing order of their lengths, and exit the current recursive stage (if and) as soon as significance is declared, rather than moving on to longer intervals. This aligns with the objective of looking for the shortest intervals (so the examination of longer intervals is unnecessary if shorter significant intervals have been found). Any non-bottom-up implementation of NSP would therefore unnecessarily be wasting computational resources. This is in contrast to, for example, the region-based multiple testing method of \cite{mkg15}, in which the successive $p$-value adjustments (which lead to power improvements) are only possible because of the top-down character of that approach.

\section{Proofs of results of Section \ref{sec:nspbasic}}

\noindent {\bf Proof of Proposition \ref{prop:supnorm}.} As $[s,e]$ does not contain a change-point, there is a $\beta^*$ such that $Y_{s:e} = X_{s:e,\cdot} \beta^* + Z_{s:e}$. Therefore, 
$D_{[s, e]} = \min_{\beta}     \|  Y_{s:e} - X_{s:e, \cdot} \beta  \|_{\mathcal{I}^d_{[s, e]}} \le \|  Y_{s:e} - X_{s:e, \cdot} \beta^*  \|_{\mathcal{I}^d_{[s, e]}} = \|  Z_{s:e}  \|_{\mathcal{I}^d_{[s, e]}}$,
which completes the proof. \hfill $\square$

\vspace{10pt}

\noindent {\bf Proof of Theorem \ref{th:main}.} The second inequality is implied by (\ref{eq:simpineq}) in the main paper. We now prove the first inequality. On the set 
$\|Z \|_{\mathcal{I}^d} \le \lambda_\alpha$, each interval $S_i$ must contain a change-point as if it did not, then by Proposition \ref{prop:supnorm}, we would have to have
\begin{equation}
\label{eq:contra}
D_{S_i} \le \|Z \|_{\mathcal{I}^d} \le \lambda_\alpha.
\end{equation}
However, the fact that $S_i$ was returned by NSP means, by line 14 of the NSP algorithm, that $D_{S_i} > \lambda_\alpha$, 
which contradicts (\ref{eq:contra}). This completes the proof. \hfill $\square$

\vspace{10pt}

\noindent {\bf Proof of Proposition \ref{prop:ub}.} The inequality is true because for any fixed $\beta$, the norm $\|  Z - X \beta  \|_{\mathcal{I}^d}$ is a maximum over a larger set than the maximum in $\|  Z_{s:e} - X_{s:e,\cdot} \beta  \|_{\mathcal{I}^d_{[s,e]}}$. We now prove the equality. As $[s,e]$ does not contain a change-point, there is a $\beta^*$ such that $Y_{s:e} = X_{s:e,\cdot}\beta^* + Z_{s:e}$. We have
\begin{eqnarray*}
D_{[s,e]} & = & \min_\beta \|  Y_{s:e} - X_{s:e,\cdot}\beta  \|_{\mathcal{I}^d_{[s,e]}} = \min_\beta \|  X_{s:e,\cdot}\beta^* + Z_{s:e} - X_{s:e,\cdot}\beta  \|_{\mathcal{I}^d_{[s,e]}}\\
& = & \min_\beta \| Z_{s:e} - X_{s:e,\cdot} (\beta - \beta^*)  \|_{\mathcal{I}^d_{[s,e]}} =
\min_{\beta-\beta^*} \| Z_{s:e} - X_{s:e,\cdot} (\beta - \beta^*)  \|_{\mathcal{I}^d_{[s,e]}} = \min_{\beta} \| Z_{s:e} - X_{s:e,\cdot} \beta  \|_{\mathcal{I}^d_{[s,e]}}.
\end{eqnarray*}\hfill$\square$

\vspace{10pt}

\noindent {\bf Proof of Theorem \ref{th:main2}.} On the set 
$\min_\beta \|  Z - X \beta  \|_{\mathcal{I}^d} \le \lambda_\alpha$, each interval $S_i$ must contain a change-point as if it did not, then by Proposition \ref{prop:ub}, we would have to have
\begin{equation}
\label{eq:contra2}
D_{S_i} \le \min_\beta \|  Z - X \beta  \|_{\mathcal{I}^d} \le \lambda_\alpha.
\end{equation}
However, the fact that $S_i$ was returned by NSP means, by line 14 of the NSP algorithm, that $D_{S_i} > \lambda_\alpha$, 
which contradicts (\ref{eq:contra2}). This completes the proof. \hfill $\square$

\section{Estimated $\sigma^2$, and other light-tailed distributions}
\label{sec:lt}

We first show under what condition Theorem \ref{lem:k07} in the main paper remains valid with an estimated variance $\sigma^2$, and give an estimator of $\sigma^2$ that satisfies this condition for certain
matrices $X$ and parameter vectors $\beta^{(j)}$. Similar considerations are possible for the light-tailed distributions from the latter part of this section, but we omit them here.
With $\{Z_t\}_{t=1}^T \sim N(0, \sigma^2)$ rather than $N(0, 1)$, the statement of Theorem \ref{lem:k07} of the main paper trivially modifies to $\lim_{T\to\infty} P\left( \max_{1\le s\le e\le T}\,\, U_{s,e}(Z) \le \sigma(a_T + b_T\, \gamma  )   \right) = \exp(-e^{-\gamma})$.
From the form of the limiting distribution, it is clear that the theorem remains valid if $\gamma_T \underset{T\to\infty}{\longrightarrow} \gamma$ is used in place of $\gamma$, yielding
\begin{equation}
\label{eq:gammaT}
\lim_{T\to\infty} P\left( \max_{1\le s\le e\le T}\,\, U_{s,e}(Z) \le \sigma(a_T + b_T\, \gamma_T  )   \right) = \exp(-e^{-\gamma}).
\end{equation}
With $\sigma$ estimated via a generic estimator $\hat{\sigma}$, we ask under what circumstances
\begin{equation}
\label{eq:hatgamma}
\lim_{T\to\infty} P\left( \max_{1\le s\le e\le T}\,\, U_{s,e}(Z) \le \hat{\sigma}(a_T + b_T\, \gamma  )   \right) = \exp(-e^{-\gamma}).
\end{equation}
In light of (\ref{eq:gammaT}), it is enough to solve for $\gamma_T$ in $\sigma(a_T + b_T\, \gamma_T  ) = \hat{\sigma}(a_T + b_T\, \gamma  )$, yielding
$\gamma_T = \frac{a_T}{b_T}\left( \frac{\hat{\sigma}}{\sigma} - 1  \right) + \frac{\hat{\sigma}}{\sigma} \gamma$.
In view of the form of $a_T$ and $b_T$ defined in Theorem \ref{lem:k07} of the main paper, we have 
$\gamma_T \underset{T\to\infty}{\longrightarrow} \gamma$ on a set large enough for (\ref{eq:hatgamma})
to hold if
\begin{equation}
\label{eq:oreq}
\left|\frac{\hat{\sigma}}{\sigma} - 1\right| = o_P(\log^{-1}\,T),\quad\text{or equivalently}\quad \left|\frac{\hat{\sigma}^2}{\sigma^2} - 1\right| = o_P(\log^{-1}\,T).
\end{equation}
After \cite{r84} and \cite{ds01}, define $\hat{\sigma}_R^2 = \frac{1}{2(T-1)} \sum_{t=1}^{T-1} (Y_{t+1} - Y_{t})^2$.
Define the signal in model (\ref{eq:reg}) of the main paper by $f_t = X_{t,\cdot} \beta^{(j)}$ for $t = \eta_j+1, \ldots, \eta_{j+1}$, for $j = 0, \ldots, N$. The total variation
of a vector $\{f_t\}_{t=1}^T$ is defined by $TV(f) = \sum_{t=1}^{T-1}|f_{t+1} - f_t|$. As in \cite{ds01}, we have
$\mathbb{E}\{  (\hat{\sigma}_R^2 / \sigma^2 - 1)^2     \} = O(T^{-1}\{1 + TV^{2}(f)\})$,
from which (\ref{eq:oreq}) follows, by Markov inequality, if
\begin{equation}
\label{eq:tvf}
TV(f) = o(T^{1/2}\log^{-1}T).
\end{equation}
By way of a simple example, in Scenario 1, $TV(f) = \sum_{j=1}^N |f_{\eta_j} - f_{\eta_j+1}|$, and therefore (\ref{eq:tvf}) is satisfied if the sum of jump magnitudes
in $f$ is $o(T^{1/2}\log^{-1}T)$. Note that if $f$ is bounded with a number of change-points that is finite in $T$, then $TV(f) = \text{const}(T)$. Similar arguments
apply in Scenario 2, and in Scenario 3 for some matrices $X$.

Without formal theoretical justifications, we also mention two further estimators of $\sigma^2$ (or $\sigma$) which we use in our numerical work.
In Scenarios 1 and 2, we use $\hat{\sigma}_{MAD}$, the Median Absolute Deviation (MAD) estimator as implemented in the R routine \verb+mad+, computed on the sequence $\{2^{-1/2}(Y_{t+1} - Y_t)\}_{t=1}^{T-1}$.
Empirically, $\hat{\sigma}_{MAD}$ is more robust than $\hat{\sigma}_R$ to the presence of change-points in $f_t$, but is also more sensitive to departures from the
Gaussianity of $Z_t$.
In Scenario 3, in settings outside Scenarios 1 and 2, we use the following estimator. In model (\ref{eq:reg}) of the main paper, we estimate $\sigma$ via least squares, on a rolling window
basis, using the window of size $w = \min\{T, \max([T^{1/2}], 20)\}$, to obtain the sequence of estimators $\hat{\sigma}_1, \ldots, \hat{\sigma}_{T-w+1}$. We take
$\hat{\sigma}_{MOLS} = \text{median}(\hat{\sigma}_1, \ldots, \hat{\sigma}_{T-w+1})$, where MOLS stands for `Median of OLS estimators'. The hope is that 
most of the local estimators $\hat{\sigma}_1, \ldots, \hat{\sigma}_{T-w+1}$ are computed on change-point-free sections
of the data, and therefore the median of these local estimators should serve as an accurate estimator of the true $\sigma$. Empirically, $\hat{\sigma}_{MOLS}$ is a useful
alternative to $\hat{\sigma}_R$ in settings in which condition (\ref{eq:tvf}) is not satisfied.

\cite{kw14} provide a result similar to Theorem \ref{lem:k07} of the main paper for distributions of $Z$ dominated by the Gaussian in a sense specified below. These include, 
after scaling so that $\mathbb{E}(Z) = 0$ and $\mbox{Var}(Z) = 1$, the symmetric Bernoulli, symmetric binomial and uniform distributions, amongst others. We now briefly
summarise it. Consider the cumulant-generating function of $Z$ defined by $\varphi(u) = \log\mathbb{E}(e^{uZ})$ and assume that for
some $\sigma_0 > 0$, we have $\varphi(u) < \infty$ for all $u \ge -\sigma_0$. Assume further that for all $\varepsilon > 0$, $\sup_{u\ge\varepsilon} \varphi(u)/(u^2/2) < 1$.
Finally, assume
\[
\varphi(u) = \frac{u^2}{2} - \kappa u^d + o(u^d),\quad u \downarrow 0,
\]
for some $d\in \{3, 4, \ldots  \}$ and $\kappa > 0$. Typical values of $d$ for non-symmetric and symmetric distributions, respectively, are 3 and 4. Under these assumptions,
we have
\[
\lim_{T\to\infty} P\left( \frac{1}{2} \left\{\max_{1\le s\le e\le T}\,\, U_{s,e}(Z)\right\}^2 \le \log\left\{ T\log^{\frac{d-6}{2(d-2)}} T       \right\} + \gamma   \right) = \exp(-\Lambda_{d,\kappa} e^{-\gamma}),
\]
for all $\gamma\in\mathbb{R}$, where $\Lambda_{d,\kappa} = \pi^{-1/2}\Gamma(d/(d-2)) (2\kappa)^{2/(d-2)}$. After simple algebraic manipulations, this result permits a selection
of $\lambda_\alpha$ for use in Theorem \ref{th:main} of the main paper, similarly to Section \ref{sec:gauss} of the main paper.

\section{Importance of two-stage search for shortest interval of significance}
\label{sec:imp2stage}

\begin{figure}
\centering
\begin{minipage}{.33\textwidth}
  \centering
  \includegraphics[width=\linewidth]{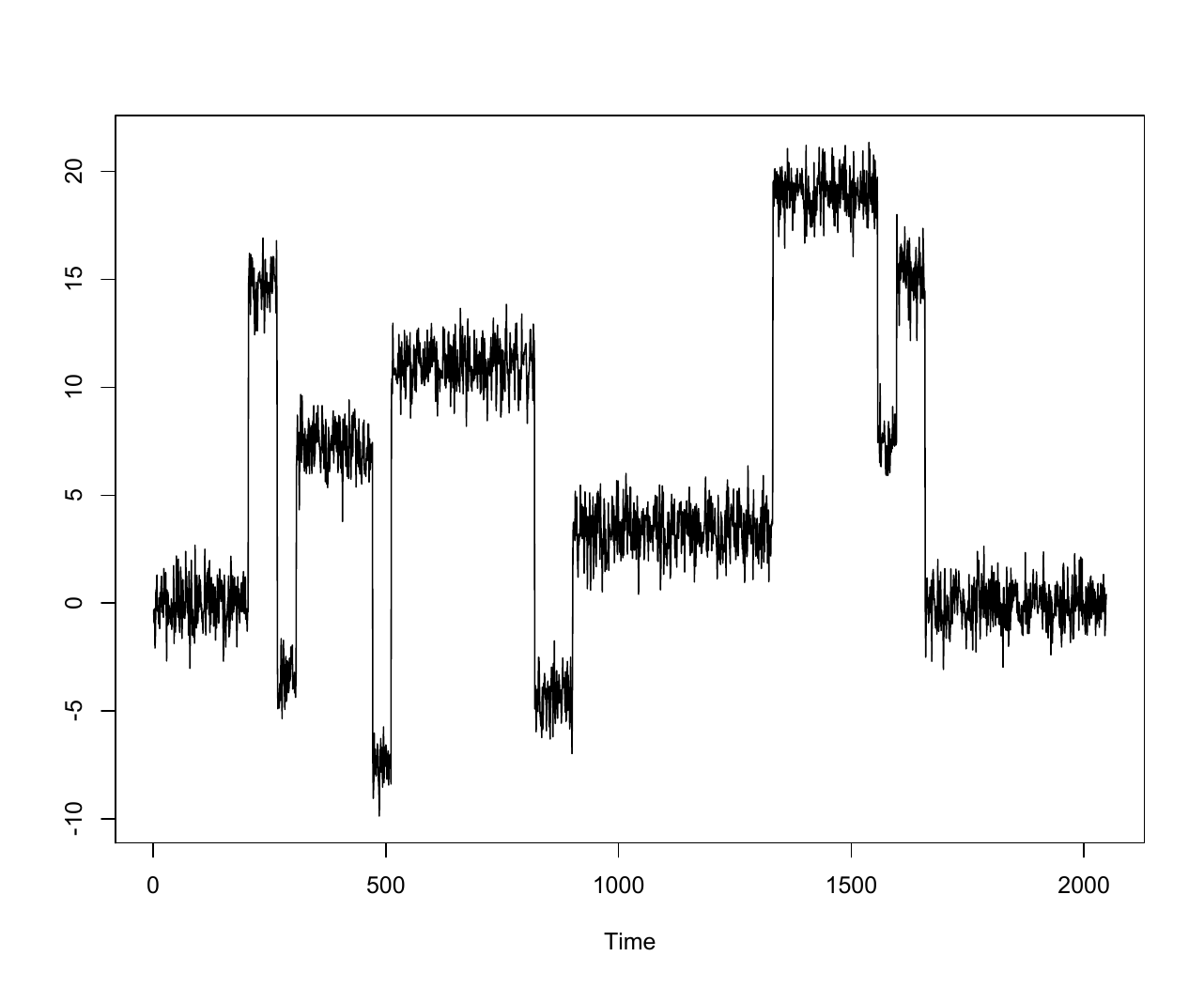}
\end{minipage}%
\begin{minipage}{.33\textwidth}
  \centering
  \includegraphics[width=\linewidth]{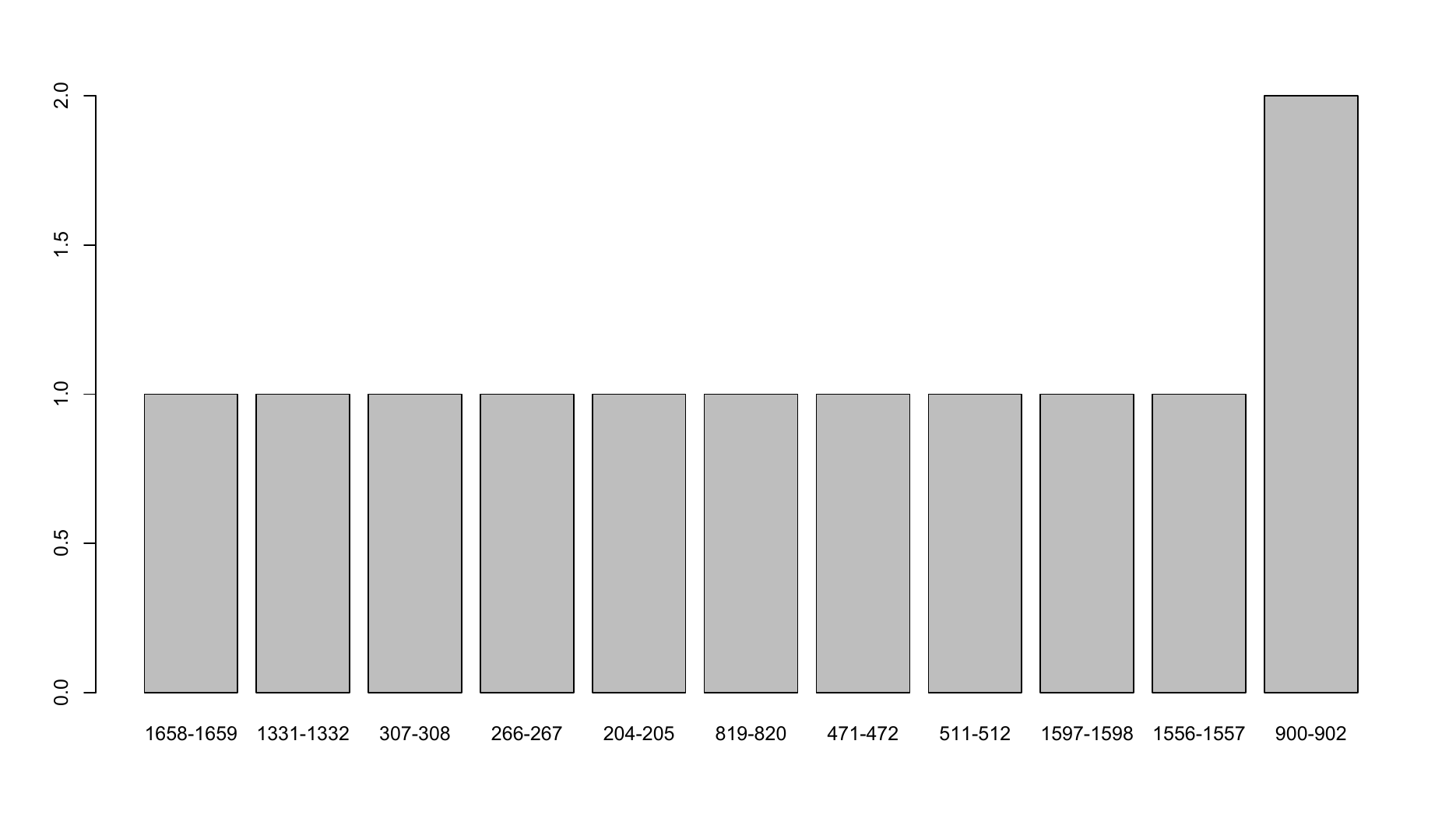}
\end{minipage}%
\begin{minipage}{.33\textwidth}
  \centering
  \includegraphics[width=\linewidth]{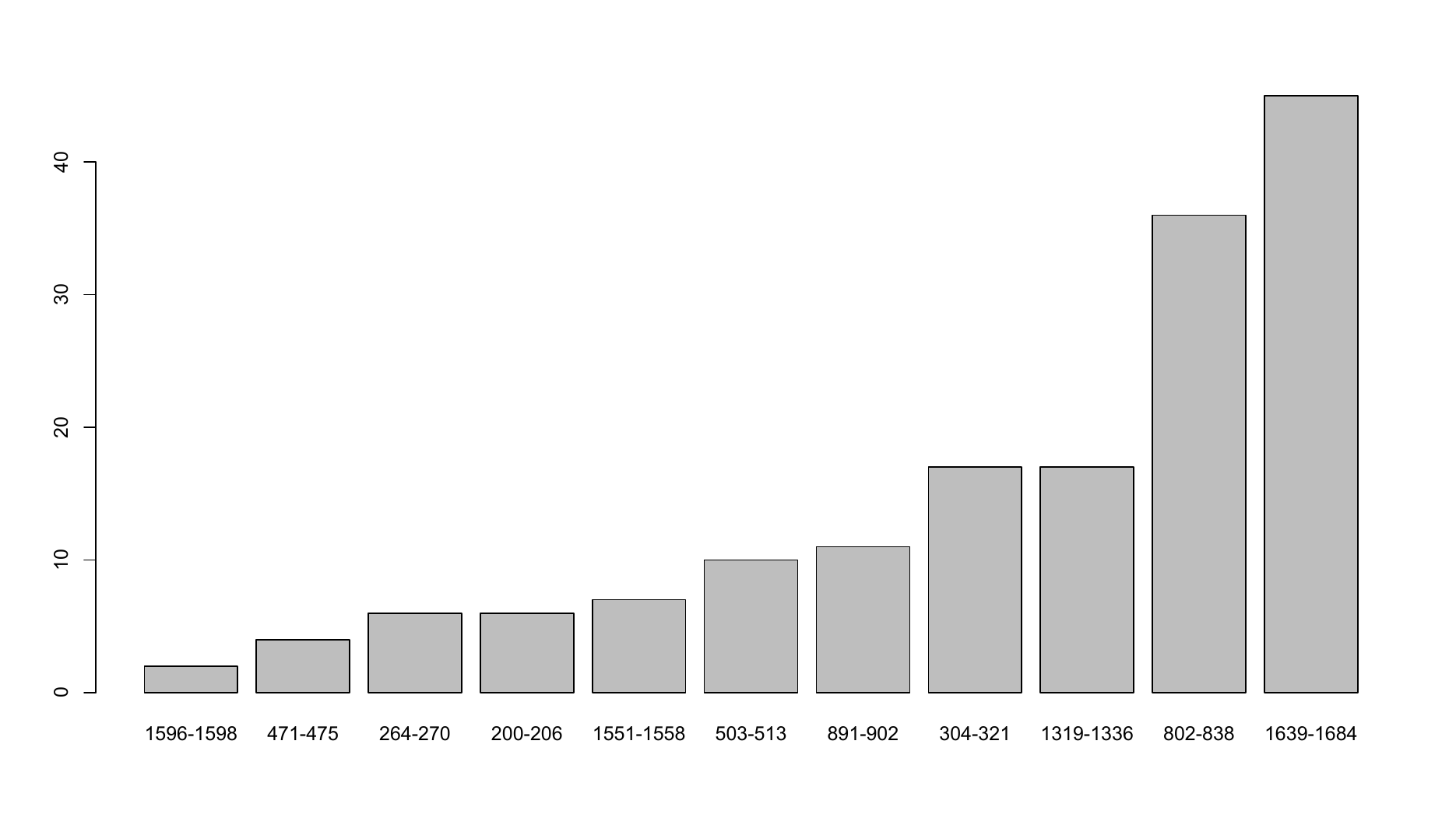}
\end{minipage}%
\caption{Left: realisation $Y_t$ of noisy {\tt blocks} with $\sigma=1$. Middle: prominence plot of NSP-detected intervals. Right: the same for NSP(1). See Section \ref{sec:imp2stage} for more details. \label{fig:blocks_easy}}
\end{figure}

We next illustrate the importance of the two-stage search for the shortest interval of significance, whose stage two is performed in line 19 of the NSP algorithm via the call
\[
[\tilde{s}, \tilde{e}] := \textsc{ShortestSignificantSubinterval}(s_{m_0}, e_{m_0}, Y, X, M, \lambda_\alpha).
\]
Consider the \verb+blocks+ signal referred to in the main paper but with the much smaller noise standard deviation $\sigma=1$. A realisation $Y_t$ is shown in the left plot of Figure \ref{fig:blocks_easy}.
All $N=11$ change-points are visually obvious and hence we would expect NSP to return 11 intervals $[\tilde{s}_i, \tilde{e}_i]$, exactly covering the true change-points, for which we would 
have $\tilde{e}_i-\tilde{s}_i = 1$ for most if not all $i$.
As shown in the middle plot of Figure \ref{fig:blocks_easy}, the NSP procedure with no overlap and with the same parameters as in Section \ref{sec:csim} of the main paper returns 11 intervals of significance with
$\tilde{e}_i-\tilde{s}_i = 1$ for $i=1,\ldots,10$ and $\tilde{e}_{11}-\tilde{s}_{11} = 2$. The 11 intervals of significance cover the true change-points.

However, consider now an alternative version of NSP, labelled NSP(1), which only performs a one-stage search for the shortest interval of significance. NSP(1) proceeds by replacing line 19 of
the NSP algorithm by
\[
[\tilde{s}, \tilde{e}] := [s_{m_0}, e_{m_0}].
\]
In other words, $[s_{m_0}, e_{m_0}]$ is not searched for its shortest sub-interval of significance, but is added to $\mathcal{S}$ as it is. The output of NSP(1) on $Y_t$ is shown in the right plot
of Figure \ref{fig:blocks_easy}. The intervals of significance returned by NSP(1) are unreasonably long from the statistical point of view, with $\tilde{e}_i - \tilde{s}_i$ varying from 2 to 45. However, 
this has a clear explanation from the point
of view of the algorithmic construction of NSP(1). For example, in the first recursive stage, in which $[s,e] = [1,T]$, the spacing of the (approximately) equispaced grid from which the candidate intervals $[s_m, e_m]$ are drawn varies between
45 and 46. Therefore, it is unsurprising that the first detection performed by NSP(1) is such that $\tilde{e}_i - \tilde{s}_i = 45$.

This issue would not arise in NSP, as NSP would then search this detection interval for its shortest significant sub-interval. From the output of the NSP procedure, we can see that this second-stage 
search drastically reduced the length
of this detection interval, which is unsurprising given how obvious the change-points
are in this example. This illustrates the importance of the two-stage search in NSP.

For very long signals, it is conceivable that an analogous three-stage search may be a better option, possibly combined with a reduction in $M$ to enhance the speed of the procedure.

\section{Self-normalised NSP -- further discussion}
\label{sec:suppl:sn}

We now outline the construction of $\hat{Z}^{(k)}$ for $k = 1, 2, 3$ so that (\ref{eq:sntarget}) in the main paper is guaranteed, and
propose a suitable estimator of $V_T^2$ for use in (\ref{eq:sntarget}) in the main paper.

\vspace{10pt}

\noindent {\em $k = 1$.}
Let $(\hat{Z}_{i+1}^{(1)}, \ldots, \hat{Z}_j^{(1)})$ be the ordinary least-squares residuals from regressing $Y_{(i+1):j}$ on $X_{(i+1):j,\cdot}$, where $j-i > p$. As $[s,e]$ contains
no change-point, we have $(\hat{Z}^{(1)}_{i+1})^2+\ldots+(\hat{Z}^{(1)}_j)^2 \le Z_{i+1}^2+\ldots+Z_j^2$ and hence
$\log^{1/2+\epsilon} \{cV_T^2/((\hat{Z}^{(1)}_{i+1})^2+\ldots+(\hat{Z}^{(1)}_j)^2)\} \ge \log^{1/2+\epsilon} \{cV_T^2/(Z_{i+1}^2+\ldots+Z_j^2)\}$.

\vspace{10pt}

\noindent {\em $k = 2$.}
We use
\begin{equation}
\label{eq:revbound}
(\hat{Z}_{i+1}^{(2)}, \ldots, \hat{Z}_j^{(2)}) = (1 + \epsilon) (\hat{Z}_{i+1}^{(1)}, \ldots, \hat{Z}_j^{(1)}),
\end{equation}
which guarantees
$(\hat{Z}^{(2)}_{i+1})^2+\ldots+(\hat{Z}^{(2)}_j)^2 \ge Z_{i+1}^2+\ldots+Z_j^2$ for $\epsilon$ and $j-i$ suitably large, for a range of distributions of
$Z_t$ and design matrices $X$. We now briefly sketch the argument justifying this for Scenario 1; similar considerations are possible in Scenario 2 but are notationally
much more involved and we omit them here. The argument relies again on self-normalisation. From standard least-squares theory (in any Scenario), we have
$(\hat{Z}^{(1)}_{(i+1):j})^\top \hat{Z}^{(1)}_{(i+1):j} = Z_{(i+1):j}^\top Z_{(i+1):j} - Z_{(i+1):j}^\top X_{(i+1):j,\cdot} (X_{(i+1):j,\cdot}^\top X_{(i+1):j,\cdot})^{-1} X_{(i+1):j,\cdot}^\top Z_{(i+1):j}$.
In Scenario 1, $(X_{(i+1):j,\cdot}^\top X_{(i+1):j,\cdot})^{-1} = (j-i)^{-1}$, and hence\\
$Z_{(i+1):j}^\top X_{(i+1):j,\cdot} (X_{(i+1):j,\cdot}^\top X_{(i+1):j,\cdot})^{-1} X_{(i+1):j,\cdot}^\top Z_{(i+1):j} = U_{i+1,j}(Z)^2$.
From the above, we obtain
\begin{eqnarray}
(\hat{Z}^{(1)}_{(i+1):j})^\top \hat{Z}^{(1)}_{(i+1):j} & = & Z_{(i+1):j}^\top Z_{(i+1):j} \left(1 -  \frac{U_{i+1,j}(Z)^2}{Z_{(i+1):j}^\top Z_{(i+1):j}}    \right)\nonumber\\
& = & Z_{(i+1):j}^\top Z_{(i+1):j} \left(1 -  \frac{1}{j-i}\log^{1+2\epsilon} \{cV_T^2/(Z_{i+1}^2+\ldots+Z_j^2)\}\right.\nonumber\\
\label{eq:lowbound}
& \times & \left. I^2_{\rho_{1/2, 1/2+\epsilon, c}}(\zeta_T^{\text{se}}, V_i^2/V_T^2, V_j^2/V_T^2) \right).
\end{eqnarray}
In light of the distributional result (\ref{eq:anylevel}) of the main paper, the relationship between the statistic $I_{\rho_{1/2, 1/2+\epsilon, c}}(W, u, v)$
and \cite{rs04}'s statistic $\text{UI}(\rho_{1/2, 1/2+\epsilon, c})$, as well as their Remark 5, we are able to bound
$\sup_{0\le i < j \le T} I^2_{\rho_{1/2, 1/2+\epsilon, c}}(\zeta_T^{\text{se}}, V_i^2/V_T^2, V_j^2/V_T^2)$ by a term of order $O(\log\,T)$ on a set
of probability $1 - O(T^{-1})$. Making the mild assumption that $\sup_{0\le i < j \le T} \log^{1+2\epsilon} \{cV_T^2/(Z_{i+1}^2+\ldots+Z_j^2)\} \asymp
l_T = o_P(T \log^{-1} T)$ and continuing from (\ref{eq:lowbound}), we obtain 
$(\hat{Z}^{(1)}_{(i+1):j})^\top \hat{Z}^{(1)}_{(i+1):j} \ge Z_{(i+1):j}^\top Z_{(i+1):j} \left(1 -  C(j-i)^{-1} l_T \log\,T  \right)$
for a certain constant $C > 0$, which can be bounded from below by $Z_{(i+1):j}^\top Z_{(i+1):j} (1 + \epsilon)^{-2}$, uniformly over those $i,j$ for
which $(j-i)^{-1} l_T \log\,T \to 0$. This justifies (\ref{eq:revbound}) and completes the argument.

\vspace{10pt}

\noindent {\em $k = 3$.}
Having obtained $\hat{Z}^{(1)}_{(i+1):j}$ and $\hat{Z}^{(2)}_{(i+1):j}$ as above, the problem of obtaining $\hat{Z}_{s:e}^{(3)}$ to guarantee
\begin{eqnarray}
\lefteqn{\sup_{s-1 \le i < j \le e} \frac{|\hat{Z}^{(3)}_{i+1}+\ldots+\hat{Z}^{(3)}_j|}{\sqrt{(\hat{Z}^{(2)}_{i+1})^2+\ldots+(\hat{Z}^{(2)}_j)^2} \log^{1/2+\epsilon} \{cV_T^2/((\hat{Z}^{(1)}_{i+1})^2+\ldots+(\hat{Z}^{(1)}_j)^2)\}}}\nonumber\\
\label{eq:k3}
&& \le \sup_{s-1 \le i < j \le e} \frac{|Z_{i+1}+\ldots+Z_j|}{\sqrt{(\hat{Z}^{(2)}_{i+1})^2+\ldots+(\hat{Z}^{(2)}_j)^2} \log^{1/2+\epsilon} \{cV_T^2/((\hat{Z}^{(1)}_{i+1})^2+\ldots+(\hat{Z}^{(1)}_j)^2)\}},
\end{eqnarray}
which in turn guarantees the bound (\ref{eq:sntarget}) in the main paper, is practically equivalent to the multiresolution norm minimisation solved in Step 1 of Section \ref{sec:iidg} of the main paper except
it now uses a weighted version of the norm $\|  \cdot  \|_{\mathcal{I}^a_{[s,e]}}$, where the weights are given in the denominator of (\ref{eq:k3}). This weighted
problem is solved via linear programming  just as easily as Step 1 of Section \ref{sec:iidg} of the main paper, the only difference being that the relevant constraints are multiplied by the corresponding weights.

We now discuss further practicalities of the self-normalisation.
In the exposition of the main paper, we use all intervals $[i+1,j] \subseteq [s,e]$, i.e. the set $\mathcal{I}_{[s,e]}^a$. In practice, for
computational reasons, we compute the supremum on the LHS of (\ref{eq:sntarget}) in the main paper over the dyadic set $\mathcal{I}_{[s,e]}^d$, which does not alter the validity of the bound.
Our empirical experience is that the statistic on the LHS of (\ref{eq:sntarget}) of the main paper is fairly robust to the choice of $V_T^2$, as the latter only enters through
the (close to) square-root logarithmic term in the denominator. In addition, over-estimation of $V_T^2$ for use on the LHS of (\ref{eq:sntarget}) of the main paper is permitted as it only strengthens
the bound in (\ref{eq:sntarget}) of the main paper. For these reasons, we do not dwell on the accurate estimation of $V_T^2$ here, but use the rough estimate
$\hat{V}_T^2 = \frac{T}{T-w+1} \sum_{t=1}^{T-w+1} \hat{\sigma}_t^2$,
where the $\hat{\sigma}_t$'s are the constituents of the $\hat{\sigma}_{MOLS}$ estimator from Section \ref{sec:lt}.
As clarified earlier, the use of (\ref{eq:revbound}) requires that small values of $j-i$ do not enter in the computation
of the supremum on the LHS of (\ref{eq:sntarget}) of the main paper.
In practice, however, we use all $[i+1,j] \in \mathcal{I}_{[s,e]}^d$.
This is because the function $I_{\rho_{1/2, 1/2+\epsilon, c}}(\zeta_T^{\text{se}}, V_i^2/V_T^2, V_j^2/V_T^2)$ naturally penalises small scales (i.e. short
intervals $[i+1, j]$) through the use of the logarithmic term in the denominator. Therefore, in practice, short intervals $[i+1,j]$ do not tend to achieve
the supremum on the LHS of (\ref{eq:sntarget}) of the main paper and as a result, we have found further exclusion of such short intervals unnecessary.
Finally, we have experimented with $\epsilon$ in the range $[0.03, 0.1]$ and found little difference in practical performance. Our code
uses $\epsilon = 0.03$ as a default.

\section{NSP with autoregression}
\label{sec:arex}

\begin{figure}
\centering
\includegraphics[width=0.4\linewidth]{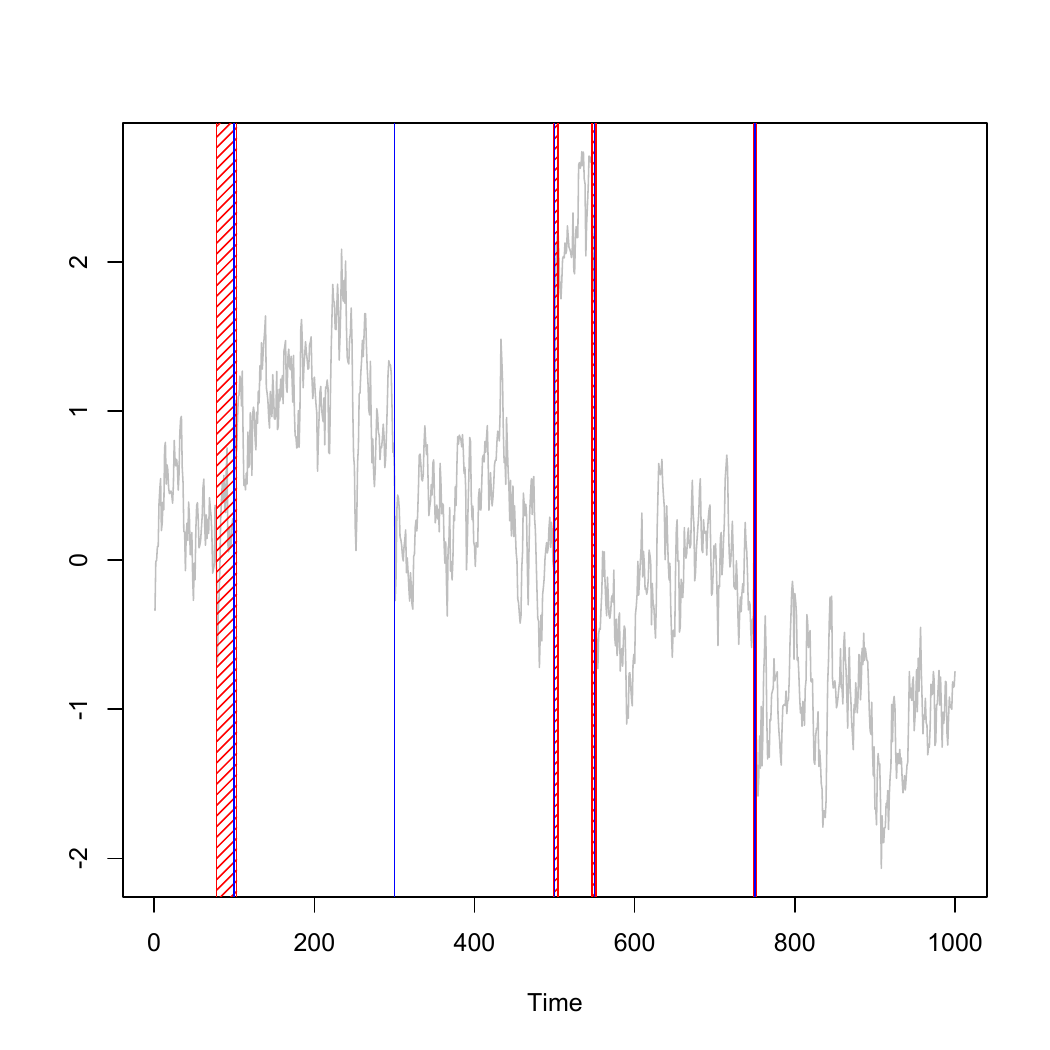}
\caption{Piecewise-constant signal from \cite{dsv18} with Gaussian AR(1) noise with coefficient 0.9 and standard deviation $(1-0.9^2)^{-1/2}/5$ (light grey), NSP intervals of significance (shaded red), true change-points (blue); see Section \ref{sec:arex} for details. \label{fig:nspar}}
\end{figure}

We use the piecewise-constant signal of length $T=1000$ from the first simulation setting in \cite{dsv18}, contaminated with
Gaussian AR(1) noise with coefficient 0.9 and standard deviation $(1-0.9^2)^{-1/2}/5$. A sample path, together with the true change-point locations,
is shown in Figure \ref{fig:nspar}.

We run the AR version of the NSP algorithm (as outlined in Section \ref{sec:ar} of the main paper), with the following parameters: a deterministic equispaced interval sampling grid, $M=100$, $\alpha = 0.1$, no overlap, $\hat{\sigma}^2_{MOLS}$ estimator of the residual variance. The resulting intervals are shown in Figure \ref{fig:nspar}; NSP intervals cover four out of the five true change-points, and there are no spurious intervals.

We simulate from this model 100 times and obtain the following results. In 100\% of the sample paths, each NSP interval of significance covers one true change-point (which fulfils the promise of Theorem \ref{th:main} of the main paper). The distribution of the detected numbers of intervals is as in Table \ref{tab:no}; we recall that NSP, with a fixed significance level, does not promise to detect the number of intervals equal to the number of true change-points in the underlying process.

\begin{table}
\caption{Percentage of sample paths with the given numbers of NSP-detected intervals in the autoregressive example of Section \ref{sec:arex}.\label{tab:no}}
\centering
\begin{tabular}{ |c|c|c|c|c| } 
\hline
no. of intervals of significance & 2 & 3 & 4 & 5 \\
 \hline
percentage of sample paths & 11 & 32 & 42 & 15\\
 \hline
\end{tabular}
\end{table}

\section{Computation of the NSP threshold by simulation}
\label{sec:suppl:sim}

In a number of locations in the main paper, we mention the possibility of obtaining the NSP thresholds by simulation. We now clarify how this is done. For example, to solve
\[
P(\| Z \|_{\mathcal{I}^d} > \lambda_\alpha) = \alpha
\]
for $\lambda_\alpha$ (see e.g. Theorem \ref{th:main} of the main paper) by simulation, we would simulate multiple realisations of $\| Z \|_{\mathcal{I}^d}$ and choose $\lambda_\alpha$ as the 
$100(1-\alpha)\%$ empirical quantile of the sample. We proceed similarly in Section \ref{sec:simthresh}, in which the task is to approximate the distribution of
$\min_\beta \|  Z - X \beta  \|_{\mathcal{I}^d}$. It is important to note that this can easily be done for any distribution of $Z$ (assumed known), not just Gaussian.
(If there is uncertainty regarding the distribution of $Z$ and there are a few plausible candidates, the corresponding threshold can be computed for each of them and the
largest one among them chosen for use in the NSP algorithm.)

This threshold selected as the empirical quantile of $\min_\beta \|  Z - X \beta  \|_{\mathcal{I}^d}$, for the Gaussian case in Scenarios 1 and 2, is implemented in the R package \verb+nsp+ and can be used upon setting \verb+thresh.type = "sim"+ in the
\verb+nsp_poly+ routine.

One remaining question is whether it is possible to use the standard (non-self-normalised) NSP without knowledge of the distribution of the innovations $Z$. Here, the following simple practical procedure for determining the threshold via simulation may help.
\begin{enumerate}
\item
Pre-estimate the time-varying signal $X\beta$ via a localised moving-window fit; then pre-estimate the innovations $\hat{Z}$.
\item
Re-sample the innovations to estimate the distribution of the multiscale deviation measure $\| \hat{Z} \|_{\mathcal{I}^d}$.
\item
Use a suitable empirical quantile of this distribution as the NSP threshold.
\end{enumerate}

\section{Detection consistency and lengths of NSP intervals -- proofs and discussion}

\noindent {\bf Proof of Theorem \ref{th:mainconsist} (main paper).}
Assume initially that $f_t$ has a single change-point $\eta_1$. As NSP considers all intervals by the assumption of the theorem, it will
certainly consider intervals symmetric about the true change-point, i.e. $[\eta_1 - d + 1, \eta_1 + d]$, for all appropriate $d$. In Scenario 1, there is an explicit formula for the deviation measure $D_{[s,e]}$ on any interval $[s,e]$, given by
\begin{equation}
\label{eq:dexpl}
D_{[s,e]} = \max_{\tau\in\{1, \ldots, e-s+1\}} \frac{1}{2\sqrt{\tau}} \left(  \max_{s_1\in\{ s, \ldots, e+1-\tau  \}} \sum_{t=s_1}^{s_1+\tau-1} Y_t - \min_{s_1\in\{ s, \ldots, e+1-\tau  \}} \sum_{t=s_1}^{s_1+\tau-1} Y_t      \right).
\end{equation}
Without loss of generality, assume $f_{\eta_1} > f_{\eta_1+1}$. Representation (\ref{eq:dexpl}) implies
\begin{eqnarray}
D_{[\eta_1 - d + 1, \eta_1 + d]} & \ge & \frac{1}{2\sqrt{d}} \left(  \max_{s_1 \in \{\eta_1-d+1, \ldots, \eta_1+1\}}   \sum_{t=s_1}^{s_1+d-1} Y_t   -  \min_{s_1 \in \{\eta_1-d+1, \ldots, \eta_1+1\}}   \sum_{t=s_1}^{s_1+d-1} Y_t\right)\nonumber\\
& \ge & \frac{1}{2\sqrt{d}} \left(    \sum_{t=\eta_1-d+1}^{\eta_1}    Y_t    - \sum_{t=\eta_1+1}^{\eta_1+d}     Y_t   \right)\nonumber\\
\label{eq:bfb}
& \ge & \frac{1}{2}|f_{\eta_1+1} - f_{\eta_1}|   \sqrt{d} - \|  Z   \|_{\mathcal{I}^a}.
\end{eqnarray}
On the set $\|  Z   \|_{\mathcal{I}^a} \le \lambda_\alpha$, (\ref{eq:bfb}) is further bounded from below by $\frac{1}{2}|f_{\eta_1+1} - f_{\eta_1}|   \sqrt{d} - \lambda_\alpha$.
From the definition of the NSP algorithm, detection on $[s,e]$ is triggered by the event $D_{[s,e]} > \lambda_\alpha$, so detection on $[\eta_1 - d + 1, \eta_1 + d]$ is triggered if (note: not ``only if'' as we are using lower bounds here) $\frac{1}{2}|f_{\eta_1+1} - f_{\eta_1}|   \sqrt{d} - \lambda_\alpha > \lambda_\alpha$,
or
\begin{equation}
\label{eq:det}
|f_{\eta_1+1} - f_{\eta_1}|   \sqrt{d} > 4 \lambda_\alpha.
\end{equation}
As NSP looks for shortest intervals of detection, the NSP interval of significance around $\eta_1$ will definitely be no longer than $2d = |[\eta_1-d+1, \eta_1 + d]|$. However, from (\ref{eq:det}), it is sufficient for detection to be triggered if $d > \frac{16 \lambda_\alpha^2}{|f_{\eta_1+1} - f_{\eta_1}|^2}$.
This shows that the maximum length of an NSP interval of significance will not exceed $2\bar{d}$, where
$\bar{d} = \left\lceil \frac{16 \lambda_\alpha^2}{|f_{\eta_1+1} - f_{\eta_1}|^2} \right\rceil + 1$.
We now turn our attention to the multiple change-point case. For each change-point $\eta_j$, define its corresponding $\bar{d}_j$ as in formula (\ref{eq:dj}) of the main paper.
Recall we are on the set $\|  Z   \|_{\mathcal{I}^a} \le \lambda_\alpha$. Note first that even though the NSP interval of significance around $\eta_j$ is guaranteed to be of length at most $2\bar{d}_j$, it will not necessarily be a subinterval of $[\eta_j - \bar{d}_j +1, \eta_j + \bar{d}_j]$ (as NSP simply looks for the shortest intervals of significance and interval symmetry around the true change-point is not explicitly promoted). Therefore, in order that an interval detection around $\eta_j$ does not interfere with detections around $\eta_{j-1}$ and $\eta_{j+1}$, the distances $\eta_j - \eta_{j-1}$ and $\eta_{j+1} - \eta_{j-1}$ must be suitably long, but this is guaranteed by Assumption \ref{ass:dist} from the main paper. This completes the proof.\hfill$\square$

\vspace{10pt}

As an aside, note in addition that in the Gaussian case $Z_t \sim N(0, 1)$, Theorem \ref{lem:k07} of the main paper implies 
$\lambda_\alpha = O(\log^{1/2}T)$; in fact for $\alpha = 0.05$, we have $\lambda_\alpha \le 1.33 \sqrt{2\log\,T}$ for $T \ge 100$, for $\alpha = 0.1$, we have $\lambda_\alpha \le 1.25 \sqrt{2\log\,T}$ over the same range of $T$.

\vspace{10pt}

\noindent {\bf Proof of Corollary \ref{cor:consist} (main paper).}
From Lemma 1 in \cite{y88}, we have
\[
P(\|  Z   \|_{\mathcal{I}^a} \le \sigma (1 + \Delta) \sqrt{2\log\,T}) \to 1
\]
as $T \to \infty$. This combined with the statement of Theorem \ref{th:mainconsist} in the main paper proves the result.\hfill$\square$

\vspace{10pt}

\noindent {\bf Proof of Theorem \ref{th:linconsist} (main paper).}
Assume initially that $f_t$ has a single change-point $\eta_1$.
In the same way in which the NSP procedure is ``blind'' to constant shifts in the data in Scenario 1, it is also invariant to the addition of linear trends in the piecewise-linear Scenario 2. Assume, therefore, that we have added a linear trend to $Y_t$ in such a way that the true signal is symmetric around the true change-point $\eta_1$. The case that will lead to the longest interval is one in which the change-point leads to a trapezoid shape of the true signal (as in, for example, $1, 2, 3, 3, 2, 1$) rather than one with a single peak or trough (e.g. $1, 2, 3, 2, 1$). Therefore we assume the former case as the ``worst case'' (whether this is or is not assumed will only lead to $O(1)$ differences in the length of the NSP intervals, so is irrelevant from the point of view of rates). Note that for such a trapezoid signal, the location of $\eta_1$ is unambiguous (in the cartoon example above, it must be at the first 3).
For such a transformed signal (a transformation which does not change the output of the NSP algorithm), consider intervals symmetric around the true change-point, i.e. $[\eta_1-d+1, \eta_1+d]$, which will be considered by this version of NSP as it considers all intervals. We have
\begin{equation}
\label{eq:dlin}
D_{[\eta_1-d+1, \eta_1+d]} = \min_{\tilde{f}_{(\eta_1-d+1):(\eta_1+d)}}\|   Y_{(\eta_1-d+1):(\eta_1+d)} - \tilde{f}_{(\eta_1-d+1):(\eta_1+d)}     \|_{\mathcal{I}^a_{[\eta_1-d+1, \eta_1+d]}},
\end{equation}
where the minimum is taken with respect to all linear fits on $[\eta_1-d+1, \eta+d]$. Consider a single scale $\tau$. Observing that taking moving partial sums does not change the linearity of $\tilde{f}$, and continuing from $(\ref{eq:dlin})$, we have
\begin{eqnarray}
D_{[\eta_1-d+1, \eta_1+d]} & \ge & \min_{\tilde{f}_{(\eta_1-d+1):(\eta_1+d)}}     \max_{s_1\in\{ \eta_1-d+1, \ldots, \eta_1+d+1-\tau  \}} \left|\tau^{-1/2}\sum_{t=s_1}^{s_1+\tau-1} Y_t - \tilde{f}_{(\eta_1-d+1):(\eta_1+d)}  \right|\nonumber\\
& \ge & \min_{\tilde{f}_{(\eta_1-d+1):(\eta_1+d)}}     \max_{s_1\in\{ \eta_1-d+1, \ldots, \eta_1+d+1-\tau  \}} \left|\tau^{-1/2}\sum_{t=s_1}^{s_1+\tau-1} f_t - \tilde{f}_{(\eta_1-d+1):(\eta_1+d)}  \right|\nonumber\\
\label{eq:dlinbound}
& - & \|   Z \|_{\mathcal{I}^a}.
\end{eqnarray}
Observe now that since $f_t$ is symmetric around $\eta_1$, the minimising $\tilde{f}$ must be constant. So restrict the class of candidate fits $\tilde{f}$ to constant. Denote the slope of $f_t$ before the change-point by $\xi$. We have
\begin{eqnarray}
\lefteqn{\min_{\tilde{f}_{(\eta_1-d+1):(\eta_1+d)}}     \max_{s_1\in\{ \eta_1-d+1, \ldots, \eta_1+d+1-\tau  \}} \left|\tau^{-1/2}\sum_{t=s_1}^{s_1+\tau-1} f_t - \tilde{f}_{(\eta_1-d+1):(\eta_1+d)}  \right| =}\nonumber\\ 
&& \frac{\tau^{1/2}}{2} \left( \frac{1}{\tau} \max_{s_1\in\{ \eta_1-d+1, \ldots, \eta_1+d+1-\tau  \}}  \sum_{t=s_1}^{s_1+\tau-1} f_t -  \frac{1}{\tau}
\min_{s_1\in\{ \eta_1-d+1, \ldots, \eta_1+d+1-\tau  \}}  \sum_{t=s_1}^{s_1+\tau-1} f_t \right) = \nonumber\\
\label{eq:dlinbound2}
&& \frac{\tau^{1/2}}{2}   \xi (  d - \tau ).
\end{eqnarray}
Take $\tau = C d$ for $C \in (0, 1)$. (\ref{eq:dlinbound}) and (\ref{eq:dlinbound2})
together imply $D_{[\eta_1-d+1, \eta_1+d]} \ge C_1 \xi d^{3/2} - \|   Z \|_{\mathcal{I}^a}$
for a certain universal constant $C_1$. Therefore, on $\|  Z  \|_{\mathcal{I}^a} \le \lambda_\alpha$, detection on $[\eta_1-d+1, \eta_1+d]$ will be triggered if
$C_1 \xi d^{3/2} > 2\lambda_\alpha$, or in other words if $d \ge C_2 \lambda_\alpha^{2/3} \xi^{-2/3}$,
for a large enough constant $C_2$. This shows that the NSP interval of significance will be of length $O(\lambda_\alpha^{2/3} \xi^{-2/3})$.

We now discuss the slope $\xi$. Suppose before the symmetrisation the slopes around $\eta_1$ were $\xi_1$ and $\xi_2$. After the symmetrisation, they are now $\xi_1 + \xi_3$ and
$\xi_2 + \xi_3$ where $\xi_1 + \xi_3 = -(\xi_2 + \xi_3)$, which means $\xi = |\xi_1 - \xi_2|/2$ (w.l.o.g., $\xi > 0$). Typically, if $f_t = f(t/T)$ for a certain piecewise-linear function $f(u) : (0, 1] \to \mathbb{R}$, then $\xi = O(T^{-1})$.
In the Gaussian case, we have $\lambda_\alpha = O(\sqrt{\log\,T})$. Therefore, if $\xi = O(T^{-1})$, then the NSP interval of significance will have the length $O(T^{2/3} \log^{1/3}T )$.

In the multiple change-point case, the argument about the relevance of Assumption \ref{ass:dist} from the proof of Theorem \ref{th:mainconsist} (main paper) still applies here, and this completes the proof of the theorem.\hfill$\square$

\vspace{10pt}

\noindent {\bf Proof of Corollary \ref{cor:consistlin} (main paper).} The argument is identical to the proof of Corollary \ref{cor:consist} from the main paper.\hfill$\square$

\section{NSP with autocorrelated innovations}
\label{sec:ai}

Scenario 4 permits the use of NSP in settings in which autocorrelation is present, but this is done through the use of the lagged response as an additional covariate, rather than through allowing the innovations $Z_t$ to be autocorrelated. We now briefly explore the case in which the $Z_t$'s themselves are serially correlated. This presents an alternative to the discussion of Section \ref{sec:gauss}
of the main paper.

Suppose that $Z_t$ can be modelled as an autoregressive process as follows.
\[
U_t = Z_t - a_1 Z_{t-1} - \ldots - a_r Z_{t-r} =: a(L) Z_t,
\]
where $U_t$ is independent (not necessarily identically distributed) noise distribution acceptable to NSP in Scenarios 1, 2 or 3, and $L$ is the lag operator. We propose the following iterative scheme which builds on the NSP procedure for independent innovations. We use the (most general) language of Scenario 3.

Clearly, if the user knew $r$ and $(a_1, \ldots, a_r)$, they would be able to transform the regression problem (\ref{eq:reg}) from the main paper into
\begin{eqnarray}
a(L)Y_t & = & a(L)X_{t,\cdot} \beta^{(j)} + U_t\quad \text{for}\quad t = \eta_j + 1 + r,\ldots, \eta_{j+1},\nonumber\\
\label{eq:filtered}
a(L)Y_t & = & a(L)X_{t,\cdot} \beta^{(j,t)} + U_t\quad \text{for}\quad t = \eta_j + 1, \ldots, \eta_j+r.
\end{eqnarray}
Due to the smoothing action of the filter $a(L)$, this now only approximates a piecewise-constant parameter regression setting, as it features the short ``smooth transition'' sections indexed $t = \eta_j + 1, \ldots, \eta_j+r$. However, the presence of these smooth transitions does not spoil the applicability of NSP, with the intervals of significance obtained on the regression problem (\ref{eq:filtered}) having a similar interpretation as in the case of exactly abrupt transitions.

In practice, $r$ or $(a_1, \ldots, a_r)$ will be unknown to the analyst. We suggest the following scheme, in which these are treated as nuisance parameters and estimated from the data, as in \cite{fs20}.
\begin{enumerate}
\item
Similarly to \cite{fs20}, estimate $r$ and $(a_1, \ldots, a_r)$ (to obtain, respectively, $\hat{r}$ and $\hat{a} = (\hat{a}_1, \ldots, \hat{a}_{\hat{r}})$) on a stretch of the data believed to contain no change-points.
\item
\label{eq:transf}
Transform the regression problem using the estimated operator $\hat{a}(L)$ to obtain a problem of the form (\ref{eq:filtered}).
\item
Run NSP suitable for independent innovations on the transformed problem, to obtain a set $\mathcal{S}$ of the NSP intervals of significance.
\item
Re-estimate $r$ and $(a_1, \ldots, a_r)$ on the longest stretch of data outside the NSP intervals of significance.
\item
Go back to step \ref{eq:transf}. and iterate until no changes are seen in the NSP intervals of significance.
\end{enumerate}

\section{Additional arguments regarding the real-data analysis}
\label{sec:suppl:data}

\begin{figure}
\centering
\begin{minipage}{.3\textwidth}
  \centering
  \includegraphics[width=\linewidth]{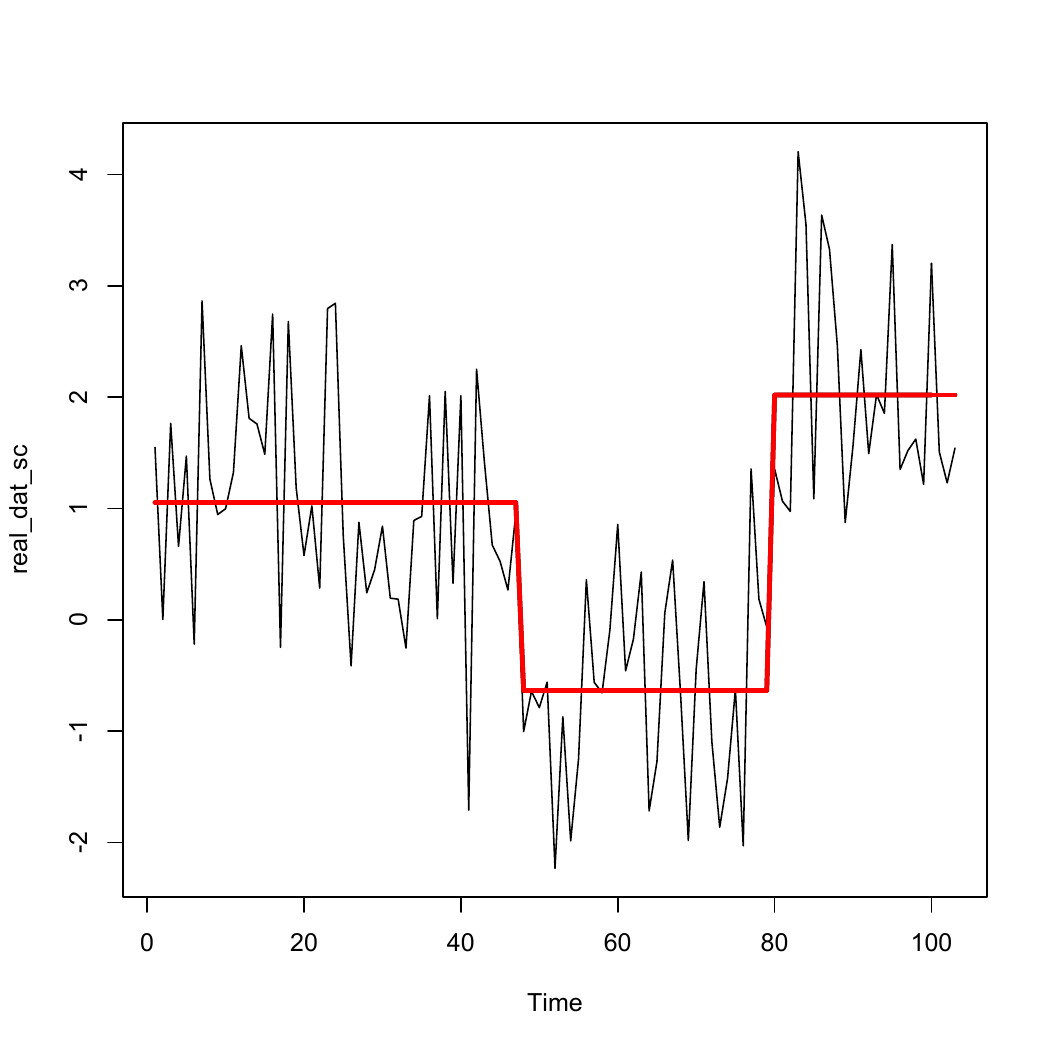}
\end{minipage}
\begin{minipage}{.3\textwidth}
  \centering
  \includegraphics[width=\linewidth]{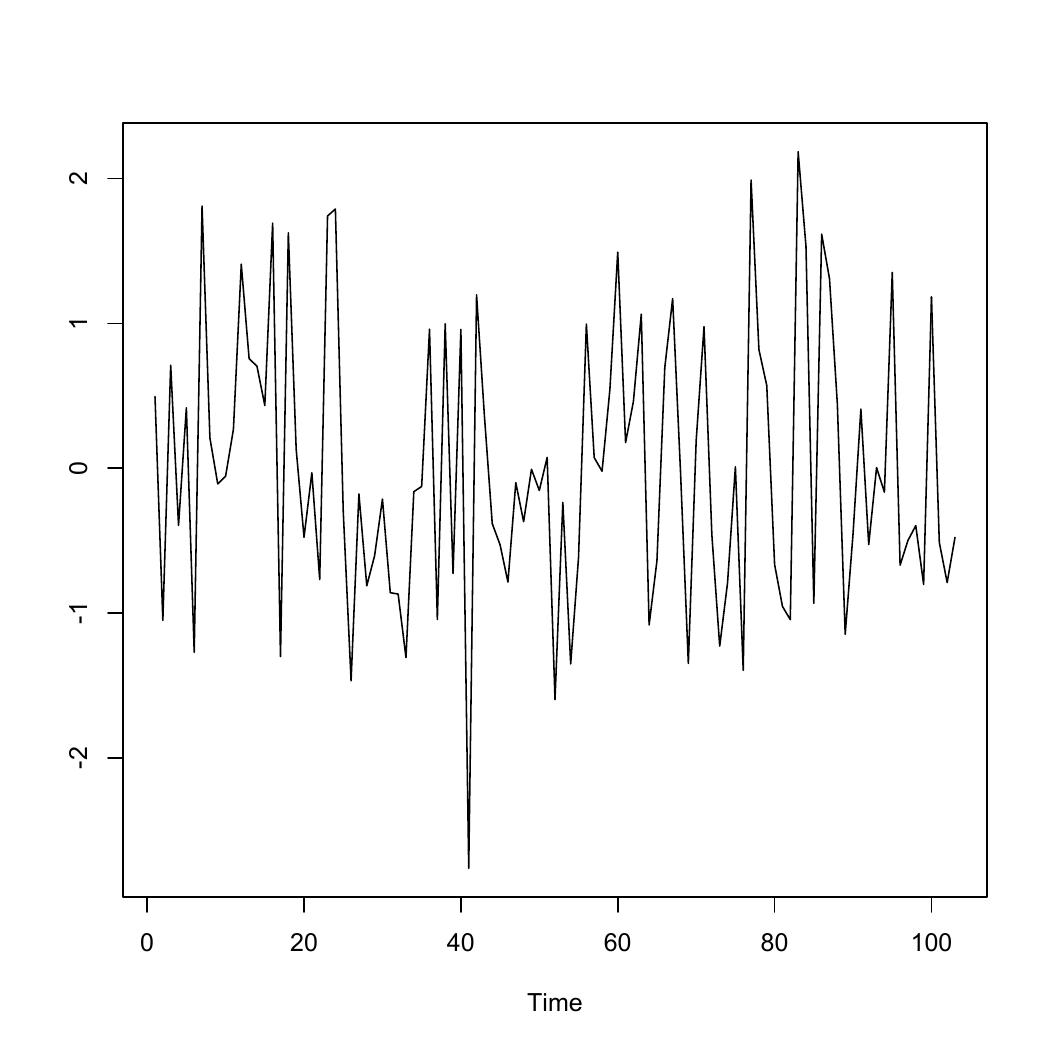}
\end{minipage}
\begin{minipage}{.3\textwidth}
  \centering
  \includegraphics[width=\linewidth]{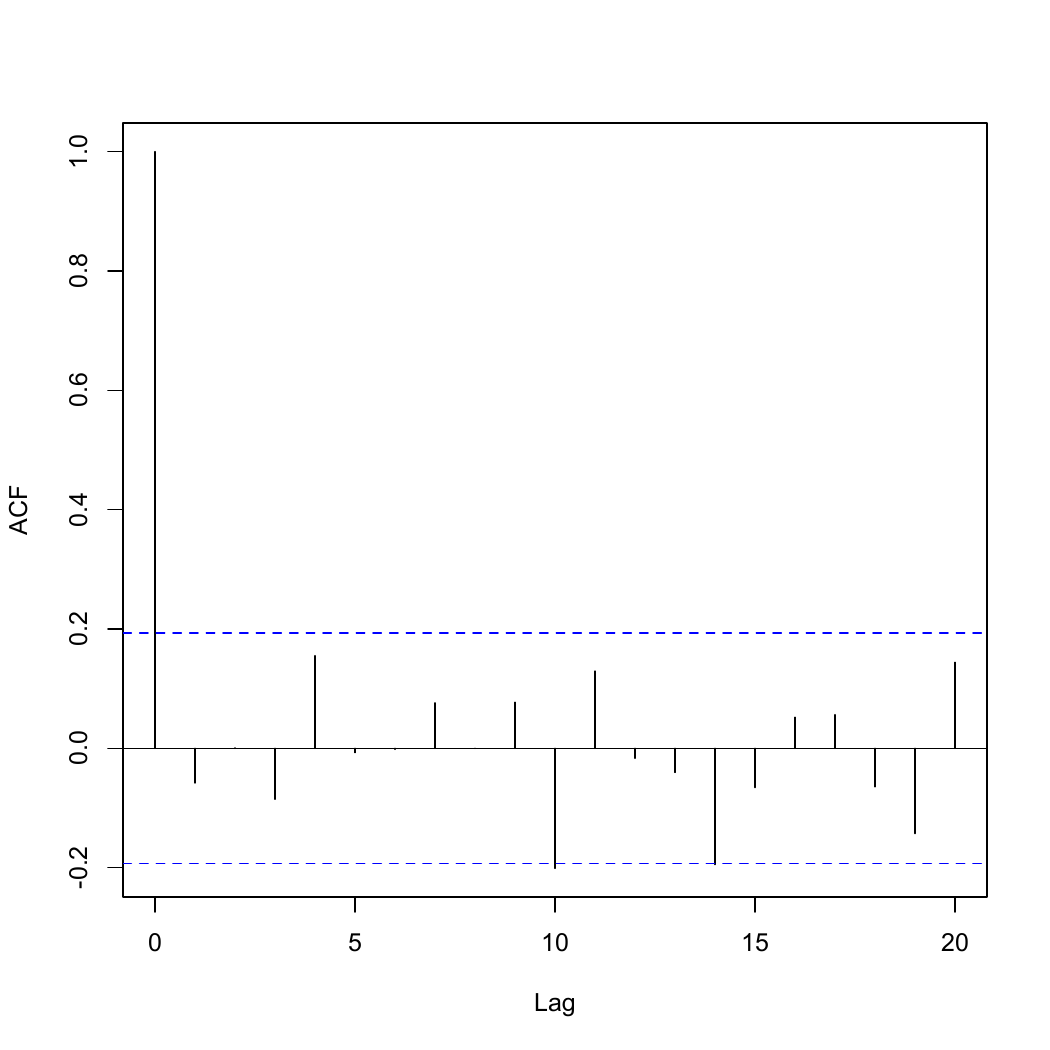}
\end{minipage}
\caption{Left: scaled interest rate data (black) with a change-point fit obtained in R package {\tt breakfast} (red); middle: residuals from the fit; right: their sample acf. \label{fig:rate_resid}}
\end{figure}

\begin{figure}
\centering
\begin{minipage}{.45\textwidth}
  \centering
  \includegraphics[width=\linewidth]{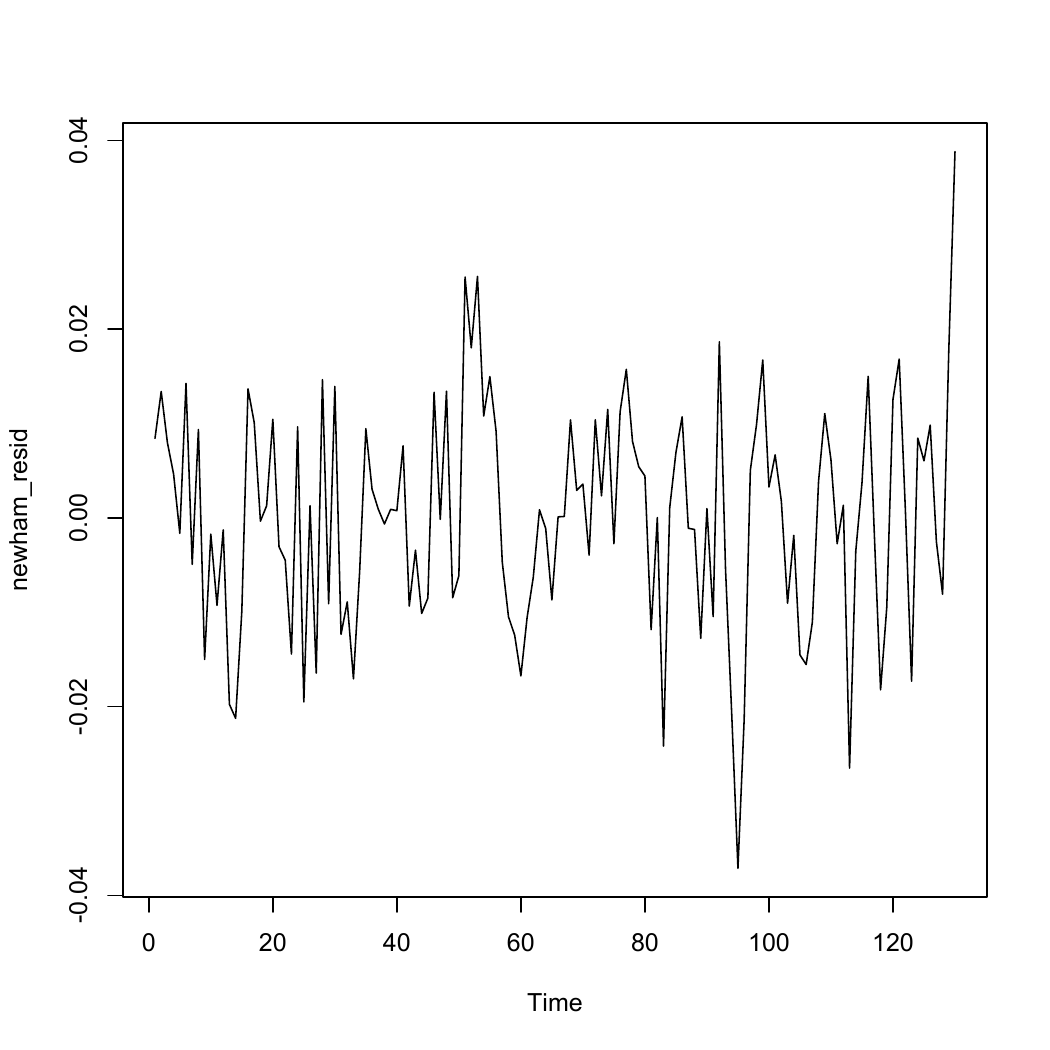}
\end{minipage}
\begin{minipage}{.45\textwidth}
  \centering
  \includegraphics[width=\linewidth]{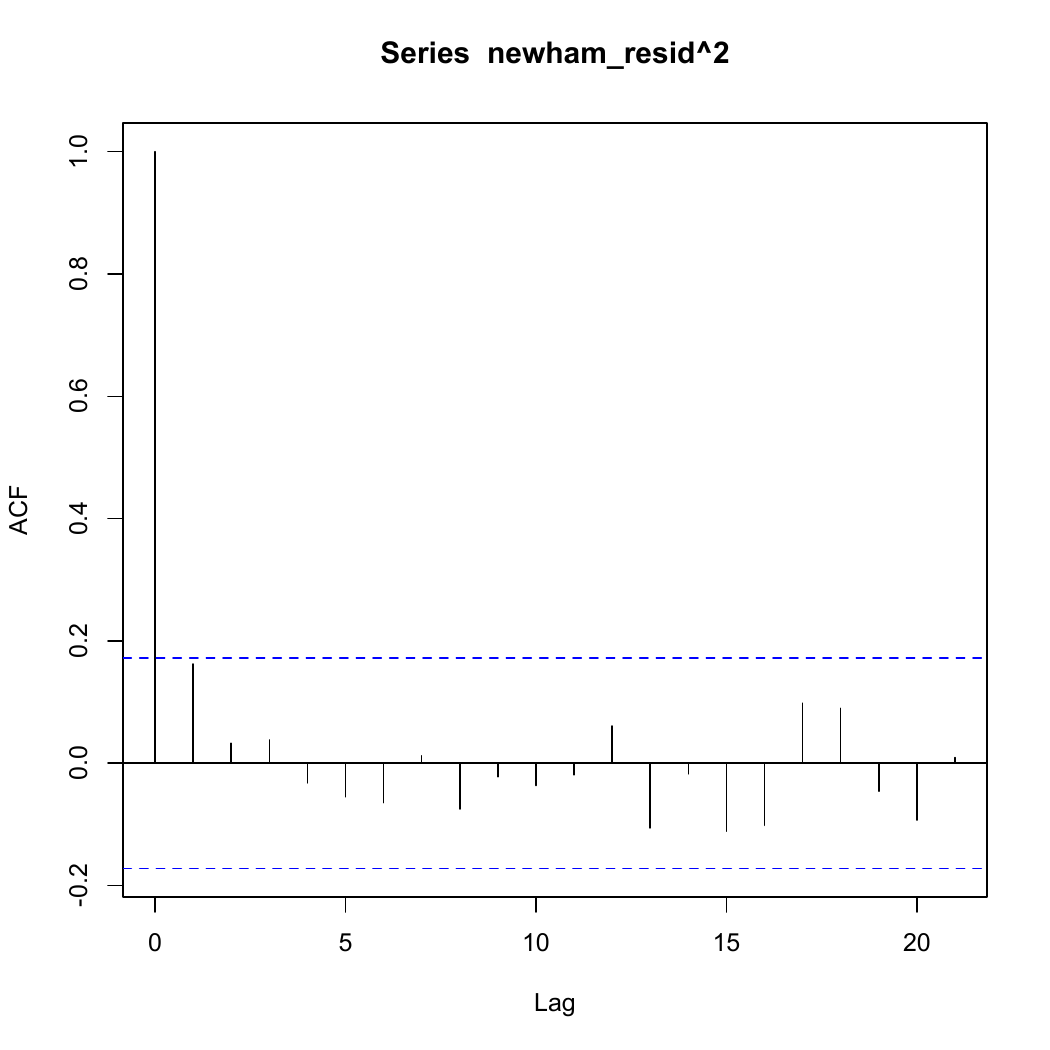}
\end{minipage}
\caption{Left: concatenated residuals from two linear regression fits, before and after the change-point (time $t=60$, as in the paper), in the Newham house price data; right: the sample acf of their squares. \label{fig:newham_resid}}
\end{figure}

In this section, we show that the application of NSP to the real-data examples of Section \ref{sec:data} of the main paper is justified as the errors do not exhibit significant serial correlation in the interest rate case or conditional heteroskedasticity in the price series case. Figure \ref{fig:rate_resid} demonstrates this for the interest rate data (note NSP was used on the scaled data shown in Figure \ref{fig:rate_resid}, where the scaling had been performed to remove heteroscedasticity). Figure \ref{fig:newham_resid} shows this for the Newham house price data example (the presence of significant autocorrelation in the squared empirical residuals could have been indicative of heteroscedasticity).

\section{Discussion}
\label{sec:disc}

We conclude with a brief discussion of a few speculative aspects of NSP.

\paragraph{Possible use of NSP in online monitoring for changes}

NSP can in principle be used in the online setting, in which `alarm' should be raised as soon as $Y$ starts deviating from linearity with respect to $X$. In particular,
consider the following simple construction: having observed $(Y_t, X_t)$, $t = 1, \ldots, T$, successively run NSP on the intervals $[T-1, T]$, $[T-2, T]$, \ldots, until either the first interval of significance is discovered, or $[1,T]$
is reached. This will provide an answer to the question of whether the most recently observed data deviates from linearity and if so, over what time interval.

\paragraph{Using and interpreting NSP in the presence of gradual change}

If NSP is used in the absence of change-points but in the presence of gradual change, obtaining a significant interval means that it must (at global significance
level $\alpha$) contain some of the period of gradual change. However, this does not necessarily mean that the entire period of gradual change is contained within the given interval of significance. Note that this is the situation portrayed in Section \ref{sec:pl} of the main paper, in which the simulation model used is a `gradual change' model from the point of view of the NSP$_0$ method, but an `abrupt change' model from the point of view of NSP$_1$ and NSP$_2$.

\paragraph{Possible use of NSP in testing for time series stationarity}

It is tempting to ask whether NSP can serve as a tool in the problem of testing for second-order stationarity of a time series. In this problem, the response $Y_t$ would be the time series in question, while the covariates $X_t$ would be the Fourier basis. The performance of NSP in this setting will be reported in future work.

\paragraph{Does the principle of NSP extend to other settings?}

NSP is an instance of a statistical procedure which produces intervals of significance (rather than point estimators) as an output. It is an interesting open question to what extent this emphasis on ``intervals of significance before point estimators'' may extend to other settings, e.g. the problem of parameter inference in high-dimensional regression.

\section*{Acknowledgements and disclosure of interests}
I wish to thank Yining Chen, Paul Fearnhead, Shakeel Gavioli-Akilagun, Zakhar Kabluchko and David Siegmund for helpful discussions. Research partially supported by EPSRC grant EP/V053639/1. There are no competing interests to declare.

\bibliographystyle{plainnat}
\end{document}